\documentstyle[aps,eqsecnum,twocolumn,amsfonts,amssymb,prd,epsfig,amsmath]{revtex}

\begin{document}
\title{Stochastic Theory of Relativistic Particles Moving in a Quantum Field: I.
Influence Functional and Langevin Equation}
\author{Philip R. Johnson\thanks{%
Electronic address: {\tt philipj@physics.umd.edu}} and B. L. Hu\thanks{%
Electronic address: {\tt hub@physics.umd.edu}}}
\address{Department of Physics, University of Maryland\\
College Park, Maryland 20742-4111}
\date{May 31, 2001}
\maketitle

\begin{abstract}
We treat a relativistically moving particle interacting with a quantum field
from an open system viewpoint of quantum field theory by the method of
influence functionals or closed-time-path coarse-grained effective actions.
The particle trajectory is not prescribed but is determined by the
backreaction of the quantum field in a self-consistent way. Coarse-graining
the quantum field imparts stochastic behavior in the particle trajectory.
The formalism is set up here as a precursor to a first principles derivation
of the Abraham-Lorentz-Dirac (ALD) equation from quantum field theory as the
correct equation of motion valid in the semiclassical limit. This approach
also discerns classical radiation reaction from quantum dissipation in the
motion of a charged particle; only the latter is related to vacuum
fluctuations in the quantum field by a fluctuation-dissipation relation,
which we show to exist for nonequilibrim processes under this type of
nonlinear coupling. This formalism leads naturally to a set of Langevin
equations associated with a generalized ALD equation. These multiparticle
stochastic differential equations feature local dissipation (for massless
quantum fields), multiplicative noise, and nonlocal particle-particle
correlations, interrelated in ways characteristic of nonlinear theories,
through generalized fluctuation-dissipation relations.
\end{abstract}

\section{Introduction}

\subsection{Particles and fields}

Charged particles moving in a quantum field is an old topic in
electromagnetic radiation theory, plasma physics and quantum/atom optics.
Interesting features include the relation between quantum fluctuations and
radiation reaction \cite{Atom-Quantum-Optics}, and, for the case of
uniformly accelerated detectors, thermal radiance in the detector (known as
the Unruh effect \cite{Unruh76(Original-discovery-Unruh-effect)}). For
relativistic particles there are also bremsstrahlung, synchrotron radiation,
and pair creation. The motion of congruences of charged particles shows up
in particle beam, plasma and nuclear (e.g. heavy-ion collision) physics.
Their collective or transport properties must be treated by additional
statistical mechanical considerations beyond those applied to single
particles.

Theoretically, since particle-field interaction is in principle described by
quantum field theory, one may get the impression that ordinary field
theoretic methods are both necessary and sufficient for the study of this
problem. Quantum field theory, in the way we usually learn it, is
customarily formulated in the context of, and with special emphasis on, how
to answer questions posed in particle physics. For example, the S matrix is
constructed for calculating scattering amplitudes, a perturbation expansion
(e.g. Feynman diagram techniques) is used for theories with a small coupling
constant (e.g., QED), and even the effective action is usually based on
in-out (Schwinger-DeWitt) boundary conditions (the vacuum persistence
amplitude). But, when one wants to find the evolution equation of a particle
one needs an in-in boundary condition and an initial value formulation of
quantum field theory \cite{Schwinger-Keldysh,CalzettaHu87(CTP/backrection)}.
This is the starting point for constructing theories describing the {\em %
nonequilibrium dynamics} of many-particle systems \cite{Calzetta-Hu-QKFT88}
where both the statistical mechanics depicting the collective behavior of a
congruence of particles as well as the statistical mechanical properties of
interacting quantum field theory need to be included in our consideration 
\cite{Calzetta-Hu-cddn,Calzetta-Hu(00):stobol}.

The relationship between the contrasting paradigms of particles and fields
is at the heart of our inquiry. The concept of a particle is very different
from that of a quantum field: a particle moving in real space cannot be
completely described by, say, a number representation in Fock space, which
underlies the second quantization formulation of canonical field theory.
Background field decomposition into a classical field and its quantum
fluctuations is a useful scheme, where one can follow the development of the
classical background field with influences from the quantum fluctuations.
While these methods have been applied extensively to the semiclassical
evolution of quantum fields, the classical picture of particle motion is
still quite remote. The Feynman path integral formulation (for particles
worldlines as opposed to fields) makes it easier to introduce classical
particle trajectories as the path giving the extremal contribution to the
quantum action, the stationary phase defining one condition of classicality.
One can then add on radiative corrections by carrying out the loop
expansion. This approach puts a natural emphasis on particle trajectories
rather than scattering amplitudes between momentum eigenstates. But how does
one go further to incorporate or explain stochastic behavior of the
particle? We shall see that when the coupling between the particle and the
field is non-negligible, backreaction of the field (with all its activities
such as vacuum polarization, pair creation, etc.) on the particle (beyond
simple radiative corrections) imparts stochastic components to its classical
trajectory.

At a deeper level, which we will tend to in the next series of papers, even
a simple quest to understand the detailed behavior of quantum relativistic
trajectories (i.e. worldlines) is a complex challenge, with problems
connected with many central issues of quantum physics. For example,
understanding how a particle moves through time (or more accurately, how a
property that can be recognized as time emerges for a single particle)
provides technical and conceptual insights into the larger questions
concerning the nature of time in general. Here we focus on the regime where
the notion of trajectory and particle identity is well-defined- this implies
that both particle creation/annihilation and quantum exchange statistics
(e.g. Fermi exclusion or Bose condensation effects) are of negligible
importance.

Is there a unified framework to account for all these aspects of the
problem? This is what we strive to develop in this series of papers. We see
from the above cursory queries that to address this deceptively simple
problem in its full complexity one cannot simply apply the standard textbook
recipes of quantum field theory. Even a ``simple'' problem such as particles
moving in a quantum field, when full backreaction is mandated, involves new
ways of conceptualization and formulation. In terms of new
conceptualization, it requires an understanding of the relation of quantum,
stochastic, classical (as ingrained in the process of decoherence)
behaviors, and of the manifestation of statistical mechanical properties of
particles and fields, in addressing where dissipation and fluctuations arise
and how the correlations of the quantum field enter. In terms of new
formulation, it involves the adaptation of the open-systems concepts and
techniques in nonequilibrium statistical mechanics, and an initial value
(in-in) path integral formulation of quantum field (and quantized worldline)
theory for deriving the evolution equations. We will discuss these two
aspects to highlight the basic issues involved in this problem.

In this investigation we take a microscopic view, using quantum field theory
as the starting point. This is in variance to, say, starting at the
stochastic level with a phenomenological noise term often-times put in by
hand in the Langevin equation. We want to give a first-principles derivation
of moving particles interacting with a quantum field from an open-systems
perspective. A consequence of coarse-graining the environment (quantum
field) is the appearance of noise which is instrumental to the decoherence
of the system and the emergence of a classical particle picture. At the
semiclassical level, where a classical particle is treated self-consistently
with backreaction from the quantum field, an equation of motion for the {\it %
mean} coordinates of the particle trajectory is obtained. This is identical
in form to the classical equations of motion for the particle since
higher-order quantum effects, which arise when there are nonlinear
interactions, are suppressed by decoherence.

Backreaction of radiation emitted by the particle on the particle itself is
called {\it radiation reaction}. For the special case of uniform
acceleration it is equal to zero. This well-known, but at first-sight
surprising, result is consistent owing to the interplay of the so-called
acceleration field and radiation field \cite{Classical-Radiation-Reaction}.
Radiation reaction (RR) is often regarded as balanced by {\it vacuum
fluctuations} (VF) via a fluctuation dissipation relation (FDR). This leads
to a common misconception: RR exists already at the classical level, whereas
VF is of quantum nature. There is, as we shall see in this paper,
nonetheless a FDR at work balancing quantum dissipation (the part which is
over and above the classical radiation reaction) and vacuum fluctuations.
But it first appears only at the stochastic-level, when self consistent
backreaction of the {\it fluctuations} in the quantum field is included in
our consideration. In addition to providing a source for decoherence in the
quantum system making it possible to give a classical description such as
particle trajectories, fluctuations in the quantum field are also
responsible for the added {\it dissipation} (beyond the classical RR), and a
stochastic component in the particle trajectory (beyond the mean). Their
balance is embodied in a set of generalized fluctuation-dissipation
relations, the precise conditions for their existence we will demonstrate in
this paper.

Ultimately a satisfactory description of the particle-field system would
have to come from a full quantum theory treatment. One needs to stipulate
and demonstrate clearly what successive approximations one introduces to
some larger interacting quantum system will enable us to begin to see the
precursor or progenitor of the particle, and the residual or background
quantum field, depicted so simply in the end as the particle-field system in
the conventional field theory idiom. The imprint of the particle's motion is
left in the altered correlations of the quantum field \cite
{RavalHu:StochasticAcceleratedDetectors}.

Later we shall see that while the problem of the nonequilibrium quantum
dynamics of relativistic particles (or fields) is important in its own
right; the rich physics contained is also closely connected with other
problems of fundamental significance such as Quantum Gravity and String
theory, including semiclassical gravity and quantum fields in black hole and
early universe spacetimes. The remainder of this introduction summarizes the
main ideas in this work and places it in the diverse range of prior research
on this subject matter. We try to present a viewpoint and method which may
provide a comprehensive and unified account while pointing out specific
areas requiring further attention.

\subsection{Quantum, Stochastic and Semiclassical Regimes}

Our treatment in this first series emphasizes the semiclassical level and
stochastic regimes. Our approach is sharply distinguished from the case
where the motion of the particle is prescribed, i.e., a stipulated
trajectory, that is characteristic of many treatments of
`particle-detectors' in quantum fields. Here, the trajectory is determined
by the quantum field in a self-consistent manner. In the former case, there
is a tacit presence of an agent which supplies the energy to keep the
trajectory on a prescribed course, its effect showing up in the radiation
given off by a detector/particle undergoing, say, acceleration. The field
configuration would have to adjust to this prescribed particle motion
accordingly, without affecting the particle motion. In the latter case, the
only source of energy sustaining the particle's motion arises from the
quantum field, and both the field and particle adjust to each other in a
self-consistent manner. The former case has been treated in detail by Raval,
Hu, and Anglin (RHA) \cite{RavalHu:StochasticAcceleratedDetectors} (see
references therein for prior work) using the concept of quantum open-systems
and the influence functional technique. We investigate the second class of
problems now with the same open-system methodology.

A closed quantum system can be partitioned into several subsystems according
to the relevant physical scales. If one is interested in the details of one
such subsystem, call it the distinguished (relevant) system, and decides to
ignore certain details of the other subsystems, comprising the environment,
the distinguished subsystem is thereby rendered as an open-system \cite
{HuPhysica}. The overall effect of the coarse-grained environment on the
open-system can be captured by the influence functional technique of Feynman
and Vernon \cite{Feynman-Vernon-Hibbs}, or the closely related
closed-time-path effective action method of Schwinger and Keldysh \cite
{Schwinger-Keldysh}. These are initial value formulations. For the model of
particle-field interactions we study, this approach yields an exact,
nonlocal, coarse-grained effective action (CGEA) for the particle motion 
\cite{Hu91-silarg}. The CGEA may be used to treat the complete quantum
dynamics of interacting particles. However, only when the particle
trajectory becomes largely well-defined (with some degree of stochasticity
caused by noise) as a result of effective decoherence due to interactions
with the field can the CGEA be meaningfully transcribed into a stochastic
effective action, describing stochastic particle motion \cite
{RavalHu:StochasticAcceleratedDetectors,Gell-MannHartle93(DecoherenceFunctionalEquationsM,Hu91(Tsukuba)}%
.

The effect of the environment (the coarse-grained subsystems) on the system
(the distinguished subsystem) is known as backreaction. One form of
classical backreaction in the context of a moving charged particle in a
quantum field is radiation reaction which exerts a damping effect on the
particle motion. As remarked before, it should not be mistaken to be
balanced directly by an FDR with vacuum fluctuations of the quantum field.
The latter induce quantum dissipation and are instrumental to decohering the
classical particle. Let us now further examine the role of dissipation,
fluctuations, noise, decoherence \cite
{Calzetta-Hu-cddn,Calzetta-Hu(00):stobol} and the relation between quantum,
stochastic and classical behavior \cite
{Gell-MannHartle93(DecoherenceFunctionalEquationsM,Hu91(Tsukuba)}.

Decoherence or dephasing refers to the loss of phase coherence in the
quantum open system arising from the interaction of the systems with the
environment. Effective decoherence brings about the emergence of classical
behavior in the system which generally carries also stochastic features.
Under certain conditions quantum fluctuations in the environment act
effectively as a classical stochastic source, or noise. While noise in the
environment is instrumental in decoherence, decoherence is a necessary
condition for the appearance of a classical trajectory. In this emergent
picture of the quantum to classical transition, there is always some degree
of resultant stochasticity in the system dynamics \cite
{Gell-MannHartle93(DecoherenceFunctionalEquationsM}. When sufficiently
coarse-grained descriptions of the microscopic degrees of freedom are
considered, nearly complete decoherence leads to negligible noise, and the
classical description of the world is complete. Moving back from
classicality towards a description of more finely-grained histories\footnote{%
The finest-grained histories are just the skeletonized paths in the path
integral formulation. Coarse-graining is achieved by integrating over
subsets of these paths.}, the quantum effects suppressed by decoherence on
macroscopic scales reveal themselves in the emergence of stochasticity. In
this realm, decoherence, noise, dissipation, particle creation, and
backreaction are seen as aspects of the same basic quantum-open-system
processes \cite
{Gell-MannHartle93(DecoherenceFunctionalEquationsM,Hu91(Tsukuba)}. If
`minimal' additional smearing is not adequate to decohere the particle
trajectories, the stochastic limit is not physically meaningful because the
quantum interference between particle histories continues to play a
significant role. Even without a field's presence as an environment, the
particle's own quantum fluctuations (represented by its higher order
correlation functions) may be treated as an effective environment for the
lower-order particle correlation functions (particularly the mean
trajectories), so that a stochastic regime may still be realized when
appropriately coarse-grained histories are considered \cite
{Calzetta-Hu-cddn,Calzetta-Hu(00):stobol,Hu91(Tsukuba)}. When averaged
descriptions are considered such that the coarse-grained set of histories
has substantial inertia, the quantum-fluctuation induced noise is
negligible, and one moves from the stochastic to semiclassical domains. Note
that there is now growing recognition that the transition from quantum to
classical behavior \cite
{Gell-MannHartle93(DecoherenceFunctionalEquationsM,Joos-Zeh-etc} is
characterized by a rich diversity of scales and phenomena, and it should
therefore not be thought of as a single transition, but rather a succession
of regimes that may, or may not, be well-separated in practice \cite
{Hu94(Time.asymetry),DowkerKent}.

The view of the emergence of semiclassical solutions as decoherent histories 
\cite{Gell-MannHartle93(DecoherenceFunctionalEquationsM} also suggests a new
way to look at the radiation-reaction problem for charged particles. The
classical equations of motion with backreaction are known as the
Abraham-Lorentz-Dirac (ALD) equations \cite{LD-Equation}. The solutions to
the ALD equations have prompted a long history of controversy due to such
puzzling features as pre-accelerations, runaways, non-uniqueness of
solutions, and the need for non-Newtonian initial data \cite
{Classical-Radiation-Reaction}. It has long been felt that the resolution of
these problems must lie in quantum theory. But, this still leaves open the
questions of when, if ever, the Abraham-Lorentz-Dirac equation appropriately
characterizes the classical limit of particle backreaction; how the
classical limit emerges; and what imprints the correlations of the quantum
field environment leave. We show that these questions, and the traditional
paradoxes, both technical and conceptual, can be resolved in the context of
the initial value quantum open system approach.

Indeed, since every possible fine-grained history is included in the path
integral for the quantum evolution, there is no a priori reason to reject
fine-grained runaway solutions as unphysical, nor is there any sense in
which a particular fine-grained history pre-accelerates, since the
fine-grained paths that appear in the path integral aren't causally
determined by earlier events in any case. From this perspective, the
appropriate questions to address are, which `quasi-classical' coarse-grained
solutions decohere more readily. Certainly it would be strange if runaway
(or pre-accelerating) decoherent coarse-grained histories occurred with any
appreciable associated probability. So the question that should be asked, in
the context of how classical solutions arise from the quantum realm, is
whether decohered particle-histories are 1) solutions to the ALD equation,
2) unique and runaway free, and 3) without preacceleration on scales larger
than the coarse-graining scale at which the trajectories decohere. We shall
see that, in fact, the semiclassical solutions do satisfy these criteria,
and therefore, one is entirely justified in using the ALD equation in the
classical limit. One also sees that because this approach describes
coarse-grained histories, the ALD equation as an approximation fails in the
finest-grained quantum limit.

Therefore, a fundamental understanding of the stochastic and semiclassical
limits must be set in the context of the full quantum theory of particles
and fields. In the second series of papers, we further develop the worldline
plus field framework to describe fully quantized relativistic particles in
motion through spacetime interacting with quantum fields and highlight its
semiclassical, stochastic and quantum features. This framework for
understanding relativistic systems is important because, in the
nonequilibrium dynamics of real particles, the localized nature of the
particle state is a prominent characteristic of the semiclassical limit, and
this fact is not most naturally described by the usual perturbative field
theory in momentum space. Furthermore, interaction, correlation, measurement
and decoherence invariably take place in both space and time. The incorrect
treatment of ``measurement'' as an instantaneous occurrence leads to
fundamental inconsistencies in the description of physical initial states
which continues to be a substantial obstacle to a deeper understanding of
decoherence and nonlocal correlation. For this reason, addressing the
stochastic (and quantum) limits in a relativistic framework treating space
and time covariantly is a vital element in fully understanding the
stochastic-semiclassical limit. By developing a new framework for
relativistic particle-field quantum dynamics we will show how this
alternative approach provides powerful tools for addressing both
conventional issues, and for exploring new questions.

\subsection{Coarse-graining the particle}

Our problem, as well as many others from quantum and atom optics, provides a
good example of where a quantum field (e.g. photons) acts as an environment
in its influence on an atom or electron system. There are, of course,
physical contexts for which it is more appropriate to coarse-grain the
particle degrees of freedom, whereby one obtains a coarse-grained effective
action for the field. In this complementary view, matter plays the role of
an environment for the field as a system. When both particle and field
coarse-grainings converge to mutually decoherent sets of particle and field
histories, one recovers the classical limit \cite
{Hartle-Dowker-Halliwell-Brun}. These regimes are illustrated schematically
in Figure 1, where the field degrees of freedom are denoted by $\varphi,$
and the particle degrees of freedom are denoted by $z.$

\subsection{Nonlinear Coupling}

Comparing with prior work on this subject matter using the same approach,
the most closely related being that of Raval, Hu and Anglin \cite
{RavalHu:StochasticAcceleratedDetectors}, who derived the influence
functional for n-detectors moving in a quantum field, the main distinct
feature here is that the particle trajectory is not prescribed, but is a
dynamical degree of freedom determined self-consistently by the field. As
such the coupling between the particle(s) and the field is of a more
fundamental nonlinear nature. One type of nonlinear coupling between the
system and the environment in the quantum Brownian motion models \cite
{CaldeiraLeggett,QBMII,Initial-correlations} has been considered by Hu, Paz,
and Zhang \cite{QBMII}. Their model contains nonlinear couplings that are
polynomial in the field variables, whereas here we consider the opposite
example, where the field variables are linear, but the system variables are
not. This is the case for QED, which is our ultimate goal (to be discussed
in Paper III \cite{JH3} of this series).

Our treatment of nonlinear particle-field interactions is in sharp contrast
with the existing body of work on the semiclassical and stochastic limit for
linear QBM. We find that the requirement of self-consistency, together with
the nonlinearity of the fundamental interactions, leads to a description of
the stochastic limit that is more unified and tightly constrained than that
which has been described up to now. In fact, confronting nonlinear systems
is a crucial next step in exploring the stochastic limit, and the deeper
relationship between noise, decoherence, and the intricate evolution of
correlations for systems with many degrees of freedom. Most studies of
decoherence have invoked linear systems with an (externally) predetermined
basis of states which are then shown to decohere. But a complete description
of decoherence must include the mechanism by which a decoherent basis of
states is self-consistently (e.g. not externally) selected. Understanding
this mechanism as a dynamical feature of nonlinear theories is an important
problem that cannot be fully addressed within linear models.

Another important difference between linear and nonlinear theories is in
their respective equations of motion for correlation functions. For linear
theories, these are just given by the classical equations of motion, and the
quantum {\it evolution} may be reproduced by a probabilistic description of
initial conditions. An example of this is the Wigner function evolution for
linear systems with initially positive definite distributions in phase
spaces. The equations of motion are then the classical ones, and negative
values of the Wigner function never evolve in the future. One may then view
the theory as dynamically equivalent to classical statistical dynamics
(though of course the {\it meaning }of the Wigner functions is still quantum 
\cite{CalzettaHabibHu:QKineticFieldTheory}). This fact is what makes the
construction of a stochastic limit that bridges linear classical and quantum
theories fairly straightforward. The same is not true of nonlinear theories.
Their quantum equations of motion (even for the mean-field) are not
equivalent to the classical ones. This implies that there are important {\it %
qualitative }differences in the quantum dynamics of linear versus nonlinear
theories that are significant even in the semiclassical regime, and that
will be missed by the study of linear systems alone.

The important result in this first paper is the derivation of
self-consistent Langevin equations for relativistic particle motion in a
quantum field starting from relativistic quantum mechanics. A stochastic
description of the limit of nonlinear theories requires great care, and, in
its most general form, will involve nonlinear stochastic differential
equations where the statistics of the induced particle fluctuations are {\it %
externally} determined by the quantum statistics of the fields.

Because there are inherent dangers in a phenomenological approach to
nonlinear stochastic equations, it is crucial to work from first principles
stressing the microscopic origin of fluctuations. In this work, we treat an
example of a nonlinear stochastic system in the regime where decoherence
allows the expansion around the semiclassical solution. The decoherent
semiclassical solutions need not, and usually are not, equilibrium
solutions. Hence, compared with the traditional derivation of the Langevin
equations in the context of near-equilibrium linear response theory, this
method is significantly more general.

For Langevin equations the effect of the field is registered in the
stochastic properties of the particles. The stochastic mean of the equations
of motion corresponds to the mean-field semi-classical limit; but in
addition, the symmetrized n-point correlation functions for the particles
are given by the higher order stochastic moments. Therefore, the stochastic
particles may be thought of as being ``dressed'' by the non-local statistics
of the field. These equations impart stochastic features to the particle and
field properties beyond the semiclassical limit, leading towards the quantum
domain. The stochastic equations of motion featuring nonlocal noise, causal
particle-particle interactions, and nonlocal particle-particle correlations
make the evolution highly non-Markovian (i.e. history dependent). Only in
the single particle semiclassical limit (with local dissipation and white
noise) is the evolution strictly Markovian. For well separated particles, in
the high temperature limit, the Langevin equations may be approximated by
Markovian dynamics; this is the regime in which the nonlocal role of the
quantum vacuum has been essentially washed out. But as one moves more deeply
into the stochastic regime (especially at low temperatures), the Markovian
approximation is no longer adequate. It is in this realm that our methods
are markedly distinct from the Hamiltonian methods more commonly used in the
derivation of Markovian master equations.

\subsection{Prior work in relation to ours}

The treatment of a quantum field as a bath of harmonic oscillators has a
long history. Many authors take a semi-phenomenological approach in adding a
noise to the quantum equation of motion by hand to get the Quantum Langevin
equations (QLE). This practice works reasonably well in the linear response
regime for a system in equilibrium, but can otherwise violate the
fluctuation-dissipation relation and bring about pathologies (such as
a-causal evolution). There are many ways to derive the QLE describing the
dissipative dynamics of a quantum system in contact with a quantum
environment, such as modified (Heisenberg picture) Hamilton equations of
motion or path integrals. Caldeira and Leggett's revival of the
Feynman-Vernon influence functional in the study of quantum Brownian motion
(QBM) has led to an extensive literature on QBM \cite{CaldeiraLeggett},
particularly in regard to decoherence \cite{QBMII,QBM(Decoherence)}.

Ford, Lewis, and O'Connell have done systematic and comprehensive work on
the problem of charge motion in an electromagnetic field as a thermal bath
in the linear (dipole coupling) regime \cite{FOL}. They have detailed the
conditions for causality and a good thermodynamic limit, and have further
used the QLE paradigm to find stochastic equations of motion for
non-relativistic charged particles in the equilibrium limit. A crucial point
of their analysis is that the solutions can be (depending on the cutoff of
the field spectral density) runaway free and causal in the late-time limit 
\cite{FO(Runaways)}. In \cite{FO(FDR)}, they suggest a form of the equations
of motion that give fluctuations without dissipation for a free electron,
but this result arises from the particular choice they made of a field
cutoff. In contrast, we take the effective theory point of view which
emphasizes the typical insensitivity of low-energy phenomena to unobserved
high-energy structure. The following two papers in this series make clear
that a special value for the cutoff is unnecessary for consistency of the
low-energy behavior\footnote{%
Though there is a maximum value of the cutoff beyond which the dynamics
become unstable.}. Further distinctions are briefed as follows: First, our
general method extends to the nonequilibrium regime, allowing consideration
of specific initial states at a finite time in the past. Second, we do not
make the dipole approximation ab initio because we are especially interested
in the nonequilibrium dynamics of nonlinear systems. When our equations of
motion are linearized, we pay special attention to the constraints that the
original nonlinear theory imposes. Third, by adopting the worldline
quantization framework, we are able to derive equations of motion from fully
relativistic quantum mechanics. Fourth, we highlight the role of decoherence
in the emergence of a stochastic regime. Fifth, we show how nonlocality may
still be a feature in the stochastic regime, unless sufficient noise washes
out the nonlocal correlations between separate particles.

Our results also represent an extension of Barone and Caldeira's analysis of
decoherence for an electron in an quantum electromagnetic field \cite
{BaroneCaldeira}. Their work employs the path integral method except they
are limited to non-relativistic particles and dipole coupling, and it
focuses mainly on the reduced density matrix. Rather we give a full
description of the stochastic dynamics of this nonlocal nonlinear theory. An
advantage of Barone and Caldeira's work is that it is not limited to
initially factorized states; they use the preparation functional method
which allows the inclusion of initial particle field correlations. Despite
this, Romero and Paz have pointed out that the preparation function method
still suffers from an unphysical depiction of the initial state \cite
{Romero.Paz96(QBM.with.Initial.Correlations)}. For this reason, a completely
satisfactory treatment of particle decoherence in a field-environment that
accounts for particle-field correlations in more realistic (i.e. physically
prepared) states remains to be given.

Using the influence functional, Di\'{o}si \cite{Diosi} derives a Markovian
master equation in non-relativistic quantum mechanics. In contrast, it is
our intent to emphasize the non-Markovian and nonequilibrium regimes with
special attention paid to self-consistency. The work of \cite{Diosi} differs
from ours in the treatment of the influence functional as a functional of
particle trajectories in the relativistic worldline quantization framework.
Ford has considered the loss of electron coherence from vacuum fluctuation
induced noise with the same noise kernel that we employ \cite
{Ford(ElectronCoherence)}. However, his application concerns the case of
fixed or predetermined trajectories.

\subsection{Organization, notations and units}

In Section 2, we begin with a review on how a quantum field can be treated
as a bath of harmonic oscillators and we connect with the well-studied
quantum Brownian motion model (QBM) and quantum Langevin equations (QLE) 
\cite
{Feynman-Vernon-Hibbs,CaldeiraLeggett,QBMII,FOL,BaroneCaldeira,HuMataczQBMParametricOsc94,CaldeiraWeissKleinert}%
. We write down the form of the influence functional (IF) for nonlinear
particle-field interactions, and we further review the influence functional
formalism in Appendix A, where we discuss the necessary assumptions for
deriving an evolution propagator for the kind of hybrid particle/field model
that we employ. Underlying the formulation in Appendix A is the worldline
quantization method for relativistic particles where the particle coordinate
is the relevant particle degree of freedom. Use of this framework, rather
than the conventional quantum field description for the charged particles,
is central to this work. However, the detailed development of this approach
is not needed for the semiclassical and stochastic regimes considered in
this first series.

In Sec. 3, we derive the coarse-grained effective action (CGEA). The
background material in our approach consists of the Schwinger-Keldysh
closed-time-path (CTP) effective action applied to an open system, which
results in a CTP CGEA that is closely related to the influence action in the
Feynman-Vernon influence functional method. For easy access and
identification of notations and conventions, we review these necessary
formalisms in the Appendices, which can be read independently. These
techniques provide a general and powerful framework for studying
nonequilibrium quantum processes, especially for non-Markovian dynamics,
which are prevalent when backreaction of the environment on the system is
fully and correctly accounted for. We introduce the reduced-density-matrix
for a system after assuming a system-environment partition. The reduced
density matrix evolution operator is found in terms of the IF, and the CGEA.
In Sec. 4, we define the stochastic effective action, and show the close
relationship of noise, dissipation, and decoherence in the stochastic limit.
In Sec. 5, we discuss stochastic fields. Sec. 6 presents a general (i.e.,
not restricted to a specific model of the particles or field) derivation of
the stochastic equations in the form of nonlinear Langevin equations for
general nonlinear particle-field coupling. We discuss when the traditional
linear Langevin equations may be recovered. In Sec. 7 we derive a
generalized fluctuation-dissipation relation for the noise and
radiation-reaction in these expressions. This provides the necessary
framework of a stochastic theory approach to investigate relativistic
particle motion in quantum fields.

In Paper II \cite{JH2}, we shall apply these results to a specific model of
relativistic particles in a scalar field. We derive the influence functional
and stochastic effective action for this nonlinear model, and then discuss
in detail the semiclassical limit. The main result is the derivation of
equations of motion for the semiclassical limit that, to lowest order, are
modified (time-dependent) Abraham-Lorentz-Dirac (ALD) equations for
charged-particle radiation reaction. This work demonstrates that the ALD
equation is a good approximation to the semiclassical limit of scalar QED,
in the regime where the particles are effectively classical. This derivation
is more general, however, and reveals the time-dependent renormalization and
radiation-reaction typical of nonequilibrium quantum field theory. These
features play a crucial role in establishing the consistency of this limit
by showing that runaway solutions and acausal effects do not occur when the
field is suitably regulated\footnote{%
This demonstration of causality addresses the early-time, nonequilirium
setting rather than the late-time equilibrium limit that is analyzed in \cite
{FO(Runaways)}. At late times, the conclusions in \cite{FO(Runaways)} also
apply to the linearized version of our results.}. We then derive
(time-dependent) Abraham-Lorentz--Dirac-Langevin (ALDL) equations, which
describe the fluctuations (and radiation reaction) of relativistic particles
in the stochastic limit. We use these results to explore the question of
Brownian motion for a free particle in a quantum scalar field.

In Paper III \cite{JH3}, we extend these results to the electromagnetic
field by deriving the influence functional for QED in the Lorentz gauge,
from which we find the interacting photon-particle stochastic action. In the
semiclassical limit we obtain a modified, time-dependent, ALD solution for
QED. We then demonstrate a direction dependent Unruh effect with the vacuum
fluctuations of the field appearing thermal (at the Unruh temperature) for
charged particles in a constant electric field. This general approach
clearly illustrates the distinct natures of ALD radiation reaction, the
associated ordinary radiation into infinity, and the Unruh effect, while at
the same time deriving all these effects in a unified and self-consistent
manner. As an interesting application, the Langevin equation may be viewed
as a model for a stochastic particle event horizon, with implications for a
nonequilibrium treatment of black hole Hawking radiation and horizon
fluctuations.

Our second series of papers (IV and V \cite{JH4-5}) will be on relativistic
quantum particle-field interaction. It is here that the pre-cursors of the
quantized worldline framework introduced in Appendix A are fully developed.
In Paper IV, we `first' quantize the free relativistic particle using the
path integral method. The worldline path integral representation is used to
construct the ``in-out'' and ``in-in'' relativistic particle generating
functional for worldline coordinate correlation functions. In Paper V, we
use the influence functional to construct an interacting quantum theory of
both relativistic particles and fields. This involves the introduction of a
nonlocal worldline kernel that provides a route for exploring the role of
correlation, dissipation, and nonequilibrium open-system phenomena in
particle creation and other relativistic processes. This work is also a
model for how open-system methodology may be introduced into String theory
since the worldline path integral is the point particle limit of the String
case. This formalism provides a powerful approach to a new set of
quantum-relativistic-statistical processes where one may simultaneously take
advantage of the influence functional and first-quantization (worldline)
techniques.

We shall usually set $c=1$, but keep $\hbar$ explicit as a marker for
quantum effects. The metric tensor $g_{\mu\nu}$ is diag$(+1,-1,-1,-1)$.
Greek letters will (usually) denote spacetime indices, and Latin letters $%
a,b,c$ will indicate the CTP time-branch (see Appendix C). The mixed
function/functional notation $f[z;x)$ will be used when $f$ is a functional
of $z,$ but a function of $x.$ The Einstein summation convention is employed
except when otherwise noted. Particle degrees of freedom will be
collectively denoted $z=\left\{ z_{n}^{\mu}\left( \tau_{n}\right) \right\} $
with $\tau_{n}$ as their worldline parameters. The path integral measures
are denoted $Dz$ and $D\varphi,$ and are defined in Appendix A.

\section{The effective action, effective field theory, and particle reduced
density-matrix}

Particles moving in quantum fields may be cast in the form of nonlinear
quantum Brownian motion. In this paper, a set of $N$ particles (with action $%
S_{z})$ constitutes the system and a scalar field $\hat{\varphi}\left(
x\right) $ (with the action $S_{\varphi})$ constitutes the environment. The
particles and field interact nonlinearly as determined by an interaction
action $S_{int}.$ The n$^{th}$ particle degree of freedom is its spacetime
trajectories $z_{n}^{\mu}\left( \tau_{n}\right) $ parametrized by $\tau_{n}$%
, which is not necessarily the particle's proper time. In this first series,
our focus is on the stochastic dynamics of the particle worldlines; for
simplicity, we neglect additional particle degrees of freedom such as spin.
We also work in the regime where particles have approximately well-defined
(emergent) trajectories, though with statistical fluctuations arising from
the particles' contact with the quantum field. Consequently, the forces that
arise from spin-statistics (e.g. Fermi-exclusion or Bose-condensation) are
negligible. We discuss how these quantum effects are suppressed by
decoherence (the same decoherence that gives the particles stochastic
trajectories) in our second series.

In this relativistic treatment, both the particle's space and time
coordinates will emerge as stochastic processes, with correlation functions
of the form $\langle z_{1}^{\mu }\left( \tau _{1}\right) ...z_{N}^{\nu
}\left( \tau _{N}\right) \rangle _{s}.$ The notation $\langle \rangle _{s}$
denotes the stochastic average with respect to a derived probability
distribution. The statistics of these processes are found from the
underlying quantum statistics of the quantum field acting as an environment.
The semiclassical regime is characterized by the particle trajectories being
well-defined classical variables. The stochastic regime is intermediate
between the semiclassical and quantum regimes, where the particle worldlines
are classical stochastic processes.

The quantum open-system of particle degrees of freedom is described by the
reduced density-matrix 
\begin{equation}
\hat{\rho}_{r}\left( t\right) =Tr_{\varphi }\hat{\rho}_{U}\left( t\right) ,
\end{equation}
where $\hat{p}_{U}\left( t\right) $ is the unitarily evolving state of the
universe (i.e. the particles plus field) at time $t.$ We assume that the
initial state (at time $t_{i})$ of the universe can be represented in the
factorized (tensor-product) form 
\begin{align}
\hat{\rho}_{U}\left( t_{i}\right) & =\hat{\rho}_{z}\left( t_{i}\right)
\otimes \hat{\rho}_{\varphi }\left( t_{i}\right)   \label{density matrix} \\
& =\int d\varphi _{i}d\varphi _{i}^{\prime }d{\bf z}_{i}d{\bf z}_{i}^{\prime
}\rho _{\varphi }\left( \varphi _{i},\varphi _{i}^{\prime }\right)  
\nonumber \\
& \times \rho _{z}\left( z_{i},z_{i}^{\prime }\right) |\varphi
_{i},z_{i}\rangle \langle \varphi _{i}^{\prime },z_{i}^{\prime }|,  \nonumber
\end{align}
where $\hat{\rho}_{z}$ and $\hat{\rho}_{\varphi }$ have been expanded in
terms of basis states $|\varphi _{i},z_{i}\rangle =|\varphi _{i}\rangle
\otimes |{\bf z}_{i},t_{i}\rangle .$ The initial particles states are the
Lorentz invariant relativistic configuration-space state defined as 
\begin{align}
|z_{i}\rangle & \equiv |{\bf z}_{i},t_{i}\rangle  \\
& =\int \frac{d^{4}p}{\left( 2\pi \right) ^{4}}\delta ^{\left( 4\right)
}\left( p^{2}-m^{2}\right) \theta \left( p^{0}\right) e^{\frac{i}{\hslash }%
p_{\mu }z^{\mu }}|p\rangle ,  \nonumber
\end{align}
where we have assumed that the initial particle state is positive frequency
(this assumption may be relaxed).

The particle's reduced density-matrix at later times is given by (see
Appendix A) 
\begin{align}
\rho _{r}\left( z_{f},z_{f}^{\prime }\right) & =\int d{\bf z}_{i}d{\bf z}%
_{i}^{\prime }J_{r}\left( z_{f},z_{f}^{\prime };z_{i},z_{i}^{\prime }\right) 
\label{Evolution of reduced density matrix} \\
& \times \rho _{z}\left( z_{i},z_{i}^{\prime }\right) .  \nonumber
\end{align}
The variable $z$ stands for the entire collection of particle coordinates $%
\left\{ z_{n}^{\mu }\right\} .$ The open-system evolution operator $J_{r}$
is given by 
\begin{align}
J_{r}\left( z_{f},z_{f}^{\prime };z_{i},z_{i}^{\prime }\right) & =\int_{{\bf %
z}_{i},{\bf z}_{i}^{\prime }}^{{\bf z}_{f},{\bf z}_{f}^{\prime
}}DzDz^{\prime }e^{\frac{i}{\hslash }\left( S_{z}\left[ z\right] -S_{z}\left[
z^{\prime }\right] \right) }  \nonumber \\
& \times F\left[ z,z^{\prime }\right] .  \label{jr}
\end{align}
$F\left[ z,z^{\prime }\right] $ is the influence functional introduced by
Feynman and Vernon \cite{Feynman-Vernon-Hibbs}. The measure $Dz$ depends on
the details of how the particle action is defined and quantized, and it may
implicitly include additional gauge degrees of freedom (such as a lapse
variable $N$). In the semiclassical/stochastic treatment presented in this
paper, we will in fact assume that the particle degrees of freedom have been
gauge-fixed so that the parameters $\tau $ are the proper-times. The
detailed treatment of the influence functional for relativistic particle
worldlines is found in our second series \cite{JH4-5}, though a brief
discussion may be found in Appendix A and in paper III \cite{JH3}.

The influence functional is given by 
\begin{align}
F\left[ z,z^{\prime }\right] & =\int d\varphi _{f}d\varphi _{i}d\varphi
_{i}^{\prime }\int_{\varphi _{i},\varphi _{i}^{\prime }}^{\varphi
_{f}=\varphi _{f}^{\prime }}D\varphi D\varphi ^{\prime }
\label{definition of influence functional} \\
& \times e^{\frac{i}{\hslash }\left( S_{\varphi }\left[ \varphi \right]
-S_{\varphi }\left[ \varphi ^{\prime }\right] +S_{int}\left[ \varphi ,z%
\right] -S_{int}\left[ \varphi ^{\prime },z^{\prime }\right] \right) } 
\nonumber \\
& \times \rho _{\varphi }\left( \varphi _{i},\varphi _{i}^{\prime
};t_{i}\right)   \nonumber \\
& =\exp \left\{ \frac{i}{\hslash }S_{IF}\left[ z.z^{\prime }\right] \right\}
,  \nonumber
\end{align}
where $S_{IF}\left[ z,z^{\prime }\right] $ is called the influence action.
Notice that the definition of the influence functional itself involves only
path integrals over the fields $\varphi ,$ and hence, does not require doing
particle worldline path integrals. Indeed, the results of this paper follow
from effectively treating the worldlines as classical variables coupled to a
quantum field, which is the usual semiclassical paradigm. In the next
section we indicate the connection between the worldlines as classical
variables and the quantum equations of motion for the quantum-average
trajectories.

Defining the coarse-grained effective action 
\begin{equation}
S_{CGEA}=S_{z}\left[ z\right] -S_{z}\left[ z^{\prime}\right] +S_{IF}\left[
z,z^{\prime}\right] ,
\end{equation}
$J_{r}$ may, alternatively, be expressed as 
\begin{equation}
J_{r}\left( z_{f},z_{f}^{\prime};z_{i},z\right) =\int_{{\bf z}_{i},{\bf z}%
_{i}^{\prime}}^{{\bf z}_{f},{\bf z}_{f}^{\prime}}DzDz^{\prime}e^{\frac{i}{%
\hslash}S_{CGEA}\left[ z,z^{\prime}\right] }.   \label{evolution kernel}
\end{equation}
The evolution kernel $J_{r}$ gives all information about the quantum
open-system dynamics of the particles.

If the particle-field interaction is characterized by a coupling constant $e$
and is linear in the field variables, the influence action is of order $%
e^{2}.$ One approximation scheme for Eq. (\ref{evolution kernel}) is to
expand the influence functional as 
\begin{equation}
e^{\frac{i}{\hslash}S_{IF}\left[ z,z^{\prime}\right] }=1+\frac{i}{\hslash }%
S_{IF}\left[ z,z^{\prime}\right] +{\cal O}\left( e^{4}\right) , 
\label{ordinary perturbation theory}
\end{equation}
giving a perturbation theory based on the order of a coupling constant. $%
J_{r}$ then describes free quantized-particle evolution punctuated by
discrete interactions with the field. This is just ordinary perturbative
quantum field theory in the ``in-in'' formulation. Unlike ``in-out'' field
theory, the lowest order (non-trivial) terms are of order $e^{2},$ which is
natural since $J_{r}$ gives the evolution of the density matrix, not the
wavefunction. But, particularly when there is a background field $\bar {%
\varphi}=\langle\hat{\varphi}\rangle,$ (\ref{ordinary perturbation theory})
does not give an accurate depiction of the classical limit of particle
trajectories, say, as given by the mean-values $\langle\hat{z}_{n}\rangle =%
\bar{z}$, where $\hat{z}_{n}$ is the quantum operator for the particle's
coordinates. The infinite set of background field terms in the expansion
from (\ref{ordinary perturbation theory}) may be summed, giving a far more
accurate approximation to the particle evolution. The resulting
re-organization of Feynman diagrams is automatically obtained by performing
a saddle point expansion of (\ref{evolution kernel}).

It is convenient to define sum and difference variables 
\begin{align}
z^{-}& =\left( z-z^{\prime }\right)  \\
z^{+}& =\left( z+z^{\prime }\right) /2.  \nonumber
\end{align}
Evaluating the evolution kernel in Eq. (\ref{evolution kernel}) for linear
systems is a standard step in the derivation of the master equation using
path integral methods. One begins by reparametrizing the worldlines defining
the fluctuation and classical worldlines, $\breve{z}_{n}$ and $\bar{z}_{n},$
respectively, by 
\begin{equation}
\breve{z}_{n}^{\pm }=\hat{z}_{n}^{\pm }-\bar{z}_{n}^{\pm },
\end{equation}
The extremal solutions to the effective action, denoted $\bar{z}_{n},$ are
defined to be the classical solutions to the real part of $S_{CGEA}\left[
z^{\pm }\right] :$ 
\begin{equation}
\left. \frac{\delta S_{CGEA}^{R}\left[ z^{\pm }\right] }{\delta z_{n}^{\pm }}%
\right| _{z^{\pm }=\bar{z}_{n}^{\pm }}=0,
\end{equation}
satisfying the boundary conditions 
\begin{equation}
{\bf \bar{z}}^{\pm }(t_{i})={\bf z}_{i}^{\pm }
\end{equation}
This definition is made because the imaginary part of the effective action
does not modify the stationary phase solution.

The evolution operator for the open system is then given by 
\begin{align}
& J_{r}\left( z_{f}^{\pm},z_{i}^{\pm}\right) =e^{\frac{i}{\hslash}%
S_{CGEA}^{R}\left[ \bar{z}^{\pm}\right] }\int_{{\bf \breve{z}}_{i}^{\pm
}=0}^{{\bf \breve{z}}_{f}^{\pm}=0}D\breve{z}^{\pm} \\
& \times e^{\frac{i}{\hslash}\int d\tau\frac{\delta S_{CGEA}^{I}}{\delta 
\bar{z}^{a}\left( \tau\right) }\breve{z}^{a}\left( \tau\right) +\frac {i}{%
2\hslash}\int\int d\tau d\tau^{\prime}\frac{\delta^{2}S_{CGEA}}{\delta \bar{z%
}^{a}\delta\bar{z}^{b}}\breve{z}^{a}\left( \tau\right) \breve{z}^{b}\left(
\tau^{\prime}\right) }  \nonumber \\
& \times\left( 1+{\cal O(}\breve{z}^{3}\right) ),  \nonumber
\end{align}
where we have factored out the ``zero-loop'' term in front, and expanded the
exponential of the non-quadratic terms in the effective action. $%
S_{CGEA}^{I} $ is the imaginary part of the effective action.

The higher-order quantum corrections depend on the solutions $\bar{z}%
_{n}^{\pm }$ unless the effective action is quadratic; and hence, the model
is purely linear. For the linear case, $J_{r}$ may be found exactly since
all the path integrals are Gaussian\footnote{%
Actually, the integrals are nearly, but not quite, Gaussian because the
limits on the time-integrals is not $\left[ -\infty ,+\infty \right] $ but
rather $\left[ t_{i},+\infty \right] ,$ where $t_{i}$ is the initial time.}.
It is given by 
\begin{equation}
J_{r}\left( z_{f}^{\pm };z_{i}^{\pm }\right) =F\left( t_{f},t_{i}\right)
\times e^{\frac{i}{\hslash }S_{CGEA}\left[ \bar{z}^{\pm }\right] }
\end{equation}
where $F$ may be found by the normalization condition 
\begin{align}
& \int dz_{f}dz_{f}^{\prime }\delta (z_{f}-z_{f}^{\prime })\left.
J_{r}\left( z_{f},z_{f}^{\prime };{\bf z}_{i}^{\pm },t_{i}\right) \right|
_{t_{f}=t_{f}^{\prime }}  \nonumber \\
& =\delta \left( z_{i}-z_{i}^{\prime }\right) .
\end{align}
From $J_{r}$ it is straightforward to find the master equation from the
expression 
\begin{equation}
\frac{\partial }{\partial t_{f}}\rho _{r}\left( z_{f}^{\pm }\right) =\int d%
{\bf z}_{i}^{+}d{\bf z}_{i}^{-}\frac{\partial }{\partial t_{f}}J_{r}\left(
z_{f}^{\pm };z_{i}^{\pm }\right) \rho _{r}\left( z_{i}^{\pm }\right) . 
\nonumber
\end{equation}
Because the one-loop (fluctuation) term is independent of $\bar{z}$ for
linear theories, the resulting noise and dissipation are independent of the
system history. It is only for linear theories that the noise induced by the
environment is completely extrinsic in this sense.

For non-linear theories, the one-loop (and higher-order) corrections will
depend on $\bar{z}^{\pm}$ and hence will modify the master equation, giving
system-history dependent (colored) noise and dissipation. But, rather than
using the master equation as our primary tool, we instead follow the
approach in \cite{RavalHu:StochasticAcceleratedDetectors,Einstein-Langevin}
for deriving quantum Langevin equations \cite{Feynman-Vernon-Hibbs}, to find
a stochastic description of the particle dynamics from the influence
functional.

\section{Influence functional for quantum scalar field environment}

We now derive the influence functional assuming a system of spinless
particles locally coupled to a scalar field. The free scalar field action is 
\begin{equation}
S_{B}[\varphi]=\int dt\,d{\bf x}\;\frac{1}{2}\left[ \left( \partial
_{t}\varphi\right) ^{2}-\left( {\bf \nabla}\varphi\right)
^{2}+m^{2}\varphi^{2}\right] .
\end{equation}
We assume that the interaction term is linear in the field variables $%
\varphi:$ 
\begin{equation}
S_{int}=\int dt\,d{\bf x}\,\,j[z_{n}\left( \tau_{n}\right) ;{\bf x,}%
t)\,\varphi(t,{\bf x}),
\end{equation}
but $j[z;{\bf x,}t)$ is an arbitrary (nonlinear) functional of the particle
coordinates $z_{n}$. The construction of a perturbation theory to derive the
influence functional for nonlinear environments (e.g. $\varphi^{4}$) has
been developed by Hu, Paz and Zhang in \cite{QBMII}. We consider couplings
that are nonlinear in the system variables, as is the case for $QED.$ For a
system interacting with a quantum field, one can follow the method
introduced by Hu and Matacz \cite{HuMataczQBMParametricOsc94} for deriving
the influence functional in terms of the amplitude functions of the
parametric oscillators which are the normal modes of the field. We expand
the field in terms of its real-valued, normal modes 
\begin{equation}
u_{\lambda k}({\bf x})=\left\{ 
\begin{array}{c}
\text{$\sin$}{\bf k}\cdot{\bf x\;;\;\;}\lambda=1 \\ 
\text{$\cos$}{\bf k}\cdot{\bf x;\;\;}\lambda=2
\end{array}
\right. ;\;\;\omega_{{\bf k}}=+\sqrt{{\bf k}^{2}+m^{2}}, 
\label{Normal modes}
\end{equation}
as 
\begin{align}
\varphi(x) & =\left( 2/L\right) ^{3/2}%
\mathop{\textstyle\sum}%
_{{\bf k}\lambda}\,\varphi_{{\bf k\lambda}}(t)\,u_{{\bf k\lambda}}({\bf x)}
\label{definition of  phi(t)} \\
& =\left( 2/L\right) ^{3/2}%
\mathop{\textstyle\sum}%
_{\alpha}\,\varphi_{{\bf \alpha}}(t)\,u_{{\bf \alpha}}({\bf x),}  \nonumber
\end{align}
where $\alpha=\left( {\bf k},\lambda\right) .$ We assume that the field is
in a box of size $L$ with periodic boundary conditions so that the mode wave
vectors are given by ${\bf k}=2\pi{\bf n}/L$ for positive integers $n_{i}$.
We may take the limit $L\rightarrow\infty$ to recover the continuum limit,
though in doing so we must be mindful of both Lorentz invariance and
infrared divergences.

With this normal-mode decomposition, the environment becomes a collection of
real harmonic oscillators $\varphi_{{\bf \alpha}}(t)$ that are linearly
coupled to the system with the quantum Brownian motion action 
\begin{equation}
S_{B}+S_{int}=%
\mathop{\textstyle\sum}%
_{\alpha}\int dt\,[\left\{ \dot{\varphi}_{\alpha}^{2}/2+\omega_{\alpha}^{2}%
\varphi_{\alpha}^{2}/2+\varphi_{\alpha}f_{\alpha}(t)\right\} , 
\label{action in terms of harmonic oscillators}
\end{equation}
where 
\begin{equation}
f_{\alpha}(t)=\left( 2/L\right) ^{3/2}\int d{\bf x}\;\,j[z(t);{\bf x,}%
t)\,u_{\alpha}({\bf x}).\text{\ }   \label{f_alpha(t)}
\end{equation}
The functional integrals (in the path integral representation of the
influence functional Eq. (\ref{definition of influence functional})) then
become integrals over the oscillator's coordinates. The Jacobian for the
change of measure $\int\Pi_{{\bf x,}t}\,d\varphi(t,{\bf x})\rightarrow\int
\Pi_{\alpha}\,d\varphi_{\alpha}(t)$ is $1$. We may therefore use the
well-known results for the influence functional \cite
{Feynman-Vernon-Hibbs,HuMataczQBMParametricOsc94}. Initially correlated
states may be treated by the preparation function method \cite
{Initial-correlations}, but the qualitative results are largely the same as
those for initially uncorrelated states after fast transients have subsided 
\cite{Romero.Paz96(QBM.with.Initial.Correlations)}.

We note that it is a consequence of the fact that the influence functional
is derived in the functional Schr\"{o}dinger picture that there is an
associated choice of reference frame and fiducial time. Hence, even in the $%
L\rightarrow \infty$ limit, the influence functional is not manifestly
covariant, but instead depends explicitly and implicitly on the initial time
at which the particle and field states are defined. This initial time is
also identified with the spacelike hypersurface on which the initial state
takes a factorized form, which is likewise not a covariant construction (See
appendix C for more discussion on this point). Despite these facts, we see
in Paper II\ that the equations of motion for the particles quickly `forget'
the initial time $t_{i},$ and therefore become well-approximated by
manifestly covariant equations of motion at later times.

The influence functional factorizes as a product of terms $F_{\alpha}$ for
each field mode $\alpha:$

\begin{align}
F[f^{\pm}] & =\text{$\exp$}\left\{ \frac{i}{\hbar}S_{IF}[f^{-},f^{+}]\right%
\} =\Pi_{\alpha}\,F_{\alpha}[f_{\alpha}^{\pm}]
\label{influence functional as product factors} \\
& =\text{$\exp${\LARGE \{}}-\frac{1}{\hslash}\sum_{\alpha}\int dt\int
^{t}dt^{\prime}[2f_{\alpha}^{-}(t)\mu_{\alpha}(t,t^{\prime})f_{%
\alpha}^{+}(t^{\prime})  \nonumber \\
& -\frac{1}{\hslash}\,f_{\alpha}^{-}(t)\nu_{\alpha}(t,t^{\prime})\,f_{\alpha
}^{-}(t^{\prime})]\text{{\LARGE \}}},   \label{IF}
\end{align}
We have defined sum and difference variables 
\begin{align}
f_{\alpha}^{+}(t) & =\left[ f_{\alpha}(t)+f_{\alpha}^{\prime}(t)\right] /2 
\nonumber \\
f_{\alpha}^{-}(t) & =\left[ f_{\alpha}(t)-f_{\alpha}^{\prime}(t)\right] ,
\end{align}
in terms of which $S_{IF}$ takes a particularly simple form. The influence
kernel 
\begin{equation}
\zeta_{{\bf \alpha}}\left( t,t^{\prime}\right) =\nu_{\alpha}\left(
t,t^{\prime}\right) +i\mu_{\alpha}\left( t,t^{\prime}\right)
\end{equation}
for a field initially in a thermal state 
\begin{equation}
\hat{\rho}_{\varphi}=\exp\left\{ -\beta\hat{H}_{\varphi}\right\}
\end{equation}
(at inverse temperature $\beta=1/k_{B}T)$ is given by 
\begin{equation}
\zeta_{n}=\frac{1}{2\omega_{n}}\text{{\Large \{}coth}\left( \hbar\beta
\omega_{n}/2\right) \text{$\cos$}\omega_{n}(t-t^{\prime})-i\text{sin}%
\omega_{n}(t-t^{\prime})\text{{\Large \}}}.
\end{equation}
Substitution of (\ref{f_alpha(t)}) into (\ref{IF}) allows us to write the
influence functional as 
\begin{align}
& F[j(z),j(z^{\prime})]  \nonumber \\
& =\text{$\exp${\LARGE \{}}\frac{-1}{\hbar}\int dx\int dx^{\prime }\,\text{%
{\large (}}j[z;x)-j[z^{\prime};x))\times \\
& (G^{+}(x,x^{\prime})j[z;x^{\prime})+G^{+}(x,x^{\prime})^{\ast}j[z^{\prime
};x^{\prime})\text{{\large )}{\LARGE \}}}  \nonumber \\
& =\text{$\exp$}\left\{ \frac{i}{\hbar}S_{IF}[j,j^{\prime}]\right\} , 
\label{F[j'j'] Influence functional}
\end{align}
where $G^{+}\left( x,x^{\prime}\right) $ is the positive frequency Wightman
functions given by the mode expansion 
\begin{align}
G^{+}(x,x^{\prime})\, & =\frac{1}{\sqrt{2L^{3}}}\sum_{\alpha}\theta
_{tt^{\prime}}\,u_{\alpha}({\bf x})\,u_{\alpha}({\bf x}^{\prime})\zeta_{{\bf %
k}}(t,t^{\prime})  \label{Delta(x,x') influence functional kernel} \\
& =\frac{1}{\sqrt{2L^{3}}}\sum_{{\bf k}}\theta_{tt^{\prime}}\text{$\cos$ }%
{\bf k(x}-{\bf x}^{\prime})\zeta_{{\bf k}}(t,t^{\prime}). 
\label{second line}
\end{align}
In going from (\ref{Delta(x,x') influence functional kernel}) to (\ref
{second line}) we have used $\sum_{\lambda}\,u_{{\bf k\lambda}}({\bf x})\,u_{%
{\bf k\lambda}}({\bf x}^{\prime})=\cos$ ${\bf k(x}-{\bf x}^{\prime}).$ We
abbreviate $\theta\left( t,t^{\prime}\right) =\theta_{tt^{\prime}}.$ The
factor of $1/2$ comes from double counting the equivalent modes ${\bf k}$
and ${\bf -k}.$ In the limit $L\rightarrow \infty,$ we may go from discrete
to continuous modes by replacing $\Sigma_{{\bf k>0}}\rightarrow\left(
L^{3}/2\right) ^{1/2}\int _{0}^{\infty}d{\bf k}$. The Green's function $%
G^{+}\left( x,x^{\prime }\right) $ is given by the free field correlation
function 
\begin{equation}
G^{+}(x,x^{\prime})=Tr\left( \hat{\varphi}(x)\hat{\varphi}(x^{\prime})\hat{%
\rho}_{\varphi}\left( t_{i}\right) \right)
\end{equation}
where $\hat{\rho}_{\varphi}\left( t_{i}\right) $ is the initial state of the
field. The real part of $G^{+}$ is the field commutator (also called the
Schwinger, or causal, Green's function $G^{C}$): 
\begin{equation}
G^{C}=%
\mathop{\rm Re}%
G^{+}(x,x^{\prime})=\left\langle \left[ \hat{\varphi }(x),\hat{\varphi}%
(x^{\prime})\right] \right\rangle .
\end{equation}
The imaginary part of $G^{+}$ is the field anticommutator (also called the
Hadamard Green's function): 
\begin{equation}
G^{H}=%
\mathop{\rm Im}%
G^{+}(x,x^{\prime})=\left\langle \left\{ \hat{\varphi }(x),\hat{\varphi}%
(x^{\prime})\right\} \right\rangle .
\end{equation}
The commutator is a quantum state independent function; it determines the
causal radiation fields in both the quantum and classical theory, and is
responsible for dissipation in the equations of motion. $G^{H}$ is quantum
state dependent; it determines the form of quantum correlations, and is
responsible for noise in the stochastic limit. Indeed, because the
field-part of the action is quadratic, the Hadamard Green's function
contains the full information about the field statistics; though (as we
shall see) the field-induced noise is determined by the quantum field
statistics plus the particle kinematics.

Using the symmetry properties of the Green's functions and the form of the
influence functional, only the retarded part of the commutator appears in
the equations of motion, where 
\[
G_{R}(x,x^{\prime})=\theta(t,t^{\prime})%
\mathop{\rm Re}%
G_{+}(x,x^{\prime}). 
\]
Defining sum and difference variables 
\begin{align}
j^{+} & =\left[ j[z,x)+j[z^{\prime},x)\right] /2  \nonumber \\
j^{-} & =\left[ j[z,x)-j[z^{\prime},x)\right] ,
\end{align}
the influence functional takes on the form 
\begin{align}
F[j_{z,z^{\prime}}^{\pm}] & =\text{$\exp${\LARGE \{}}\frac{-i}{\hbar}\int
dx\int dx^{\prime}\text{{\large [}}2j^{-}(x)\,G_{R}(x,x^{\prime})j^{+}(x^{%
\prime})  \nonumber \\
& -ij^{-}(x)G_{H}(x^{\prime},x^{\prime})j^{-}(x^{\prime})\text{{\large ]}%
{\LARGE \}}}.   \label{F[j,j'] in terms of Green's functions}
\end{align}

When the function $j(z,x)$ factors as $j(z,$ $x)=\sum_{n}z_{n}h_{n}(x),$ we
recover the influence functional for linear quantum Brownian motion of $n$
particles following arbitrary yet {\it prescribed} (i.e. not dynamically nor
self-consistently determined) trajectories. The functions $h_{n}(t,{\bf x})$
provide effective time-dependent coupling constants \cite
{HuMataczQBMParametricOsc94}, the influence functional for this case is
reviewed in Appendix B.{\large \ }The closely related CTP coarse-grained
effective action, and its adaptation to the particle-field models we
consider, is reviewed in Appendix C. Further use of the CTP formalism,
together with the nonlinear particle-field influence functional, is made in
the second series, where the quantum regime of relativistic particles is
developed.

\subsection{The stochastic generating functional}

Rather than developing the theory at the level of the master equation, we
return to Eq. (\ref{Evolution of reduced density matrix}) and (\ref
{evolution kernel}) but now add the new source terms 
\begin{align}
& \int d\tau \text{{\Large (}}J_{\mu }\left( \tau \right) z^{\mu }\left(
\tau \right) -J_{\mu }^{\prime }\left( \tau \right) z^{\mu \prime }\left(
\tau \right) \text{{\Large )}}  \nonumber \\
& =\int d\tau \text{{\Large (}}J_{\mu }^{+}z^{\mu -}+J_{\mu }^{-}z^{\mu +}%
\text{{\Large )}}
\end{align}
to the coarse-grained effective action (CGEA)$,$ set $z_{f}=z_{f}^{\prime },$
and integrate over $z_{f}$ (with the restriction $z^{0}>t_{i};$ see
Appendices). We find 
\begin{align}
Z\left[ J^{\pm }\right] & =\int dz_{f}\rho _{r}\left( z_{f},z_{f}\right)
_{J^{\pm }}  \nonumber \\
& =\int dz_{f}d{\bf z}_{i}d{\bf z}_{i}^{\prime }\int_{{\bf z}_{i},{\bf z}%
_{i}^{\prime }}^{z_{f}=z_{f}^{\prime }}DzDz^{\prime }  \nonumber \\
& \times e^{\frac{i}{\hslash }\left( S_{CGEA}\left[ z,z^{\prime }\right]
+\int \left( J^{+}z^{-}+J^{-}z^{+}\right) d\tau \right) }  \nonumber \\
& \times \rho \left( z_{i},z_{i}^{\prime }\right) ,
\end{align}
where $Z\left[ J^{\pm }\right] $ is the precisely the generating functional
shown in Appendix C on closed-time-path methods. $Z\left[ j^{\pm }\right] $
may be used to derive the particle correlation functions $\langle \hat{z}%
\left( \tau \right) \rangle ,$ $\langle \hat{z}\left( \tau \right) \hat{z}%
\left( \tau ^{\prime }\right) \rangle ,...$ via , e.g., (\ref{mean field ctp}%
), or as the starting point for deriving the 1PI coarse-grained effective
action via, e.g., (\ref{G=W-gz legendre transfomation}). In this section, to
keep expressions as simple as possible, we will suppress most indices (e.g.
Lorentz indices, particle labels) and return to more explicit notations
later. We also won't worry about any overall constant normalization factors
since these don't effect the resulting equations of motion.

The crucial observation for translating a stochastic description hinges on
the fact that the noise kernel (appearing in $%
\mathop{\rm Im}%
S_{IF})$ is a real, symmetric kernel with positive eigenvalues. Therefore, 
\begin{align}
\left| e^{\frac{i}{\hslash}S_{CGEA}}\right| & =e^{-\frac{1}{\hslash}%
S_{IF}^{I}}=e^{-\frac{1}{\hslash}\sum_{\alpha}\int^{T}dtdt^{\prime}\,f_{%
\alpha}^{-}(t)\,\nu_{\alpha}(t,t^{\prime})\,f_{\alpha}^{-}(t^{\prime})} 
\nonumber \\
& =\left| F\left[ z,z^{\prime}\right] \right| \leq0. 
\label{arg of exponential}
\end{align}
The inequality holds for any pair $\left( z,z^{\prime}\right) .$ Recalling
that $f^{-}=f[z]-f\left[ z^{\prime}\right] $ and $z^{-}=z-z^{\prime},$ we
may expand the argument of the exponential (\ref{arg of exponential}) in
powers of $z^{-}$ giving 
\begin{align}
& \exp\text{{\LARGE \{}}-\frac{1}{\hslash}\sum_{\alpha}\int^{T}dtdt^{\prime
}\,\text{{\Large [}}\left( \delta f_{\alpha}(t)/\delta z^{\mu}\left(
\tau\right) \,\right) \nu_{\alpha}(t,t^{\prime})  \nonumber \\
& \times\left. \left( \delta f_{\alpha}(t^{\prime})/\delta z^{\mu}\left(
\tau^{\prime}\right) \right) \text{{\Large ]}}\right| _{z^{-}=0}z^{\mu
-}\left( \tau\right) z^{\mu-}\left( \tau^{\prime}\right) d\tau d\tau^{\prime}
\nonumber \\
& +{\cal O}\left( z^{-^{3}}\right) \text{{\LARGE \}}}. 
\label{decoherence term}
\end{align}
The influence functional is exponentially small for ``large'' deviations of
the histories $z^{-}\left( \tau\right) .$ Of course, what counts as large
depends on the noise kernel and coupling represented by $f\left( t\right) .$
(This suppression of the magnitude of the influence functional is indicative
of decoherence. Off diagonal terms in the density matrix $\rho\left(
z_{f}^{+},z_{f}^{-}\right) $ will have large $z^{-}$ and hence tend to be
suppressed by (\ref{decoherence term}).)

On the assumption that decoherence is strong enough (which needs to be
verified for particular cases and models), (\ref{decoherence term})
justifies the expansion of the effective action in powers of $z^{-}.$ The
real part of the CGEA (including source terms) is 
\begin{align}
& \int d\tau\text{{\LARGE [}}\left( \left. \frac{\delta}{\delta z^{\mu
-}\left( \tau\right) }S_{CGEA}^{R}\left[ z,z^{\prime}\right] \right|
_{z^{-}=0}+J_{\mu}^{+}\left( \tau\right) \right) z^{\mu-}\left( \tau\right) 
\nonumber \\
& +\int d\tau J_{\mu}^{-}\left( \tau\right) z^{\mu+}\left( \tau\right) +%
{\cal O}\left( z^{-^{3}}\right) \text{{\LARGE ]}}.
\end{align}
Together with (\ref{decoherence term}), the generating functional is
(neglecting ${\cal O}\left( z^{-}\right) ^{3}$ terms)

\begin{align}
& Z\left[ J^{\pm }\right]   \label{generating functional path integral} \\
& =\int d{\bf z}_{i}^{+}d{\bf z}_{i}^{-}\int_{{\bf z}_{i}^{+},{\bf z}%
_{i}^{-}}^{z_{f}^{-}=0}Dz^{+}Dz^{-}  \nonumber \\
& \times \exp \frac{i}{\hslash }\text{{\LARGE \{}}\int d\tau \left( \left. 
\frac{\delta S_{CGEA}^{R}\left[ z^{\pm }\right] }{\delta z^{\mu -}\left(
\tau \right) }\right| _{z^{-}=0}+J_{\mu }^{+}\left( \tau \right) \right)  
\nonumber \\
& \times z^{\mu -}\left( \tau \right) +i\sum_{\alpha }\int^{t}dt^{\prime
}d\tau d\tau ^{\prime }\,  \nonumber \\
& \times \left[ \left( \frac{\delta f_{\alpha }(t)}{\delta z^{\mu +}\left(
\tau \right) }\right) \nu _{\alpha }(t,t^{\prime })\left( \frac{\delta
f_{\alpha }(t^{\prime })}{\delta z^{\mu +}\,\left( \tau ^{\prime }\right) }%
\right) \right] _{z^{-}=0}  \nonumber \\
& \times z^{\mu -}\left( \tau \right) z^{\mu -}\left( \tau ^{\prime }\right)
+\int d\tau J_{\mu }^{-}\left( \tau \right) z^{\mu +}\left( \tau \right) 
\text{{\LARGE \}}}\rho \left( z_{i}^{\pm }\right) .  \nonumber
\end{align}
In general, the initial state of the particle $\rho \left( z_{i}^{\pm
}\right) $ is arbitrary, but we are interested in finding a description of
the particle's stochastic trajectory when the particle position is well
localized. We therefore take the initial particle density matrix to be 
\begin{equation}
\hat{\rho}_{z}\left( t_{i}\right) =|{\bf z}_{i},t_{i}\rangle \langle {\bf z}%
_{i},t_{i}|.
\end{equation}
The generating functional takes a simpler form in this case; we find 
\begin{align}
& Z\left[ J^{\pm }\right]  \\
& =\int d{\bf z}_{i}^{+}d{\bf z}_{i}^{-}\int_{{\bf z}_{i}^{+},{\bf z}%
_{i}^{-}}^{z_{f}^{-}=0}Dz^{+}Dz^{-}  \nonumber \\
& \times \exp \frac{i}{\hslash }\text{{\LARGE \{}}\int d\tau \left( \left. 
\frac{\delta S_{CGEA}^{R}\left[ z^{\pm }\right] }{\delta z^{\mu -}\left(
\tau \right) }\right| _{z^{-}=0}+J_{\mu }^{+}\left( \tau \right) \right)  
\nonumber \\
& \times z^{\mu -}\left( \tau \right) +i\sum_{\alpha }\int^{t}dt^{\prime
}d\tau d\tau ^{\prime }\,  \nonumber \\
& \times \left[ \left( \frac{\delta f_{\alpha }(t)}{\delta z^{\mu +}\left(
\tau \right) }\right) \nu _{\alpha }(t,t^{\prime })\left( \frac{\delta
f_{\alpha }(t^{\prime })}{\delta z^{\mu +}\,\left( \tau ^{\prime }\right) }%
\right) \right] _{z^{-}=0}  \nonumber \\
& \times z^{\mu -}\left( \tau \right) z^{\mu -}\left( \tau ^{\prime }\right)
+\int d\tau J_{\mu }^{-}\left( \tau \right) z^{\mu +}\left( \tau \right) 
\text{{\LARGE \}}}{\Huge .}  \nonumber
\end{align}
In our second series of papers we discuss more general initial states.

The manipulations above are somewhat formal in that we have not explicitly
discussed how to perform the $z^{\pm}$ path integrals (discussed in our
second series). These details are important when considering higher-order
quantum corrections, but for the semiclassical/stochastic limit it is enough
that we can do standard Gaussian path integrals in $z^{\pm},$ which we can.
In the Gaussian approximation the only important modification comes from the
restriction of the integration range placed on the particle time variable: $%
t_{i}<z^{0}<t_{f}$ (see Appendix A). In the limit that initial and final
states are defined at $t_{i}=-\infty$ and $t_{f}=+\infty$ (as is the case
for asymptotic scattering processes) the $Dz^{0}$ integrations are the usual
Gaussian-type. More generally, when considering processes not too close (in
time) to the initial time hypersurfaces $\Sigma\left( t_{i}\right) ,$ the
restriction on the $z^{0}$ integration range has negligible effect on $Z%
\left[ J^{\pm}\right] $ because the integration range of $z^{0}$ may be
extended to $\pm\infty$ with negligible error. But, for very short times
after $t_{i},$ the generating functional {\it is} modified by the $z^{0}$
boundary conditions. This has the result of modifying both the noise and
dissipation in the Langevin equation derived below at very short times after 
$t_{i}$ in such a way as to preserved consistency with the given initial
state.

These modifications are in the nature of what O'Connor Stephens and Hu refer
to as finite-size effects arising from the existence of boundaries or other
topological (or more generally even dynamically-generated) constraints in
fields or spacetimes \cite{FiniteSize}. More familiar are finite size
effects coming from spatial boundaries, implying a spatial restriction on
the paths in the sum over histories. Periodicity in (imaginary) time
depicting thermal behavior may also be viewed as a kind of finite-size
effect . One advantage of the spacetime formulation (in terms of worldline
path integrals) is that it allows a natural description of finite-size
effects arising from both spatial and temporal (i.e. spacetime) boundaries
or restrictions. Hawking and Hartle's path integral calculation of black
hole radiance may be viewed as a finite size effect where the singularity
and associated black hole geometry restrict the paths leading to thermal
particle creation at the Hawking temperature \cite{HartleHawking}.
Similarly, Duru and Unal \cite{DuruUnal} use path integrals to calculate
particle production in a cosmological spacetime where the initial
singularity places a temporal restriction on the particle paths similar to
the restrictions giving in this paper. In \cite{DuruUnal}, the restriction
on worldlines ($z^{0}\left( \tau \right) >t_{i}$) is required by the initial
cosmological singularity at $t_{i},$ in our work the restriction is part of
the initial condition: we take the particle state to be known (e.g. fixed)
at, and before, $t_{i}$. Consistency then requires the restriction $%
z^{0}\left( \tau \right) >t_{i}.$

Defining the variable 
\begin{equation}
\eta_{\mu}\left( t\right) \equiv\left. \frac{\delta}{\delta z^{\mu-}\left(
t\right) }S_{CGEA}^{R}\left[ z^{\pm}\right] \right|
_{z^{-}=0}+J_{\mu}^{+}\left( t\right) ,   \label{langevin1}
\end{equation}
we may do the (approximate) Gaussian $Dz^{-}$ path integral in (\ref
{generating functional path integral}). The result is 
\begin{align}
Z\left[ J^{\pm}\right] & =\int Dz^{+}\exp\text{{\LARGE \{}}-\frac {1}{%
4\hslash}\int\int d\tau d\tau^{\prime}  \nonumber \\
& \times\eta_{\mu}\left( \tau\right) C^{-1}\left( \tau,\tau^{\prime }\right)
\eta^{\mu}\left( \tau^{\prime}\right)  \nonumber \\
& +\frac{i}{\hslash}\int d\tau J_{\mu}^{-}\left( \tau\right) z^{\mu +}\left(
\tau\right) \text{{\LARGE \}}}
\end{align}
The kernel $C^{-1}$ is defined by 
\begin{align}
\delta\left( \tau-\tau^{\prime}\right) & =\int d\tau^{\prime\prime
}dt^{\prime\prime}dt^{\prime}C^{-1}\left( \tau,\tau^{\prime\prime}\right)
\times \\
& \left[ \left( \frac{\delta f_{\alpha}(t^{\prime\prime})}{\delta z^{\mu
+}\left( \tau\right) }\right) \nu_{\alpha}(t^{\prime\prime},t^{\prime
})\left( \frac{\delta f_{\alpha}(t^{\prime})}{\delta z_{\mu}^{+}\left(
\tau^{\prime}\right) }\right) \right] _{z^{-}=0}  \nonumber \\
& \equiv\int d\tau^{\prime\prime}C^{-1}\left( \tau,\tau^{\prime\prime
}\right) C\left( \tau^{\prime\prime},\tau^{\prime}\right)  \nonumber
\end{align}
We denote the kernel $C$ (the inverse of $C^{-1})$ by $C^{\left( \eta\right)
}$ to indicate its association with $\eta\left( \tau\right) ,$ and to
distinguish it from other kernels that arise later. Next, Eq. (\ref
{langevin1}) may be solved for $z^{+}$ as a functional of $\eta\left(
\tau\right) $ and $J^{+}\left( \tau\right) $ (i.e., $z^{+}=z^{+}\left[
\eta,J^{+}\right] )$\footnote{%
In the more common linear case, the noise is independent of the system
histories ($z^{+}).$ This is a crucial difference between linear and
nonlinear theories.}$.$ We now make the change of integration variables $%
Dz^{+}\rightarrow D\eta$ using the fact that the Jacobian is equal to 1.
Then the term 
\begin{equation}
P\left[ \eta\right] =N_{0}e^{-\frac{1}{4\hslash}\int\int d\tau d\tau
^{\prime}\eta_{\mu}\left( \tau\right) C^{-1}\left( \tau,\tau^{\prime
}\right) \eta^{\mu}\left( \tau t^{\prime}\right) } 
\label{nonMarkovian prob}
\end{equation}
may be interpreted as a probability distribution for a noise $\eta\left(
\tau\right) ,$ turning (\ref{langevin1}) into a Langevin equation. $N_{0}$
is some appropriate normalization constant.

In the expression (\ref{nonMarkovian prob}) above, the kernel $C^{-1}$ is a
functional of $z^{+}.$ This fact makes the noise system-history dependent
(non-Markovian). With the change in path integration variables, we find 
\begin{equation}
Z\left[ J^{\pm}\right] =\int D\eta P\left[ \eta\right] e^{\frac{i}{\hslash}%
\int d\tau J_{\mu}^{-}\left( \tau\right) z_{\eta}^{\mu+}\left( \tau\right) },
\end{equation}
where we note that $\eta$ is a functional of $J^{+}$ through Eq. (\ref
{langevin1}). We have added the subscript $\eta$ to $z^{+}$ to indicate that
the change of variables may be inverted to give $z_{\eta}^{+}$ as a
functional of $\eta$ via the Langevin equation (\ref{langevin1}). For
simplicity, we set $J^{+}=0,$ and write $z^{+}\rightarrow z$ and $%
J^{+}\rightarrow J.$ The generating functional 
\begin{equation}
Z\left[ J\right] =\int D\eta P\left[ \eta\right] e^{\frac{i}{\hslash}\int
d\tau J_{\mu}\left( \tau\right) z_{\eta}^{\mu}\left( \tau\right) }
\end{equation}
may then be used to find the trajectory correlation functions. The
mean-trajectory is 
\begin{align}
\frac{\hslash}{i}\left. \frac{\delta Z\left[ J\right] }{\delta J_{\mu
}\left( \tau\right) }\right| _{J=0} & =\int_{{\bf z}_{0}}D\eta P\left[ \eta%
\right] \left[ z_{\eta}^{\mu}\left( \tau\right) \right] ={\large \langle}%
z^{\mu}\left( \tau_{1}\right) {\large \rangle}_{\eta }  \nonumber \\
& \equiv\bar{z}^{\mu}\left( \tau\right) .
\end{align}
The n-point correlators for the particle are given by 
\begin{align}
& \frac{\hslash^{n}}{i^{n}}\left. \frac{\delta^{n}Z\left[ J\right] }{\delta
J_{\mu}\left( \tau_{1}\right) ...\delta J_{\nu}\left( \tau _{n}\right) }%
\right| _{J=0} \\
& =\int_{z_{0}}D\eta P\left[ \eta\right] \left[ z_{\eta}^{\mu}\left(
\tau_{1}\right) ...z_{\eta}^{\nu}\left( \tau_{n}\right) \right]  \nonumber \\
& \equiv{\large \langle}z^{\mu}\left( \tau_{1}\right) ...z^{\nu}\left(
\tau_{n}\right) {\large \rangle}_{\eta}.  \nonumber
\end{align}
Brackets with a subscript $\eta$ denote the stochastic average 
\begin{equation}
\left\langle y(\eta)\right\rangle _{s}=\int D\eta P[\eta]y\left( \eta\right)
.
\end{equation}
Since $C^{-1}$ in (\ref{nonMarkovian prob}) is still a functional of $\eta$
(through Eq. (\ref{langevin1})), this is not a Gaussian probability
distribution, but is far more complicated involving the full effect of
backreaction between the system and field.

To make progress, we linearize the nonlinear Langevin equation (\ref
{langevin1}) by expanding the trajectory around the solution, $\bar {z}%
\left( \tau\right) ,$ to 
\begin{align}
& \left. \frac{\delta}{\delta z^{\mu-}\left( \tau\right) }S_{CGEA}^{R}\left[
z^{\pm}\right] \right| _{z^{-}=0,z^{+}=\bar{z}}  \label{semiclassical sol1}
\\
& =\left. \frac{\delta}{\delta z^{\mu-}\left( \tau\right) }S_{CGEA}\left[
z^{\pm}\right] \right| _{z^{-}=0,z^{+}=\bar{z}}=0.  \nonumber
\end{align}
This solution we call the semiclassical trajectory; observe that the
imaginary part of $S_{CGEA}$ plays no role at this level. Defining 
\begin{equation}
\tilde{z}^{\mu}\equiv z^{\mu+}-\bar{z}^{\mu},
\end{equation}
we find 
\begin{align}
\eta_{\mu}\left( \tau\right) & =\left. \frac{\delta S_{CGEA}\left[ z^{\pm}%
\right] }{\delta z^{\mu-}\left( \tau\right) }\right| _{z^{-}=0,z^{+}=\bar{z}%
}+\int d\tau^{\prime}\tilde{z}^{\nu}\left( \tau^{\prime }\right)
\label{1st order langevin} \\
& \times\left. \frac{\delta}{\delta z^{\nu+}\left( \tau^{\prime}\right) }%
\left( \frac{\delta S_{CGEA}\left[ z^{\pm}\right] }{\delta z^{\mu-}\left(
\tau\right) }\right) \right| _{z^{-}=0}+{\cal O}\left( \tilde{z}^{2}\right) .
\nonumber
\end{align}
Since the first term on the LHS of (\ref{1st order langevin}) vanishes by
definition of $\bar{z},$ we find the linearized Langevin equations 
\begin{equation}
\eta_{\mu}\left( \tau\right) =\int d\tau^{\prime}\tilde{z}^{\nu}\left(
\tau^{\prime}\right) \left. \frac{\delta}{\delta z^{\nu+}\left(
\tau^{\prime}\right) }\left( \frac{\delta S_{CGEA}\left[ z^{\pm}\right] }{%
\delta z^{\mu-}\left( \tau\right) }\right) \right| _{z^{-}=0}^{z^{+}=\bar{z}%
}.
\end{equation}
Expanding the probability distribution around $\bar{z}$ gives 
\begin{equation}
P\left[ \eta\right] =N_{0}e^{-\frac{1}{4\hslash}\int\int d\tau d\tau
^{\prime}\eta_{\mu}\left( \tau\right) C_{\bar{z}}^{-1}\left( \tau
,\tau^{\prime}\right) \eta^{\mu}\left( \tau^{\prime}\right) +{\cal O}\left(
\eta^{3}\right) },   \label{Gaussian noise}
\end{equation}
since 
\[
C^{-1}\left( \tau,\tau^{\prime}\right) =C_{\bar{z}}^{-1}\left( \tau
,\tau^{\prime}\right) +\left( \delta C^{-1}/\delta\bar{z}\right) \tilde {z}+%
{\cal O}\left( \tilde{z}^{2}\right) , 
\]
and $\tilde{z}$ is ${\cal O}\left( \eta\right) .$ At this point, one can
make a Gaussian approximation to the noise, which involves evaluating $C^{-1}
$ with the (self-consistently determined) semiclassical solutions $\bar{z}.$
Later we shall show how to treat non-Gaussian noise via a cumulant expansion.

Under a Gaussian approximation, it is clear from (\ref{Gaussian noise}) that
the mean noise $\langle\eta\rangle_{\eta}=0.$ The noise correlation function
becomes 
\begin{equation}
{\large \langle}\eta_{\mu}\left( \tau\right) \eta_{\nu}\left( \tau^{\prime
}\right) {\large \rangle}=2g_{\mu\nu}C_{2\bar{z}}\left( \tau,\tau^{\prime
}\right) .
\end{equation}

\subsection{The stochastic effective action}

From the preceding section we are able to justify a simpler method of
deriving the stochastic equations of motion from the stochastic effective
action. We also see more clearly the underlying role of the quantum field in
producing noise in this approach. Returning to the expression for the
influence functional in (\ref{arg of exponential}), we use a standard
functional Gaussian identity to write the imaginary part of the influence
action as 
\begin{align}
\left| F\left[ z,z^{\prime}\right] \right| & =N_{0}\int D\xi\text{$\exp $%
{\LARGE \{}$-$}\frac{1}{\hbar}\sum_{\alpha} \\
& \times\text{{\Large (}}\int
dtdt^{\prime}\xi_{\alpha}(t)\nu_{\alpha}^{-1}(t,t^{\prime})\xi_{\alpha}(t^{%
\prime})  \nonumber \\
& +\frac{i}{\hbar}\int dt\,\xi_{\alpha}(t)f_{\alpha}^{-}(t)\text{{\Large )}%
{\LARGE \}}}  \nonumber \\
& =\int D\xi P[\xi]\text{$\exp$}\left\{ \frac{i}{\hbar}\sum_{\alpha}\int
dt\,\xi_{\alpha}(t)\,f_{\alpha}^{-}(t)\right\}  \nonumber \\
& \equiv\left\langle \text{$\exp$}\left\{ \frac{i\,}{\hbar}\sum_{\alpha}\int
dt\;\xi_{\alpha}(t)\,f_{\alpha}^{-}(t)\right\} \right\rangle _{\xi}, 
\nonumber
\end{align}
where $\xi_{\alpha}(t)$ are stochastic variables, and 
\begin{equation}
P[\xi]=N_{0}\,\text{$\exp$}\left\{ -\frac{1}{2\hbar}\sum_{\alpha}\int
dtdt^{\prime}\xi_{\alpha}(t)\nu_{\alpha}^{-1}(t,t^{\prime})\xi_{\alpha
}(t^{\prime})\right\}   \label{P[xi] probability distribution}
\end{equation}
is a normalized probability distribution on the space of functions $\left\{
\xi_{\alpha}(t)\right\} .$ The kernel $\nu_{\alpha}^{-1}$ is defined by 
\begin{equation}
\int dt^{\prime}\,\nu_{\alpha}^{-1}(t,t^{\prime})\nu_{\alpha}(t^{\prime
},t^{\prime\prime})=\delta(t-t^{\prime\prime}).
\end{equation}
This allows us to write 
\begin{align}
& \exp\left[ i(S_{CGEA}^{R}+iS_{CGEA}^{I})/\hslash\right]  \nonumber \\
& =\exp\left( \frac{i}{\hslash}S_{CGEA}^{R}\right) \left| F\left[
z,z^{\prime}\right] \right|
\end{align}
as 
\begin{align}
e^{\frac{i}{\hslash}S_{CGEA}\left[ z,z^{\prime}\right] } & =\int D\xi\,P[\xi]%
\text{$\exp$}\frac{i}{\hbar}\text{{\LARGE \{}}S_{A}\left[ z\right] -S_{A}%
\left[ z^{\prime}\right]  \nonumber \\
& +\sum_{\alpha}\int dtf_{\alpha}^{-}(t)\text{{\Large (}}\xi_{\alpha }(t)\, 
\nonumber \\
& +2\int^{t}dt^{\prime}\mu_{\alpha}(t,t^{\prime})f_{\alpha}^{+}(t^{\prime })%
\text{{\Large )}{\LARGE \}}}  \nonumber \\
& =\int D\xi\,P[\xi]\,e^{\frac{i}{\hslash}S_{\xi}\left[ z,z^{\prime}\right]
}.
\end{align}
This expression defines the stochastic effective action 
\begin{align}
S_{\xi}\left[ z,z^{\prime}\right] & =S_{A}\left[ z\right] -S_{A}\left[
z^{\prime}\right] +\sum_{\alpha}\int dt\,f_{\alpha}^{-}(t)\text{{\Large (}}%
\xi_{\alpha}(t)  \nonumber \\
& +2\int^{t}dt^{\prime}\mu_{\alpha}(t,t^{\prime})f_{\alpha}^{+}(t^{\prime })%
\text{{\Large )}}.
\end{align}
The variables $\xi_{\alpha}(t)$ are stochastic forces (noises) coupled to
the currents $f_{\alpha}(t).$ The mean force vanishes since 
\begin{align}
& \left\langle \xi_{\alpha}(t)\right\rangle _{\xi} \\
& =N_{0}\int D\xi_{\alpha}\,\xi_{\alpha}(t)e^{\left\{ -\frac{1}{2\hbar}\int
dtdt^{\prime}\,\xi_{\alpha}(t)\,\nu_{\alpha}^{-1}(t,t^{\prime})\xi_{\alpha
}(t^{\prime})]\right\} }=0.  \nonumber
\end{align}
The $\xi$ noise correlator is

\begin{align}
\left\langle \left\{ \xi_{\alpha}(t)\,,\xi_{\alpha}(t^{\prime})\right\}
\right\rangle _{\xi} & =\int D\xi_{\alpha}\,\xi_{\alpha}(t)\,\xi_{\alpha
}(t^{\prime})\,P[\xi_{\alpha}]  \nonumber \\
& =\hbar\nu_{\alpha}(t,t^{\prime}).   \label{noise correlator}
\end{align}

Notice that in these expressions no Gaussian approximation to the noise is
made. The noise variables $\xi_{\alpha}\left( t\right) ,$ unlike the earlier
noises $\eta\left( \tau\right) ,$ are already exactly Gaussian because their
statistics are precisely those of the (assumed) Gaussian quantum field
state. On the other hand, the stochastic noises $\xi_{\alpha}\left( t\right) 
$ do not couple directly to the particle coordinates $z\left( \tau\right) ,$
as was the case for the $\eta\left( \tau\right) ,$ but instead couple to the
nonlinear function $f_{\alpha}\left( t\right) .$ We shall explore this
connection in more detail below.

We may now find the nonlinear Langevin equations of the previous section
directly from $S_{\xi}$ as the stochastic solutions to 
\begin{equation}
\left. \left( \frac{\delta S_{\xi}}{\delta z^{\mu-}}\right) \right|
_{z^{-}=0}=0,   \label{langevin2}
\end{equation}
with the noise characteristics given by (\ref{noise correlator}). As in (\ref
{semiclassical sol1}), the semiclassical solutions are found from $\left(
\delta S_{CGEA}/\delta z^{\mu-}\right) _{z^{-}=0}=0.$ The solutions, being
derived within a self-consistent initial value (``in-in'') formalism, are
real and causal \cite{DeWitt-Jordan}. The stochastic regime manifests at a
deeper level than the semiclassical limit. In our particular case, the
stochastic regime identifies the quantum field statistical correlations with
fluctuations induced in the particles' trajectories. These fluctuations
represent physical noise when the particle histories decohere\footnote{%
Whenever we speak of particle histories in the context of the semiclassical
or stochastic regime, it will be implicitly assumed that some additional
smearing of the particle histories has been done to produce truly decoherent
histories (see Sec.3.5). When the histories are so coarse-grained that they
effectively acquire substantial inertia (usually the case in the macroscopic
limit), the efffect of this noise will be comparatively small \cite
{Gell-MannHartle93(DecoherenceFunctionalEquationsM}.}.

\subsection{The cumulant expansion}

We shall find later that higher order effective noise cumulants arise due to
the nonlinearity of the particle-field coupling- we saw the first sign of
this when we derived the stochastic generating functional. To treat this
more general case it is convenient to define a generating functional for
normalized correlation functions 
\[
W\left[ f_{\alpha}^{\pm}\right] =\ln F\left[ f_{\alpha}^{\pm}\right] =\left(
i/\hbar\right) S_{IF}. 
\]
The cumulant expansion method was used by Hu and Matacz \cite
{Einstein-Langevin} in their derivation of the Einstein-Langevin equation in
stochastic semiclassical gravity. Taylor expanding $W\left[ f_{\alpha }^{\pm}%
\right] $ in powers of $f_{\alpha}^{-}(t)$ (defined in (\ref{f_alpha(t)})),
we find

\begin{align}
W[f_{\alpha}^{+},f_{\alpha}^{-}] & =\frac{i}{\hbar}%
\int_{t_{i}}^{t_{f}}dt_{1}f_{\alpha}^{-}(t_{1})C_{1\alpha}^{\left( f\right)
}(t_{1};f^{+}] \\
& -\frac{1}{2\hbar^{2}}\int_{t_{i}}^{t_{f}}dt_{1}%
\int_{t_{i}}^{t_{f}}dt_{2}f_{\alpha}^{-}(t_{1})f_{\alpha}^{-}(t_{2}) 
\nonumber \\
& \times C_{2\alpha}^{\left( f\right) }(t_{1},t_{2};f^{+}]+...  \nonumber \\
& +\frac{1}{n!}\left( \frac{i}{\hbar}\right) ^{n}\int dt_{1}\cdots
dt_{n}f_{\alpha}^{-}(t_{1})\cdots f_{\alpha}^{-}(t_{n})  \nonumber \\
& \times C_{n\alpha}^{\left( f\right) }(t_{1},...,t_{n};f^{+}]+.... 
\nonumber
\end{align}
For simplicity, we will drop the mode label $\alpha$ in the remainder of
this Section. The full stochastic action is given by the addition of
contributions from each mode (since the influence functional factors into a
product of terms for each mode). The real cumulants $C_{n}^{\left( f\right)
} $ are given by 
\begin{equation}
C_{n}^{\left( f\right) }(t_{1},...,t_{n};f^{+}]=\left( \frac{\hbar}{i}%
\right) ^{n}\left. \frac{\delta^{n}W[f^{+},f^{-}]}{\delta
f^{-}(t_{1})\cdots\delta f^{-}(t_{n})}\right| _{f^{-}=0}. 
\label{C(n) Cumulants}
\end{equation}
The $C_{n}^{\left( f\right) }$ are functionals of $f^{+}(t)$. They are,
therefore, dependent on the system history. The cumulants $C_{n}$ are of
order $e^{n}$, with $n=2$ being the Gaussian noise contribution. The
absolute value of the influence functional may be written as

\begin{align}
\left| F[f^{+},f^{-}]\right| & =\left\langle \,\text{$\exp$}\left\{ \frac{i}{%
\hbar}\int dtf^{-}(t)\Xi(t)\right\} \right\rangle _{s}
\label{Characteristic function} \\
& =\int D\Xi P[\Xi,f^{+}]\text{$\exp$}\left\{ \frac{i}{\hbar}\int
dtf^{-}(t)\Xi(t)\right\} ,  \nonumber
\end{align}
$\left| F[f^{\pm}]\right| $ is therefore the characteristic functional of
the stochastic process $\Xi$. The normalized probability functional $%
P[\Xi,f^{+}]$ may be found from the influence functional by inverting the
functional Fourier transform (\ref{Characteristic function})\footnote{%
In our particular model, the influence functional is Gaussian in $f^{\pm},$
and therefore this inversion is straightforward.}. The probability
distribution $P[\Xi,f^{+}]$ depends on the system history $f^{+}$ for
nonlinear models, $P\left[ \Xi\right] $ is independent of the $f^{+}$ for
linear models. Making a Gaussian noise approximation by neglecting all
cumulants beyond second order (i.e. by working to order $e^{2}$ in the
coupling constant), the probability distribution is 
\begin{align}
& P[\Xi,f^{+}] \\
& =P_{0}\text{$\exp$}\left\{ -\frac{1}{2\hbar}\int dt_{1}\int
dt_{2}\Xi(t_{1})C_{2}^{-1}(t_{1},t_{2};f^{+}]\Xi(t_{2})\right\} ,  \nonumber
\end{align}
and the correlator of the noise $\Xi$ is 
\[
\left\langle \Xi(t_{1})\Xi\left( t_{2}\right) \right\rangle _{s}=\hbar
C_{2}(t_{1},t_{2};f^{+}]. 
\]
In this approximation the stochastic influence action may be written 
\begin{equation}
S_{IF}[f^{\pm},\Xi]=\int dt\left\{ f^{-}(t)\left(
C_{1}(t;f^{+}]+\Xi(t)\right) \right\} .
\end{equation}

\subsection{Stochastic fields}

While our primary interest is the dynamics of the particle subsystem, it is
helpful to see how the quantum field is equivalent to a stochastic field in
terms of its effects on the particle trajectories for the model we are
considering, and when the stochastic regime is a good approximation. The
influence functional is derived by integrating out the field $\varphi$ to
achieve a description of the effective system dynamics in terms of particle
variables alone together with local noise found from the stochastic
functions $\xi_{\alpha}(t)$. We may reintroduce fields by defining the
stochastic (field) variables 
\begin{equation}
\chi_{\xi}(x)\equiv\left( 2/L\right) ^{3/2}\sum_{\alpha}\xi_{\alpha
}(t)u_{\alpha}({\bf x}),   \label{stochasic field definition}
\end{equation}
where $u_{\alpha}({\bf x})$ are the normal modes (\ref{Normal modes}). The
subscript $\xi$ indicates the functional dependence of $\chi$ on the
stochastic functions $\xi_{\alpha}(t)$. The probability distribution for the
stochastic fields $P[\chi]$ is determined by the normalization condition 
\begin{equation}
\int_{\Omega(\chi)}D\chi_{\xi}\,P[\chi_{\xi}]=\int_{\Omega(\xi)}D\xi\,P[\xi],
\end{equation}
where $\Omega(\chi)=\left\{ \chi_{\xi}|\,\xi\in\Omega(\xi)\right\} .$ Then 
\[
P[\chi_{\xi}]=\left[ \det\left( \frac{\delta\chi}{\delta\xi}\right) \right]
^{-1}P[\xi]. 
\]
But since the change of variables from $\xi$ to $\chi$ is orthogonal, the
Jacobian is $1.$ Therefore,

\begin{align}
P[\chi] & =\text{$\exp${\LARGE \{}}-\frac{1}{2\hbar}%
\mathop{\textstyle\sum}%
_{\alpha}\int dxdx^{\prime}\,(\chi(x)\,u_{\alpha}({\bf x})\,  \nonumber \\
& \times\nu_{\alpha}^{-1}(t,t^{\prime})\,u_{\alpha}({\bf x}^{\prime
})\,\chi(x^{\prime}))\text{{\LARGE \}}}  \nonumber \\
& =\text{$\exp$}\left\{ -\frac{1}{2\hbar}\int dxdx^{\prime}\,\chi
(x)\,\Upsilon(x,x^{\prime})\,\chi(x^{\prime})\right\} , 
\label{chi prop dist}
\end{align}
where 
\begin{equation}
\Upsilon(x,x^{\prime})=\sum_{\alpha}u_{\alpha}({\bf x})\,\nu_{%
\alpha}^{-1}(t,t^{\prime})\,u_{\alpha}({\bf x}^{\prime}).
\end{equation}
For spacelike separated points $\left( x-x^{\prime}\right) ,$ $\Upsilon
(x,x^{\prime})$ does not vanish, which reflects the nonlocal character of
the quantum field correlations, and hence the nonlocal correlations that
will be present between particles that are (even) spacelike separated.

The stochastic field $\chi(x)$ has vanishing mean, and autocorrelation
function given by 
\begin{equation}
\left\langle \chi(x)\chi(x^{\prime})\right\rangle _{s}=\hbar\left\langle
\left\{ \hat{\varphi}(x),\hat{\varphi}(x^{\prime})\right\} \right\rangle
=\hbar G^{H}(x,x^{\prime}).
\end{equation}
Thus, $\chi(x)$ encodes the same quantum statistical information as the
field anticommutator.

\section{The stochastic equations of motion}

\subsection{General Form in Terms of Stochastic Field}

From the stochastic effective action it is straightforward to derive
Langevin equations of motion for particle worldlines. We emphasize that the
exact stochastic properties of the noise {\it are }known, being derived from
the quantum statistics of the field-environment. We begin by using the
definition of $\chi$ (\ref{stochasic field definition}) and $%
f_{\alpha}^{-}(t)$ (\ref{f_alpha(t)}) to write the noise term in $S_{\xi}$ as

\begin{align}
& \sum_{\alpha}\int^{T}dt\xi_{\alpha}(t)f_{\alpha}^{-}(t)  \nonumber \\
& =\int dt\,\int d{\bf x\,}\text{{\Large \{}}j({\bf x},z(t))-j({\bf x}%
,z^{\prime}(t))\text{{\Large \}}}\sum_{\alpha}\xi_{\alpha }(t)u_{\alpha}(%
{\bf x}))  \nonumber \\
& =\int dx\,\,j^{-}(x)\chi(x).   \label{Stochastic action term}
\end{align}
Together with the dissipation term from (\ref{F[j,j'] in terms of Green's
functions}), we use (\ref{Stochastic action term}) to define 
\begin{align}
S_{\chi}[z^{\pm}] & \equiv S_{A}[z^{\pm}]+\int dx\,j^{-}(x)(\chi (x)
\label{stochastic eff action chi} \\
& +2\int dx^{\prime}G^{R}(x,x^{\prime})j^{+}(x^{\prime})).  \nonumber
\end{align}
Thus, the influence of the environment is described in terms of a current $%
j^{-}(x)$ self-interacting through the causal retarded Green's function $%
G^{R}(x,x^{\prime}),$ and with $j^{-}$ coupled multiplicatively to the
stochastic field $\chi(x)$.

Using the generating functional $W$, we find the cumulants of the noise 
\begin{equation}
C_{q}^{\left( \chi\right) }(x_{1},...x_{n};j^{+}]=\left( \frac{\hbar}{i}%
\right) ^{n}\left. \frac{\delta^{n}W[j^{+},j^{-}]}{\delta
j^{-}(x_{1})...\delta j^{-}(x_{n})}\right| _{j^{-}=0}.
\end{equation}
For the field state initially Gaussian the higher order cumulants $%
C_{q>2}^{\left( \chi\right) }(x_{1},...,x_{n};j^{+}]$ vanish, and we have
only $\left\langle \left\{ \chi\left( x\right) ,\chi\left( x^{\prime
}\right) \right\} \right\rangle =C_{2}^{\left( \chi\right)
}(x_{1},x_{2};j^{+}],$ which is proportional to the Hadamard function. We
shall assume that we have shifted the noise to a new noise with zero mean,
and correspondingly we write the first cumulant as a separate dissipative
term. The stochastic effective action (\ref{stochastic eff action chi})
gives the Langevin equations according to (\ref{langevin2}) as 
\begin{align}
-\frac{\delta S_{A}[z]}{\delta z_{n}^{\mu}(\tau)} & =2\int dx\int dx^{\prime}%
\frac{\delta j(z,x)}{\delta z_{n}^{\mu}(\tau)}G^{R}(x,x^{\prime
})j(z,x^{\prime})  \nonumber \\
& +\int dx\frac{\delta j(z,x)}{\delta z_{n}^{\mu}(\tau)}\chi(x). 
\label{field chi langevin}
\end{align}
We have used the fact that $z^{-}=0$ implies $j^{-}=0,$ $j^{+}=j,$ $\delta
j^{-}/\delta z^{-}=\delta j/\delta z,$ and $z^{+}=z.$ For a system with $N$
particles (\ref{General stochastic diff' q}) are $N\times D$ coupled
stochastic integrodifferential equations for the stochastic particle
worldline coordinates $z_{\chi}^{\mu}(\tau)$. Each realization of the
stochastic field $\chi(x)$ gives a solution $z_{\chi}^{\mu}(\tau)$ of (\ref
{field chi langevin}); that is, the system equations of motions are
stochastic functionals of $\chi(x).$ In linear models, the current $j$ is
proportional to the particle degrees of freedom (here $z),$ and therefore $%
\delta j/\delta z^{\mu}=g_{\mu}(x)$ is independent of $z.$ From this it
immediately follows that (\ref{field chi langevin}) becomes a linear
stochastic differential equation with additive noise. For such linear
models, the mean of (\ref{field chi langevin}) gives the classical limit

\begin{equation}
\frac{\delta S_{A}[\bar{z}]}{\delta\bar{z}_{n}^{\mu}(\tau)}+2\int
dxdx^{\prime}g_{n\mu}(x)G^{R}(x,x^{\prime})j(\bar{z},x^{\prime})=0,
\end{equation}
making use of $\left\langle \chi\right\rangle =0.$

\subsection{Nonlinear Coupling}

For nonlinearly coupled theories the second cumulant (under the Gaussian
noise approximation) will depend on the self-consistent solution for the
semiclassical equations of motion including the backreaction effect
represented by the first cumulant\footnote{%
For linear theories, the second cumulant is independent of the first, and
hence linear theories do not have noise that is history dependent.}. The
linearized self-consistent approximation requires expanding about the
semiclassical solutions $\bar {z}_{n}^{\mu}.$ We define $\tilde{z}_{n}\equiv
z_{n}-\bar{z}_{n}.$ The linearized equations of motion are therefore

\begin{align}
0 & =\sum_{m}\int d\tau_{m}\,\tilde{z}_{m}^{\nu}\left( \tau_{m}\right) \text{%
{\LARGE \{}}\frac{\delta^{2}S_{A}[\bar{z}]}{\delta\bar{z}_{m}^{\nu }\left(
\tau_{m}\right) \delta\bar{z}_{n}^{\mu}})  \nonumber \\
& +2\int dx\,dx^{\prime}\frac{\delta^{2}j(\bar{z},x)}{\delta\bar{z}_{m}^{\nu
}\delta\bar{z}_{n}^{\mu}}G^{R}(x,x^{\prime})j(\bar{z},x^{\prime})  \nonumber
\\
& +2\int dx\,dx^{\prime}\frac{\delta j(\bar{z},x)}{\delta\bar{z}_{n}^{\mu}}%
G^{R}(x,x^{\prime})\frac{\delta j(\bar{z},x)}{\delta\bar{z}_{m}^{\nu}} 
\nonumber \\
& +\int dx\frac{\delta j(\bar{z},x)}{\delta\bar{z}_{n}^{\mu}}\chi (x)\text{%
{\LARGE \}}}+{\cal O}\left( \tilde{z}^{2}\right) . 
\label{General stochastic diff' q}
\end{align}
The first term on the RHS is a kinetic term for the $\tilde{z}_{n}$ (there
may also be linearized force term if $S_{A}$ includes an external potential $%
V\left( z_{n}\right) $. The second and third terms give history dependent
nonlocal dissipation and multiparticle interactions. For brevity, we will
take repeated particle indices ($e.g.,$ $\tilde{z}_{m}\left( \delta/\delta%
\bar {z}_{m}\right) )$ to always imply both summation over $m$ and the
integration $\int d\tau_{m}.$ The fourth term is an additive noise. (Note
that the noise is only additive for nonlinear theories at this lowest ${\cal %
O}\left( \tilde{z}\right) $ order.)

These Langevin equations show the intimate connection between particles and
fields within the stochastic-semiclassical limit. For each realization of
the stochastic field $\chi(x),$ the Langevin equation has solutions $\tilde{z%
}_{n}[\chi,\tau)$ that are functionals of $\chi(x).$ In this sense, the
quantum statistics of the field is carried by the particle trajectories. We
could as well have treated the field as the system, and integrated out the
particles, to find a stochastic effective action $S_{stoch}[\varphi^{\pm}], $
and Langevin equations 
\begin{equation}
\left. \frac{\delta S_{stoch}[\varphi^{\pm}]}{\delta\varphi^{-}}\right|
_{\varphi^{-}=0}=0,
\end{equation}
where the stochastic field solutions are functionals of stochastic particle
trajectories. In this way, the field would encode the quantum statistics of
the particles.

If the particle-field coupling had been linear, and $j(x)$ had factored as $%
h_{n\mu}(x)z_{n}^{\mu}(\tau),$ (\ref{General stochastic diff' q}) would have
reduced to the linear stochastic differential equation

\begin{align}
& \frac{\delta S_{A}[z]}{\delta z_{n}^{\mu}(\tau)}+2\int d{\bf x}\int
dx^{\prime}\,h_{n\mu}(\tau,{\bf x})D_{R}(x,x^{\prime})  \nonumber \\
& \times h_{m\nu}(x^{\prime})z_{m}^{\nu}(\tau^{\prime})+\int dx\,h_{n\mu
}(x)\chi(x)  \label{Linear Langevin} \\
& =\frac{\delta S_{A}[z]}{\delta z_{n}^{\mu}(\tau)}+X_{n\mu}(\tau)+\int
d\tau^{\prime}\,\,\Delta_{nm,\mu\nu}(\tau,\tau^{\prime})\,z_{m}^{\nu}(\tau^{%
\prime})=0,  \nonumber
\end{align}
where 
\begin{equation}
\vartheta_{n\mu}(\tau)=\int dx\,h_{n\mu}(x)\chi(x)
\end{equation}
and 
\begin{align}
\Delta_{nm,\mu\nu}(\tau,\tau^{\prime})\, & =2\int d{\bf x}\int d{\bf x}%
^{\prime}\,h_{n\mu}(\tau,{\bf x})D_{R}(x,x^{\prime})  \nonumber \\
& \times h_{m\nu}(\tau^{\prime},{\bf x}^{\prime}).
\end{align}
Eq. (\ref{Linear Langevin}) are Langevin equations with additive noise $%
X_{n\mu}(\tau)$ and linear, though still nonlocal, dissipation kernel $%
\Delta_{nm,\mu\nu}(\tau,\tau^{\prime}).$ The equation is a general form of
the many-particle Langevin equations for Brownian motion with nonlocal
interactions and a linearly dissipative environment. It is equivalent to the
linear N-particle detector Langevin equations found by Raval, Hu and Anglin 
\cite{RavalHu:StochasticAcceleratedDetectors}.

\subsection{Higher-Order Noise Cumulants}

Rather than working with stochastic field variables that couple to $%
j^{-}[z;x)$, it is convenient to define local stochastic forces $\eta
_{n}^{\mu}\left( \tau_{n}\right) $ that couple directly to the coordinates $%
z_{n}^{\mu}\left( \tau_{n}\right) .$ These are just the stochastic forces
that appeared in section III. We shall now be more explicit in our
derivation. Using the generating functional 
\[
W[j^{\pm}\left( z^{\pm}\right) ]=-i\hbar\ln S_{IF}[j^{\pm}\left( z^{\pm
}\right) ], 
\]
we define the $q$th order cumulant (for the noise $\eta)$ to be 
\begin{align}
C_{q}^{\left( \eta\right) } & \equiv C_{q(n...m)\mu...\nu}^{\left(
\eta\right) }(\tau_{n},...\tau_{m};z^{+},...,z^{+}] \\
& =\left( \frac{\hbar}{i}\right) ^{q}\left. \frac{%
\delta^{q}S_{IF}[j^{+},j^{-}]}{\delta z_{n}^{\mu-}(\tau_{n})...\delta
z_{m}^{-\nu}(\tau_{m})}\right| _{z_{n}^{-}=0}.  \nonumber
\end{align}
The probability distribution $P[\eta;z^{+}]$ is defined by requiring

\begin{align}
F[j^{\pm}] & =\int D\eta_{n}^{\mu}P[\eta_{n}^{\mu};z^{+}] \\
& \times\text{$\exp$}\left[ \frac{i}{\hbar}\int d\tau z_{n}^{\mu-}(\tau
)\eta_{n\mu}(\tau)\right]  \nonumber \\
& =\text{$\exp${\LARGE \{}}\frac{i}{\hbar}\sum_{n}\int d\tau_{n}z_{n}^{\mu
-}C_{1(n)\mu}^{\left( \eta\right) }(\tau_{n};z^{+}]  \nonumber \\
& -\frac{1}{2\hbar^{2}}\sum_{nm}\int
d\tau_{n}d\tau_{m}z_{n}^{\mu-}z_{m}^{\nu-}  \nonumber \\
& \times C_{2(nm)\mu\nu}^{\left( \eta\right) }(\tau_{n},\tau_{m};z^{+}]+... 
\nonumber \\
& +\frac{1}{n!}\left( \frac{i}{\hbar}\right) ^{q}\sum_{n...p}\int d\tau
_{n}\cdots d\tau_{p}z_{n}^{\mu-}\cdots z_{p}^{\lambda-}  \nonumber \\
& \times C_{q(n...p)\mu\lambda}^{\left( \eta\right) }(\tau_{n},...,\tau
_{p};z^{+}]+....\text{{\LARGE \}.}}  \nonumber
\end{align}
$P[\eta;z^{+}]$ is the characteristic functional of $F\left[ j^{\pm}\right] $
with respect to $z_{n}^{\mu-}$, and $C_{q(n...m)}^{\left( \eta\right)
\mu...\nu}$are the cumulants of the noises $\eta_{n\mu}(\tau_{n}).$ Note
that there is a noise for each index $\mu$ (including the timelike
component), and for each particle $n$. Because the influence functional is
not Gaussian in $z,$ the higher order cumulants $C_{q>2}^{\left( \eta\right)
}$ do not vanish$.$ The mean $\left\langle \eta_{n\mu}\right\rangle $ of the
noise is the first cumulant $C_{1(n)\mu}^{\left( \eta\right)
}(\tau_{n};z^{+}].$ It gives the dissipative term, and as before, we will
shift the noise so that $\eta$ has zero mean. Then the stochastic effective
action becomes

\begin{align}
S_{\eta}[z^{\pm},\eta] & =S_{z}[z^{\pm}]+\sum_{n}\int d\tau\,z_{n}^{\mu
-}(\tau_{n}) \\
& \times\text{{\Large (}}C_{1(n)\mu}^{\left( \eta\right)
}(\tau,z^{+}]+\eta_{n\mu}(\tau)\text{{\Large )}}.  \nonumber
\end{align}
The stochastic equations of motion follow immediately from $\left( \delta
S_{\eta}/\delta z_{n}^{\mu-}\right) |_{z^{-}=0}=0.$ They are 
\begin{equation}
\frac{\delta S_{z}[z]}{\delta z_{n}^{\mu}(\tau_{n})}+C_{1(n)\mu}^{\left(
\eta\right) }(\tau_{n},z]+\eta_{n\mu}(\tau)=0.   \label{Additive Langevin eq}
\end{equation}

If (\ref{Additive Langevin eq}) were linear, we would interpret it as a
Langevin equation for the particle trajectories with additive, but colored
noise $\eta(\tau).$ In fact, as it stands, (\ref{Additive Langevin eq}) is
nonlinear, and the noise depends in a complicated way on the trajectories $z$
because the probability distribution $P[\eta,z^{+}]$ is a functional of the
trajectories. As before, we will assume that decoherence suppresses large
fluctuations around the mean-trajectories, allowing us to expand (\ref
{Additive Langevin eq}) in terms of the fluctuations $\tilde{z}$ around the
solutions $\bar{z}$ defined by 
\begin{equation}
\frac{\delta S_{z}[\bar{z}]}{\delta\bar{z}_{n}^{\mu}(\tau)}+C_{1(n)\mu
}^{\left( \eta\right) }(\tau,\bar{z}]=0.   \label{mean z}
\end{equation}
We find the linear stochastic integrodifferential equation for the
fluctuations $\tilde{z}(\tau)=z(\tau)-\bar{z}(\tau):$

\begin{align}
& \eta_{n\mu}(\tau;\bar{z})+\sum_{m}\int d\tau_{m}\left( \frac{\delta
^{2}S_{z}[\bar{z}]}{\delta\bar{z}_{n}^{\mu}(\tau)\delta\bar{z}%
_{m}^{\nu}(\tau^{\prime})}\right) \tilde{z}_{m}^{\nu}(\tau^{\prime})
\label{langevin eq for nonlinear systems 2} \\
& +\sum_{m}\int d\tau_{m}\left( \frac{\delta C_{1(n)\mu}^{\left( \eta\right)
}(\tau,\bar{z}]}{\delta\bar{z}_{m}^{\nu}(\tau^{\prime})}\right) \tilde{z}%
_{m}^{\nu}(\tau^{\prime})=0,  \nonumber
\end{align}
where we have used (\ref{mean z}) to eliminate the non-fluctuating terms$.$
If we let the indices $i,j,...$ denote the set of indices $\left( n,\mu
,\tau^{\prime}\right) ,$ with all integrations and summations implied by
repeated indices, we may write (\ref{langevin eq for nonlinear systems 2})
as 
\begin{equation}
(T_{ij}+R_{ij})\tilde{z}_{j}+\eta_{i}=0, 
\label{Linearized matrix langevin equation}
\end{equation}
where 
\begin{equation}
T_{ij}=\left( \delta^{2}S_{z}[\bar{z}]/\delta\bar{z}_{i}\delta\bar{z}%
_{j}\right)   \label{T matrix}
\end{equation}
and 
\begin{equation}
R_{ij}=\left( \delta C_{1i}^{\left( \eta\right) }[\bar{z}]/\delta\bar {z}%
_{j}\right) .   \label{R matrix}
\end{equation}
$T_{ij}$ is a kinetic term, plus a possible external potential term. The
diagonal terms of $R_{ij}$ represent dissipative forces, while the
off-diagonal terms are nonlocal particle-particle interactions, called
propagation terms in \cite{RavalHu:StochasticAcceleratedDetectors} for the
case of the linear particle-detector model. In (\ref{langevin eq for
nonlinear systems 2}), we have made a Gaussian approximation to the noise by
neglecting all higher order cumulants and evaluating the second order
cumulant $C_{2}^{\left( \eta\right) }(\tau_{1},\tau_{2};z^{+}]$ on the
mean-trajectory $\bar{z}.$ The noise correlator (matrix) for $\eta_{i}$ is
then

\begin{equation}
\left\langle \eta_{n\mu}(\tau)\eta_{m\nu}(\tau^{\prime})\right\rangle
=C_{2(nm)\mu\nu}^{\left( \eta\right) }(\tau,\tau^{\prime};\bar{z}]\equiv
C_{ij}.   \label{C matrix}
\end{equation}
The diagonal terms of the noise matrix gives single particle noise, of the
general type found in QBM models. The off-diagonal terms give what Raval and
Hu have called correlation noise, representing the nonlocal correlations
between different particles as mediated by the field, despite it being
coarse-grained away.

Finally, we should show that the description in terms of stochastic fields $%
\chi$ is equivalent to that in terms of local stochastic forces $\eta.$
Making use of the form of $S_{IF}[j^{\pm}],$ we may immediately evaluate the
first cumulant, which is given by 
\begin{equation}
C_{1\mu}^{\eta}(\tau,\bar{z}]=\int dx\,dx^{\prime}\frac{\delta j(x,\bar{z}]}{%
\delta\bar{z}^{\mu}(\tau)}G^{R}(x,x^{\prime})j(x^{\prime},\bar{z}].
\end{equation}
This gives the same result for dissipation as was found in Eq. (\ref{field
chi langevin}). The noise $\eta$ is 
\begin{equation}
\eta_{\mu}(\tau)=\int dx\frac{\delta j(\bar{z},x)}{\delta\bar{z}_{n}^{\mu
}(\tau)}\chi(x).   \label{noise equality}
\end{equation}
We find the correlation of $\eta$ by direct computation:

\begin{align}
\left\langle \eta_{\mu}(\tau)\eta_{\nu}(\tau^{\prime})\right\rangle &
=\left. \frac{\delta^{2}S_{IF}[j^{+},j^{-}]}{\delta z^{\mu-}(\tau)\delta
z^{\nu-}(\tau^{\prime})}\right| _{z^{-}=0}  \nonumber \\
& =\int dx\,dx^{\prime}\text{{\LARGE \{}}\frac{\delta j(x,\bar{z}]}{\delta
z^{\mu}(\tau)}\frac{\delta j(x,\bar{z}]}{\delta z^{\nu}(\tau^{\prime})}
\label{correlator equality} \\
& +\frac{\delta j(x,\bar{z}]}{\delta z^{\mu}(\tau^{\prime})}\frac{\delta j(x,%
\bar{z}]}{\delta z^{\nu}(\tau)}\text{{\LARGE \}}}G^{H}(x,x^{\prime }) 
\nonumber \\
& =\int dx\,dx^{\prime}\frac{\delta j(x,\bar{z}]}{\delta z^{\mu}(\tau)}\frac{%
\delta j(x,\bar{z}]}{\delta z^{\nu}(\tau^{\prime})}\left\langle
\chi(x)\chi(x^{\prime})\right\rangle ,  \nonumber
\end{align}
where we have used the symmetry of the Hadamard function in (\ref{correlator
equality}). This agrees with (\ref{noise equality}). Therefore, one finds
the same stochastic equations of motion whether one works in terms of a
stochastic field $\chi$, extended though space, or a stochastic force $\eta$
acting solely on the particle worldline.

\section{Generalized fluctuation-dissipation relations}

Fluctuation-dissipation relations (FDRs) play an important role in both
equilibrium and nonequilibrium statistical physics. They are vital in
demonstrating the balance necessary for statistical dynamical stability of a
system in the presences of external fluctuation forces. Because of the
intimate connection between statistical, quantum, and relativistic processes
in which fluctuations and dissipation play a role, we now address the
question of whether fluctuation-dissipation relations exist for the more
general nonlinear particle-field models under study. This inquiry will be
important for understanding the stochastic limit of particles in quantum
fields, and the relationship between dissipation, stochasticity and vacuum
fluctuations.

For linear quantum Brownian motion, the Langevin equation takes the form 
\begin{equation}
\ddot{q}+\int dt^{\prime}\mu(t,t^{\prime})\dot{q}(t^{\prime})+\omega^{2}q=%
\xi(t).
\end{equation}
The fluctuation-dissipation relation is an integrodifferential identity
relating the dissipation kernel $\mu(t,t^{\prime})$ and the noise kernel $%
\nu(t,t^{\prime})=\left\langle \xi(t)\xi(t^{\prime})\right\rangle _{s}$ by a
universal kernel $K(t,t^{\prime}$) whose properties are independent of the
spectral density of the environment and the coupling strength $e.$ For QBM,
defining the kernel $\gamma(s)$ by $\mu(s)=d\gamma(s)/ds,$the
fluctuation-dissipation relation is 
\begin{equation}
\nu(s)=\int_{-\infty}^{\infty}ds^{\prime}K(s,s^{\prime})\gamma(s)
\end{equation}
where 
\begin{equation}
K(s,s^{\prime})=\int_{0}^{\infty}\frac{kdk}{\pi}\text{coth}\left( \frac{%
\beta k\hbar}{2}\right) \text{$\cos$}k\left( s-s^{\prime}\right) .
\end{equation}
When the dissipation is approximately local, and the noise approximately
white, we recover the familiar Einstein-Kubo relation \cite{Kubo-relations}.
A special feature of the ordinary FDR for this condition is that the kernel $%
K$ is independent of the system variables. This turns out not to be true for
nonlinear theories.

For the full nonlinear Langevin equation (\ref{Additive Langevin eq}) found
in the previous section, any FDR's that are found should take the form of a
relationship between the cumulant $C_{1\left( n\right) \mu }^{\left( \eta
\right) }(\tau _{n},z],$ giving radiation reaction, and the noise correlator
for $\eta _{n\mu }\left( \tau _{n}\right) .$ But for the nonlinear Langevin
equation this connection between noise and radiation reaction is much more
complex than the familiar one characterizing the linear response regime. In
paper II, we see an example of the counter-intuitive behavior of the
nonlinear Langevin equation (\ref{Additive Langevin eq}). There it is shown
that for uniformly accelerated particles, the radiation reaction term given
by the first cumulant (which is often conveniently yet wrongly thought of as
dissipation) vanishes despite the presence of quantum fluctuation-induced
noise. The relationship between radiation reaction and the noise embodied in
any kind of general FDR is of a subtle nature, and it is not correct to
conclude, on the basis of one's experience with the linear response or
equilibrium regimes, that radiation reaction is flatly {\it balanced } by
fluctuations. What is balanced by fluctuations are the variations in the
radiation reaction force about the average (i.e. classical) force associated
with deviations of the particle trajectories around the average (i.e.
semiclassical) trajectory. We identify the stochastically varying forces
associated with fluctuations in the trajectories as the true dissipative
effect, and call this quantum dissipation because of its quantum origin
(i.e. this effect vanishes when $\hslash \rightarrow 0$). Therefore, the
proper statement is that quantum dissipation is balanced by quantum
fluctuation induced noise.

To see the emergence of the usual kind of FDR\ within the linear response
regime, we should consider the linearized Langevin equation (\ref{Linearized
matrix langevin equation}) for fluctuations $\tilde{z}^{\mu}$ around the
semiclassical solution $\bar{z}^{\mu}.$ Such a FDR will be a relation
between the matrices $R_{ij}$ and $C_{ij}$ in (\ref{R matrix}) and (\ref{C
matrix}), respectively. However, we can immediately see that there will not
be a completely general FDR involving all components of these matrices \cite
{RavalHu:StochasticAcceleratedDetectors}. The matrix $R_{ij}\ $is dependent
on the field commutator evaluated at the spacetime points $x_{i}$ and $%
x_{j}. $ For spacelike separated points, the commutator identically
vanishes. On the other hand, the matrix $C_{ij}$ is proportional to the
anticommutator evaluated at these points, which does not vanish. For
example, two oppositely charged particles in a uniform electric field
sufficiently separated will never enter each other's causal future, and
therefore $R_{12}$ will always vanish. However, the particles will be
correlated through the nonlocal vacuum fluctuations of the quantum field.
These nonlocal correlations between the two particles will be determined by
a non-vanishing $N_{12}$ (see Figure 2).

In exploring the question of what kind of FDR relations do obtain, we
generalize the treatment in RHA \cite{RavalHu:StochasticAcceleratedDetectors}
for particle detectors on fixed trajectories linearly coupled to a quantum
field. We will find that an FDR looking superficially like the one found in 
\cite{RavalHu:StochasticAcceleratedDetectors} does obtain, but that it is in
fact a much more complicated relation. We first define the retarded matrix $%
G_{ij}^{R}\equiv G^{R}\left( z_{i},z_{j}\right) .$ This matrix is the same
as $\tilde{\mu}_{ij}=%
\mathop{\rm Re}%
(Z_{ij})$ in \cite{RavalHu:StochasticAcceleratedDetectors} except for the
important difference that here the matrix is a nonlinear function of the
particle variables, whereas in RHA, the kernel $Z_{ij}$ is independent of
the particles dynamical variables.

We first consider the diagonal elements of $G_{ii}^{R},$ which, in part, are
responsible for radiation-reaction forces on the $i$th particle. When the
field is massless, radiation from the particle will travel outward on its
future lightcone$.$ Assuming that the particle trajectory $\bar{z}_{i}$ is
everywhere timelike, it will then only interact with its own field at the
instant of emission. It is straightforward to see that as a consequence of
this, $G_{ii}^{R}\propto\delta(\tau_{i}-\tau_{i}^{\prime}),$ and hence is
local. Actually, these conclusion depend on the spacetime dimension. In 2+1
dimensions, the massless, retarded Green's function is not restricted to the
lightcone, and so radiation reaction is {\it not }local. In 1+1 dimensions,
infrared divergences are a problem because the retarded Green's function
does not fall-off with distance, and therefore, radiation fields (and
radiation reaction) must be carefully defined (see \cite{Phil-Thesis}).

In 3+1 dimensions (unlike 1+1), we must confront an additional complication: 
$G_{ii}^{R}$ diverges as $\tau_{i}^{\prime}\rightarrow\tau_{i}$, and thus
does not give a finite radiation-reaction force as it stands. We follow the
approach of writing the kernel $G_{ij}^{R}$ as 
\begin{equation}
G_{ij}^{R}=\frac{1}{2}(G_{ij}^{R}-G_{ij}^{A})+\frac{1}{2}%
(G_{ij}^{R}+G_{ij}^{A}),
\end{equation}
and define the new matrices 
\begin{equation}
G_{ij}^{C}\equiv\frac{1}{2}(G_{ij}^{R}-G_{ij}^{A})
\end{equation}
and 
\begin{equation}
G_{ii}^{P}=\frac{1}{2}(G_{ij}^{R}+G_{ij}^{A}).
\end{equation}
$G_{ij}^{A}$ is the advanced Green's function evaluated at $i$ and $j.$ The
matrix $G_{ij}^{C}$ is given by the field commutator $G^{C}(x^{\prime
},x)=\left\langle \left[ \hat{\varphi}(x^{\prime}),\varphi(x)\right]
\right\rangle $ evaluated at $x^{\prime}=z_{i}$ and $x=z_{j}$. If we define
the radiation fields by $\varphi^{rad}(t_{f})\equiv\varphi^{out}(t_{f})-%
\varphi^{in}(t_{i})$, then 
\begin{equation}
\varphi^{rad}(x)=\int G^{C}(x,x^{\prime})j(x^{\prime})dx^{\prime}. 
\label{Classical radiation}
\end{equation}
The energy carried to infinity by $\varphi^{rad}$ is consistent with the
Larmor formula (or its scalar field equivalent). But while the Larmor
formula gives the net radiant energy assuming that the particle is
un-accelerated at initial and final times (or that the particle is
undergoing periodic motion), it does not account for all instantaneous
forces acting on the particles. In particular, there are both short-range
polarization type forces (including the so-called acceleration or Shott
fields) acting between particles and modifying the radiation reaction forces
giving corrections to what would be expected on the basis of the Larmor
formula alone \cite{Classical-Radiation-Reaction}. Neither of these
additional types of forces do net work, but they do modify the equations of
motion. Since we are interested in the detailed, finite-time evolution of
the particles, these type of forces should not be neglected.

In fact, only the off-diagonal elements of the matrix $G_{i\neq j}^{P}$ give
Coulomb-like (or polarization) effects. The local, but divergent diagonal
elements $G_{ii}^{P},$ renormalize the particle masses. Both of these
effects are independent of the quantum state of the field, since they depend
on the field commutator, and hence do not require a quantum mechanical
treatment to understand.

In Paper II \cite{JH2}, we will discuss renormalization effects in detail;
here, we assume that renormalization has been performed in order to focus on
the FDR question. We assume that the diagonal $G_{ii}^{P}$ dependent terms
have been combined with the kinetic terms to give mass renormalization, and
we define the dissipation matrix 
\begin{equation}
A_{ij}=G_{ij}^{R}-G_{ii}^{P}\delta_{ij}. 
\label{Z renormalization condition}
\end{equation}
Similarly, we define the noise matrix 
\[
N_{ij}\equiv G^{H}\left( z_{i},z_{j}\right) , 
\]
which is found from the field anticommutator.

Restricting to the single ($i^{\text{th}}$) particle, we define

\begin{equation}
\Gamma=-\frac{d}{d\tau}A_{ii}=-\dot{z}_{i}^{\mu}\partial_{\mu}^{i}A_{ii}, 
\label{Gamma derivative of Diss2}
\end{equation}
in analogy with the usual FDR in QBM. The chain rule has been used to write $%
d/d\tau=\dot{z}_{i}^{\mu}\left( \partial/\partial z_{i}^{\mu}\right) \equiv%
\dot{z}_{i}^{\mu}\partial_{\mu}^{i}$. In the expressions (\ref{Z
renormalization condition}) and (\ref{Gamma derivative of Diss2}), the index 
$i$ is not summed over, and the derivative is only applied to the first $%
z_{i}$ term in $A_{ii}\left( \tau,\tau^{\prime}\right) =A(z_{i}\left(
\tau\right) ,z_{i}\left( \tau^{\prime}\right) )$. Our definition and
placement of the derivative agrees with \cite
{RavalHu:StochasticAcceleratedDetectors}, where one can find a discussion on
why the appropriate placement of the derivative depends on whether the
interaction term is written as derivative coupling, or minimal coupling. Our
case is a modified example of derivative coupling.

We may now demonstrate the existence of a FDR for the diagonal elements of $%
A_{ii}$ and $N_{ii}$ by making use of the general properties of the
commutator and anticommutator for a field in an initial thermal state at
temperature T. For the diagonal elements, we have $A_{ii}\left( \tau
,\tau^{\prime}\right) =G_{ii}^{C}.$ We will now drop the $i$ index
altogether, writing $A_{ii}=A$, $N_{ii}=N,$ and $z_{i}=z,$ since we are
concerned with only one particle in the following. Defining $x^{\mu}\left(
\tau,\tau^{\prime}\right) \equiv z^{\mu}(\tau)-z^{\mu}(\tau^{\prime})$ and $%
\kappa\equiv\left| {\bf k}\right| ,$ and using the Fourier transforms of the
various Green's functions, we have

\begin{align}
N\left( \tau,\tau^{\prime}\right) & =\int
dk\,e^{-ik_{\mu}x^{\mu}}\delta(k^{2})  \label{Noise matrix integral} \\
& \times\left( \theta(k^{0})-\theta(-k^{0})+2n(\kappa)\right)  \nonumber \\
& =\int d{\bf k}e^{i{\bf k\cdot x}}\left( \frac{e^{-i\kappa
x^{0}}+e^{+i\kappa x^{0}}+2n(\kappa)}{2\kappa}\right) ,  \nonumber
\end{align}
where $n\left( \kappa\right) $ is the Bose distribution term 
\begin{equation}
n(\kappa)=\frac{1}{\text{$\exp$}\left[ \kappa/k_{B}T\right] -1}.
\end{equation}
$A\left( \tau,\tau^{\prime}\right) $ is similarly given by 
\begin{equation}
A\left( \tau,\tau^{\prime}\right) =\int d{\bf k}e^{i{\bf k\cdot x}}\left( 
\frac{e^{-i\kappa x^{0}}-e^{+i\kappa x^{0}}}{2\kappa}\right) ,
\end{equation}
without the thermal Bose distribution factor. We assume that the particle
trajectories are timelike so that $x$ is timelike. Then, after taking the
gradient of $A,$ we may choose a coordinate system so that ${\bf x}=0.$ We
find the simple result 
\begin{equation}
{\bf \nabla}_{{\bf x}}A^{3+1}(x)|_{{\bf x}=0}=\int d{\bf k\,k}\left( \frac{%
\text{$\sin$}\kappa x_{0}}{\kappa}\right) =0
\end{equation}
since the integral is odd in ${\bf k}$, while 
\begin{align}
\frac{\partial}{\partial x_{0}}A^{3+1}(x)|_{{\bf x}=0} & =4\pi\int
d\kappa\,\kappa^{2}\text{$\cos$}\left( \kappa x_{0}\right)  \nonumber \\
& =-4\pi\frac{\partial^{2}}{\partial x_{0}^{2}}\int_{0}^{\infty}d\kappa
\,\cos\left( \kappa x_{0}\right)  \nonumber \\
& =-16\pi^{2}\frac{\partial^{2}}{\partial x_{0}^{2}}\delta\left( x_{0}\right)
\end{align}
in $3+1$ dimensions. For comparison with other work, we also note that in $%
1+1$ dimensions $\partial/\partial x_{0}\left( A^{1+1}(x)|_{{\bf x}%
=0}\right) =4\pi\delta(x_{0})$ and $\nabla_{{\bf x}}A^{1+1}|_{{\bf x}=0}=0.$
Use of 
\begin{equation}
\delta(z^{0}(\tau)-z^{0}(\tau^{\prime}))=\delta(\tau-\tau^{\prime})/\left|
dz^{0}/d\tau\right| _{\tau=\tau^{\prime}},
\end{equation}
requires the trajectories to be timelike, which we assume is the case, at
least for $\bar{z}\left( \tau\right) .$ Therefore, $z^{0}\left( \tau\right) $
is a monotonic function of $\tau,$ $\dot{z}^{0}\left( \tau\right) >0,$ and
we may use 
\begin{equation}
\frac{\partial^{2}}{\partial z_{0}^{2}}=\frac{1}{\dot{z}_{0}\left(
\tau\right) }\frac{\partial}{\partial\tau}\left( \frac{1}{\dot{z}_{0}\left(
\tau\right) }\frac{\partial}{\partial\tau}\right) .
\end{equation}
In then follows that (\ref{Gamma derivative of Diss2}) is given by 
\begin{align}
\Gamma^{3+1} & =16\pi^{2}e^{2}\dot{z}^{0}\frac{\partial^{2}}{\partial
z_{0}^{2}}\delta(z^{0}(\tau)-z^{0}(\tau^{\prime}))  \nonumber \\
& =16\pi^{2}e^{2}\frac{\partial}{\partial\tau}\left[ \frac{1}{\dot{z}_{0}}%
\frac{\partial}{\partial\tau}\left( \frac{1}{\dot{z}^{0}}\delta(\tau
-\tau^{\prime})\right) \right]
\end{align}
in $3+1,$ and by 
\begin{align}
\Gamma^{1+1} & =-4\pi e^{2}\dot{z}^{0}\delta(z^{0}(\tau)-z^{0}(\tau^{\prime
}))  \label{Gamma 1+1} \\
& =-4\pi e^{2}\delta(\tau-\tau^{\prime}).   \label{Local dissipation}
\end{align}
in $1+1.$ This result holds for any state of the field since it is derived
from the field commutator. Since $\partial^{\mu}A$ is covariant, this result
is in fact coordinate independent, despite our use of ${\bf x}=0$ in the
intermediate steps of the derivation.

For the noise matrix element $N,$ we have 
\begin{equation}
N=\int d\kappa\,b(1+2n(\kappa))\left\{ e^{-i\kappa(x^{0}-\left| {\bf x}%
\right| )}+e^{+i\kappa(x^{0}+\left| {\bf x}\right| )}\right\} ,
\end{equation}
where $b=4\pi\kappa$ in $3+1,$ and $b=\kappa^{-1}$ in $1+1$ dimensions,
respectively. Finally, we define the fluctuation-dissipation kernel 
\begin{align}
K(\tau,s) & =\int_{0}^{\infty}\frac{d\kappa}{2\pi}b[1+2n(\kappa)]  \nonumber
\\
& \times\text{{\Large \{}$\cos$}\kappa\left( v(\tau\right) -v(s))  \nonumber
\\
& +\text{$\cos$}\kappa\left( u(\tau\right) -u(s))\text{{\Large \}}}
\end{align}
in terms of the out-going coordinates $u=x^{0}-\left| {\bf x}\right| $ and
the in-coming coordinates $v=x^{0}+\left| {\bf x}\right| $. We may now
verify the fluctuation dissipation relation, 
\begin{equation}
N[z;\tau,\tau^{\prime})=\int_{-\infty}^{\infty}dsK[z;\tau,s)\Gamma\lbrack
z;s,\tau^{\prime}),   \label{FDR1}
\end{equation}
by substituting $K,$ and using the locality of $\Gamma$. Eq. (\ref{FDR1})
follows immediately by virtue of the delta function in Eq. (\ref{Gamma 1+1})
for $\Gamma^{1+1}.$ In $3+1$ dimensions, we integrate by parts thereby
shifting the operator $\partial^{2}/\partial z_{0}^{2}$ from $\Gamma^{3+1}$
to $K$ in Eq. (\ref{FDR1}). The proof of the FDR then follows immediately,
once again, by virtue of the delta function. $K_{ii}$ agrees with the kernel
found by Raval, Hu and Anglin for $T=0$ in $1+1$ dimensions \cite
{RavalHu:StochasticAcceleratedDetectors}. This FDR generalizes their result
to the more physical $3+1$ dimensions, to non-zero temperatures for the
quantum field $\varphi$, and to the case of nonlinear interactions.

Of particular importance is that unlike the traditional FDR (which is a
universal kernel independent of the system dynamics), the kernel $K$ here is
a functional of the system history $z_{i}.$ Hence, (\ref{FDR1}) generally
represents a complicated (nonlinear) relationship between the noise and
dissipation matrices. We have already commented on the significant
difference between (\ref{FDR1}) and the ordinary FDR that obtains for the
linear response regime, one example being the vanishing of the (average)
radiation reaction force for uniformly accelerated charges despite the
existence of quantum fluctuations. This indicates that the formal
relationship as embodied by (\ref{FDR1}), while looking like that kind of
FDR found for linear theories, must be carefully interpreted.

\section{Conclusion}

In this paper we have presented a general framework for describing
stochastic particle motion in quantum fields. These classical nonlinear
stochastic equations of motion are good approximations to the underlying
quantum theory, in the sense that the noise average of the stochastic
variables are nearly identical in form to the expectation values of the
corresponding quantum operators for the particle spacetime coordinates, when
the coarse-grained particle trajectories are sufficiently decohered.
Specifically, the equations of motion follow from the generating functional
for worldline correlation functions. When decoherence is sufficiently
strong, quantum features associated with higher-orders quantum fluctuations
(the ${\cal O(}\breve {z}^{-^{3}})$ terms in the CGEA) are exponentially
suppressed. After doing the $\breve{z}$ path integrals to one-loop, the
remaining $z^{+}$ path-integrations are equivalent to a stochastic path
integral with respect to some system-history-dependent non-Gaussian
probability distribution. Hence, one finds that the quantum dynamics of
correlation functions are approximated by non-Markovian stochastic dynamics
with colored noise- the quality of the approximation depending on the degree
of decoherence.

The resulting nonlinear Langevin equations for the particle spacetime
trajectories may be expressed in a number of equivalent ways: in terms of a
stochastic force $\eta\left( \tau\right) $ coupling directly to the particle
coordinate, a stochastic variable $\Xi$ coupling to the functional $f[z,t),$
or a stochastic field $\chi\left( x\right) $ coupling to the conserved
current $j\left( x\right) .$ A cumulant expansion may be used to
systematically incorporate higher-order noise contributions. At order $e^{2}$
(lowest order in the cumulant expansion after the dissipation term) the
noise is Gaussian. At high temperatures, the approximately Gaussian noise is
effectively white. As one moves from the classical to stochastic regimes,
particularly at low temperature and for short-times, higher order
(non-Gaussian) noise plays an increasingly important role. Moving deeper
into the quantum regime (when decoherence is less complete), higher-order
quantum corrections become significant and the stochastic approximation
outlined here fails.

In Paper II, we will apply the formalism developed here to the specific
problem of radiation reaction for spinless particles. We derive a Langevin
equation whose average trajectory is well-approximated by the
Abraham-Lorentz-Dirac equation at late times, but where time-dependent
renormalization (arising from the nonequilibrium quantum treatment) modifies
the early time behavior making it pathology free.

\section*{Acknowledgments}

This work is supported by NSF grant PHY98-00967 and DOE grant
DEFG0296ER40949.

\appendix

\section{Initial-value methods}

The Closed-Time-Path (CTP), Schwinger-Keldysh, or `in-in' method \cite
{Schwinger-Keldysh,CalzettaHu87(CTP/backrection),Calzetta-Hu-cddn,CTP} is
designed for the computation of the evolution equations for correlation
functions of quantum operators. This is in contradistinction to the
computation of `in-out' transition amplitudes as in scattering processes, or
in dealing with equilibrium conditions, where one either knows about, or
knows what to expect in, the final state. As such, it is particularly
appropriate for nonequilibrium processes in quantum field theory,
characterized by $[\hat{\rho}_{in},\hat{H}]\neq0,$ and for which the final
state is generally unknown. Even if one knows or assumes that the system
does eventually come to equilibrium, how it approaches equilibrium requires
some knowledge of the evolution dynamics involved. Because the CTP formalism
explicitly works with true expectation values $\langle in|\hat{O}%
^{H}|in\rangle$ of a Heisenberg operator evaluated with respect to initial
states, it produces equations of motion that are both real and causal \cite
{DeWitt-Jordan}. These are essential properties that are lacking in the
equations of motion found from the ``in-out'' effective action. When there
is a natural division between system and environment (relevant and
irrelevant) degrees of freedom, the influence functional invented by Feynman
and Vernon \cite{Feynman-Vernon-Hibbs} is most appropriate as it explicitly
implements a coarse-graining of the environment. It has been shown to be
intimately related to the CTP coarse-grained effective action method (see
Appendix C).

\label{Sec-IF}The influence functional is useful when we treat a
distinguishable system (corresponding to the relevant degrees of freedom) in
contact with an environment (the irrelevant degrees of freedom). Such
situations arise frequently in practice, though from the point of view of
the microscopic degrees of freedom, the division is not always so clear-cut.
Indeed, we never directly observe the most elementary degrees of freedom,
and thus it is important to find the effective or collective variables which
capture the most relevant physics of the system at the observation scale in
question. In this work, the mass difference between the particles (e.g.
electrons) and the massless field quanta (e.g. photons) provides the basis
for a reasonable partition between system and environment. For the problems
under study here, we will choose to measure only particle observables,
leaving the field degrees of freedom as the environment; the physical
appropriateness of this choice of the partition depends on the particular
parameters of the theory and the physical questions one is asking. In other
contexts it may be more natural to measure field observables, treating the
particle degrees of freedom as an environment \cite{Calzetta-Hu-Kiefer-etc}.

Our use of the influence functional in this work involves the evolution of
initially uncorrelated states. Despite the mathematical convenience of this
assumption, we must identify the physical conditions under which this
idealization is appropriate. In fact, a sharp system and environment
division of a particle interacting with a field is unattainable: such
factorized states do not exist in infinite dimensional Hilbert spaces since
it may be shown that infinite energy is required to produce them from the
true (interacting) ground states \cite
{RavalHu:StochasticAcceleratedDetectors}. A field cutoff makes factorized
states well-defined mathematically, and reveals the need for regulating the
nonlocal particle-field correlations in the theory in addition to the local
renormalization effects that arises in more traditional field theory
applications.

Time-dependent renormalization is a generic feature of finite-time
nonequilibrium processes, because the system may not have enough time to
equilibrate with the environment. Operating at a faster scale in the
dynamics will be the time-dependent dressing of the particle \cite
{Ramsey97(thesis)}. It is not sufficient to renormalize the parameters of
the theory in the usual way, because the renormalization procedure only
provides local counterterms. Here, we are explicitly interested in the
nonlocal correlations. For example, even with the natural division of system
and environment variables given by the mass gap between the massive particle
and a massless field, the dressed system still involves extensive nonlocal
correlations in addition to the more familiar local shift in the particle
mass. Furthermore, at high energies, the energy gap between massive
particles and massless field quanta is unimportant, and the distinction
between particle and field variables becomes blurred.

\subsection{Path integral derivation of evolution kernels}

The system variables $z_{n}^{\mu}$ are the spacetime coordinates of the
worldlines of a collection of charged particles (the subscript denotes the $%
n^{th}$ particle), each independently parametrized by $\tau_{n}$. The
environment/field $\varphi\left( x\right) $ is equivalent to a collection of
harmonic oscillators. We employ a field description for the environment of
massless particles because radiation is a significant feature, even at low
energy, due to the zero-mass threshold for the production of massless field
quanta. On the other hand, at low energies, creation of massive charged
particles is strongly suppressed. Therefore, we opt for the worldline
description of massive-charged particles to take advantage of its
particle-like features. The description of particle creation, and the
handling of states with indefinite particle number (such as coherent field
states), are conveniently depicted by field-theory, but particle creation,
and related processes, can also be addressed in the worldline framework \cite
{worldline}. The naturalness of the worldline framework can be seen from a
background field perspective, where the mean particle trajectory is
analogous to a mean-field.

We call the closed system of particles plus a field the universe, and its
action will be assumed to be of the form 
\begin{equation}
S[z,\varphi]=S_{z}[z]+S_{\varphi}[\varphi]+S_{int}[z,\varphi].
\end{equation}
In most studies of quantum open systems using the influence functional the
interaction term $S_{int}$ is linear in $z$ and $\varphi,$ while the terms $%
S_{z}$ and $S_{\varphi},$ determining the isolated dynamics of the two
sub-systems, are quadratic \cite
{RavalHu:StochasticAcceleratedDetectors,CaldeiraLeggett}\footnote{%
Hu, Paz, and Zhang have considered in detail quantum Brownian motion (QBM)
with nonlinear (polynomial) particle-bath interactions using the influence
functional and perturbation theory \cite{QBMII}.}. A severe limitation of
quadratic theories with linear coupling is that they are equivalent to
non-interacting free theories in terms of appropriately redefined system and
environment variables that diagonalize the Hamiltonian. Such models do not
explain how a decoherent basis can arise dynamically, and hence are limited
in their ability to account for emergent classical and stochastic behavior
from the effect of noise-induced decoherence.

We work in a initial value formulation that is appropriate to the study of
nonequilibrium quantum open systems, with the aim of deriving the propagator
that gives the evolution of the reduced density matrix $\hat{\rho}_{r,i}$
for a system initially prepared at time $t_{i}$. (In the following, the
subscripts $i/f$ will denote initial/final.) More generally, we should
consider the evolution of $\hat{\rho}_{r,\Sigma_{i}}$ initially defined on
some initial Cauchy hypersurface $\Sigma_{i}.$ The examples in this paper
involve flat (Minkowski) space, but the methods are immediately
generalizable to curved spaces \cite
{HuMataczQBMParametricOsc94,Einstein-Langevin}. The reduced density matrix
associated with coarse-graining the environment field allows us to evaluate
the system observables at spacetime points after $\Sigma_{i}$. To derive its
propagator we begin by considering the full evolution operator for the
closed system. We assume (as part of the initial condition) that the
particles begin on the surface $\Sigma_{i}$ with parameter values $\tau_{i} $%
, where $z^{0}\left( \tau_{i}\right) =t_{i}$. The initial state of particles
plus field is therefore defined on $\Sigma_{i}.$

We assume a unitary operator $\hat{U}_{fi}$ giving the evolution of the
initial density matrix $\hat{\rho}_{U,i}$of the closed universe according to 
\begin{equation}
\hat{\rho}_{U,f}=\hat{U}_{fi}\hat{\rho}_{U,i}\hat{U}_{fi}^{\dagger }=\hat{J}%
_{fi}\hat{\rho}_{U,i},  \label{density matrix evolution eq}
\end{equation}
where $\hat{J}_{fi},$ the density matrix evolution operator, is defined by
the above expression. We take $\left| z\right\rangle $ to be the (spacetime
position) eigenstates of a Schr\"{o}dinger picture operator $\hat{z}^{\mu }$
such that $\hat{z}^{\mu }\left| z\right\rangle =$ $z^{\mu }\left|
z\right\rangle $ (we drop the particle index $n$ for brevity). The operator $%
\hat{z}^{\mu }$ is conjugate to the momentum operator $\hat{p}^{\mu },$
whose eigenstates are momentum states $|p\rangle .$ These operators satisfy
the ``equal time'' commutation relation $\left[ \hat{z}^{\mu },p^{\nu }%
\right] =i\hslash g^{\mu \nu }.$ Momentum states are physical if they are
on-shell: $p^{\mu }p_{\mu }=m^{2}.$ The spacetime position states $|z\rangle 
$ are not in the physical Hilbert space because they do not satisfy the
Hamiltonian constraint associated with the reparametrization-invariant (i.e.
generally covariant) relativistic particle. However, they do have overlap
with the physical Hilbert space, and therefore may be used as a convenient
basis despite not depicting physical particle states.

Similarly, let $\left| \varphi \right\rangle $ be the eigenstates of the Schr%
\"{o}dinger operators $\hat{\varphi}=\hat{\varphi}(t_{i})$ such that $\hat{%
\varphi}\left| \varphi \right\rangle =\varphi \left| \varphi \right\rangle .$
We may pick as a basis for the universe's Hilbert space ${\cal H}_{U}={\cal H%
}_{z}\otimes {\cal H}_{\varphi }$ the direct product states $\left| z\varphi
\right\rangle =\left| z\right\rangle \otimes \left| \varphi \right\rangle .$
We now define the transition amplitude (i.e. the matrix elements of $\hat{U}%
_{fi})$ to go from the initial state $|z_{i}\varphi _{i}\rangle $ to the
final state $|z_{f}\varphi _{f}\rangle $ by the path integral expression 
\begin{align}
\left\langle z_{f}\varphi _{f}\right| \hat{U}_{fi}\left| z_{i}\varphi
_{i}\right\rangle & =\int_{z_{i}\varphi _{i}}^{z_{f}\varphi _{f}}DzD\varphi
\exp \left\{ \frac{i}{\hbar }S[z,\varphi ]\right\}   \nonumber \\
& \equiv K(z_{f},\varphi _{f};z_{i},\varphi _{i}),
\label{K(z,phi,z,phi) in-out amplitude}
\end{align}
where the measures are defined as 
\begin{align}
\int D\varphi & =\prod_{{\bf x},t}\int_{-\infty }^{\infty }d\varphi \left( 
{\bf x,}t\right)   \label{Definition of phi measure} \\
\int Dz& =\prod_{\tau }\int_{t_{i}}^{\infty }dz^{0}\left( \tau \right)
\int_{-\infty }^{\infty }d{\bf z}\left( \tau \right) .
\label{Definition of z measure}
\end{align}
Actually, this discussion of the quantum theory leaves out many important
details because the relativistic particle has a reparametrization-invariant
action, and is therefore a gauge theory. The particulars of how the
reparametrization-invariant theory is quantized depends on the form of the
action $S_{z},$ and how the gauge redundancy is handled. As it stands (\ref
{K(z,phi,z,phi) in-out amplitude}) is ill-defined since it involves summing
over many gauge-equivalent worldlines. Since we are interested here in the
semiclassical particle behavior, we will make the simplifying assumption
that the worldline path integrals have been suitably gauge-fixed. If there
is residual gauge-freedom (such as the lapse $N$ that appears in the
proper-time path integral representation) than the measure for these
variables will be implicitly included in the definition of $Dz.$ In applying
the stationary phase method to define the semiclassical solution, we will
then fix any residual gauge-variables to the ``classical'' (i.e. extremal)
value. For instance, the proper-time path integral involves an integration
not only over worldlines as given by the measure in (\ref{Definition of z
measure}), but also an integration over all possible proper-times for a
particle to go from $z_{i}$ to $z_{f}.$ However, the stationary phase
approximation is dominated by the worldline whose parameter is the classical
proper-time satisfying 
\begin{equation}
\left( dz^{\mu }\left( \tau \right) /d\tau \right) ^{2}=1.
\label{Proper-time constraint for worldline}
\end{equation}
For simplicity, we shall fix the parametrization to the classical
proper-time value so that the semiclassical worldline solution satisfies (%
\ref{Proper-time constraint for worldline}), and we shall ignore the
integration over residual gauge freedom. Doing this therefore neglects
quantum fluctuations in the value of the lapse $N$. Since such fluctuations
may also have a stochastic character that manifests in the stochastic
regime, we should include them in our consideration for a complete
description of the stochastic regime.

There are additional subtleties to this worldline path integral regarding
the class of worldlines that are considered (this is related to the form of
the action $S_{z}$ and the choice of gauge). In our second series, concerned
with fully quantum relativistic effects, we employ the so-called proper-time
gauge. In that gauge, one includes all possible worldlines going between the
spacetime points $z_{i}$ and $z_{f},$ including spacelike and
backward-in-(Minkowski)-time worldlines. We will discuss then why, and how,
a restricted path integral must be employed where worldline paths going back
in time before $t_{i}$ must be excluded, but {\it all }other possible paths
are included. This restriction is given in the path integral measure $\int
Dz^{\mu}$ where the integration range of $z^{0}$ is $t_{i}\leq z^{0}.$ We
mentioned earlier how this temporal restriction leads to a kind of
finite-size effect modifying the particle dynamics near the initial time
hypersurface. It is a nice feature of the spacetime (worldline) formulation
that finite-size (e.g. boundary) effects coming from both spatial and
temporal restrictions of the path integral may be incorporated on equal
footing.

We may alternatively employ the worldline analog of the Coulomb gauge where $%
t=\tau.$ In this gauge, particle trajectories go strictly forward in time $t,
$ and we can set $t=\tau_{n},$ for all $n.$ Such a gauge choice, which is
inequivalent to the proper-time gauge choice, produces the Newton-Wigner
propagator for relativistic particles, rather than the Feynman propagator.
When particle creation processes are important, the difference is
significant. Here, we assume a priori that we are in a regime where
particle-creation effects are insignificant. In the Coulomb gauge ($\tau=t$%
), $z^{0}$ is no longer an independent degree of freedom, and the Langevin
equations derived in the body of this paper should be modified by $%
z^{\mu}\rightarrow{\bf z}^{i}$ (and $\tau\rightarrow t$).

In fact, insofar as this paper is concerned, we could as well have adopted
the semiclassical paradigm of treating the particle worldlines as classical
variables. We could then just take the coarse-grained effective action found
by integrating out the quantum field degrees of freedom as the definition of
a semiclassical (or stochastic) theory. If we had followed this philosophy,
the results of this paper would have been the same, but we would have lost
precious insight into the meaning (and range of applicability) of the
semiclassical or stochastic theory.

Writing the initial density matrix as in (\ref{density matrix}), the final
density matrix is then defined by 
\begin{align}
& \rho_{U,f}(z_{f},z_{f}^{\prime},\varphi_{f},\varphi_{f}^{\prime })
\label{closed density matrix equation} \\
& =\int d{\bf z}_{i}d{\bf z}_{i}^{\prime}\,d\varphi_{i}d\varphi
_{i}^{\prime}\,  \nonumber \\
& \times
J(z_{f},z_{f}^{\prime},\varphi_{f},\varphi_{f}^{\prime};z_{i},z_{i}^{%
\prime},\varphi_{i},\varphi_{i}^{\prime})  \nonumber \\
& \times\rho_{U}(z_{i},z_{i}^{\prime},\varphi_{i},\varphi_{i}^{\prime }). 
\nonumber
\end{align}
The density matrix evolution operator is given by 
\begin{align}
&
J(z_{f},z_{f}^{\prime},\varphi_{f},\varphi_{f}^{\prime};z_{i},z_{i}^{%
\prime},\varphi_{i},\varphi_{i}^{\prime}) \\
&
=K(z_{f},\varphi_{f},z_{i},\varphi_{i})\,K^{\ast}(z_{f}^{\prime},%
\varphi_{f}^{\prime},z_{i}^{\prime},\varphi_{i}^{\prime})  \nonumber \\
& =\int_{z_{i}\varphi_{i}}^{z_{f}\varphi_{f}}\int_{z_{i}^{\prime}\varphi
_{i}^{\prime}}^{z_{f}^{\prime}\varphi_{f}^{\prime}}DzD\varphi Dz^{\prime
}D\varphi^{\prime}  \nonumber \\
& \times\text{$\exp$}\left\{ \,\frac{i}{\hbar}\left( S[z,\varphi
]\,-\,S[z^{\prime},\varphi^{\prime}]\right) \right\} .  \nonumber
\end{align}

\subsection{Reduced density matrix}

Working with the density matrix evolution operator allows one to treat the
full quantum theory in an initial value framework. However, in many contexts
the \underline{full} quantum theory is not what one is really interested in.
Realistic situations involve the measurement of a subset of the universe's
degrees of freedom. One direct way to implement this physical situation is
to work with the reduced density matrix for the relevant variables (the
system) by tracing out the unobserved variables (the environment). Working
with the reduced density matrix implements one type of coarse-graining; a
discussion of more general types of coarse-graining may be found in \cite
{Hartle-Dowker-Halliwell-Brun}.

Our reduced density matrix formalism follows the effective dynamics of
coarse-grained histories whose members are fine-grained particle
trajectories but completely coarse-grained field histories\footnote{%
The finest-grained histories are the individual histories of the full set of
quantum variables that appear in the path integral for the propagator of the
universe \cite{Hartle-Dowker-Halliwell-Brun} (The classification and
enumeration of histories may also be formulated in terms of projection
operators.) Fine-grained histories are never classical since they will
always interfere with other fine-grained histories that are sufficiently
nearby. A coarse-graining is a specification of sets of fine-grained
histories that is exclusive (no fine-grained history belongs to more than
one coarse-grained history) and exhaustive (all fine-grained histories
belong to some coarse-grained history).}. Through the influence functional
one may construct an effective action $S_{CGEA}$ for these coarse-grained
histories \cite{Hu91-silarg}. It is important to keep in mind that the
coarse-graining of environment variables alone is not sufficient to allow
the description of system histories by classical dynamics -- additional
coarse-graining (smearing of the trajectories) is required to find truly
decoherent sets of histories \cite{Hartle-Dowker-Halliwell-Brun}. It is
through coarse-graining that the statistical aspects characterized by
decoherence, dissipation, and noise arise.

The path integral approach gives a convenient representation for the
evolution kernel of the reduced density matrix $J_{r}$ found by tracing over
the final time environment variables in (\ref{closed density matrix equation}%
). One obtains 
\begin{align}
& \rho _{r,f}(z_{f},z_{f}^{\prime })  \label{reduced density matrix 1} \\
& =\int d\varphi _{f}\rho _{U,f}(z_{f},z_{f}^{\prime },\varphi _{f},\varphi
_{f})  \nonumber \\
& =\int d{\bf z}_{i}d{\bf z}_{i}^{\prime }\,d\varphi _{f}\,d\varphi
_{i}d\varphi _{i}^{\prime }\,  \nonumber \\
& \times J(z_{f},z_{f}^{\prime },\varphi _{f},\varphi
_{f};z_{i},z_{i}^{\prime },\varphi _{i},\varphi _{i}^{\prime })  \nonumber \\
& \times \rho _{U,i}(z_{i},z_{i}^{\prime },\varphi _{i},\varphi _{i}^{\prime
})  \nonumber \\
& \equiv \int d{\bf z}_{i}d{\bf z}_{i}^{\prime }\,J_{r}(z_{f},z_{f}^{\prime
};z_{i},z_{i}^{\prime })\rho _{r}(z_{i},z_{i}^{\prime }).  \nonumber
\end{align}
Equation (\ref{reduced density matrix 1}) may be taken as the definition of $%
J_{r}(z_{f},z_{f}^{\prime };z_{i},z_{i}^{\prime }),$ which in general
depends on the initial state $\rho _{r}(z_{i},z_{i}^{\prime }).$ Closed form
expressions may be found for $J_{r}$ only for special initial states $\hat{%
\rho}_{U,i}.$ The most easily treated examples are systems and environments
that are initially uncorrelated, and which therefore have initially
factorized density matrices

\begin{equation}
\hat{\rho}_{U,i}=\hat{\rho}_{z,i}\otimes\hat{\rho}_{\varphi,i}
\end{equation}
[see (\ref{density matrix})] and evolution kernel $J_{r}$ given by [see (\ref
{jr})]. The influence functional is given by $F[z,z^{\prime}]$ and the
influence action is defined by $S_{IF}[z,z]=-i\hbar\ln F[z,z^{\prime}].$ The
coarse-grained effective action is defined by

\begin{equation}
S_{CGEA}[z,z^{\prime}]=S_{z}[z]-S_{z}[z^{\prime}]+S_{IF}[z,z^{\prime}]. 
\label{Effective action S=S=S'+I}
\end{equation}

An alternative form for $F[z,z^{\prime}]$ may be found by considering the
evolution operator for the environment degrees of freedom interacting with
the system variables treated as if they were a classical (c-number) source.
Defining the time-dependent Hamiltonian 
\begin{equation}
\hat{H}_{\varphi}[z]=\hat{H}_{\varphi}[\hat{\varphi}]+\hat{H}_{int}[\hat{%
\varphi},z],
\end{equation}
we use the formal solution for the evolution operator: 
\begin{equation}
\hat{U}[z]=T\,\exp\left\{ -\frac{i\,}{\hbar}\int dt\,\left( \hat{H}%
_{\varphi}[\hat{\varphi}]+\hat{H}_{int}[\hat{\varphi},z]\right) \right\} .
\end{equation}
The field evolution therefore depends on the system history $z$. Writing $%
\left| \varphi_{f},z\right\rangle =\hat{U}[t_{f},t_{i};z]\left| \varphi
_{i}\right\rangle ,$ and using the expression 
\begin{equation}
\left\langle \varphi_{f}\right| \hat{U}[z]\left| \varphi_{i}\right\rangle
=\int_{\varphi_{i}}^{\varphi_{f}}D\varphi\,\exp\left\{ \frac{i\,}{\hbar }%
\left( S_{\varphi}[\varphi]+iS_{int}[\varphi,z]\right) \right\} ,
\end{equation}
allows us to write the influence functional in a representation independent
form as 
\begin{align}
F[z,z^{\prime}] & =\int d\varphi_{f}\,d\varphi_{i}\,d\varphi_{i}^{\prime
}\,\left\langle \varphi_{f}\right| \hat{U}[z]\left| \varphi_{i}\right\rangle
\rho_{\varphi}(\varphi_{i},\varphi_{i}^{\prime},t_{i})\,  \nonumber \\
& \times\left\langle \varphi_{i}^{\prime}\right| \hat{U}[z^{\prime}]\left|
\varphi_{f}\right\rangle  \nonumber \\
& =\text{Tr}_{\varphi}(\hat{U}[z]\hat{\rho}_{\varphi}(t_{i})\hat{U}^{\dagger
}[z^{\prime}]).   \label{Representaion indep. influence functional}
\end{align}
When the environment is initially in a pure state, $\hat{\rho}%
_{\varphi}(t_{i})=\left| \psi_{i}\right\rangle \left\langle \psi_{i}\right| $%
, one obtains the simple result 
\begin{equation}
F[z,z^{\prime}]=\left\langle \psi_{i}\right| \hat{U}^{\dagger}[z^{\prime }]%
\hat{U}[z]\left| \psi_{i}\right\rangle \equiv\left\langle \psi_{f}^{\prime
},z^{\prime}|\psi_{f},z\right\rangle .
\end{equation}
showing how the influence functional measures the interference between the
different environment final states that would result through interactions
with different possible system histories $z$.

There are some general properties of the influence functional (and the
corresponding CGEA) that are worth noting. They follow from the
representation (\ref{Representaion indep. influence functional}) and the
cyclic property of the trace: 
\begin{equation}
S_{CGEA}[z,z^{\prime}]=-S_{CGEA}[z^{\prime},z]^{\ast}\text{ and }%
S_{CGEA}[z,z]=0.
\end{equation}
When the interaction between system and environment are turned off, it
immediately follows from (\ref{Representaion indep. influence functional})
that 
\begin{equation}
S_{IF}[z,z^{\prime}]=0\Rightarrow F[z,z^{\prime}]=1.
\end{equation}
More generally, it may be shown using the representation (\ref{Representaion
indep. influence functional}) that $\left| F[z,z^{\prime}\right| \leq1$ for
all $z.$ This property plays a crucial role in decoherence, and the
emergence of a semiclassical stochastic limit.

\section{Quantum Brownian motion and other models}

\subsection{Time dependent linear QBM}

An important application of the influence functional method is linear
quantum Brownian motion, for which exact results are known. We review here
the case for time-dependent frequencies and coefficients following the
treatment of Hu and Matacz \cite{HuMataczQBMParametricOsc94}. These results
are directly applicable to a number of interesting examples where particle
creation via parametric amplification is important, such as in quantum/atom
optics and quantum field theory in curved spacetime, such as Hawking and
Unruh radiation.

The most general action for which the exact influence functional may be
found is 
\begin{align}
& S[z,q]=S[z]+\sum_{n}\int ds  \nonumber \\
& \times\text{{\LARGE (}}\frac{1}{2}m_{n}(s)\left( \dot{q}%
_{n}^{2}+b_{n}(s)q_{n}\dot{q}_{n}-\omega_{n}^{2}(s)q_{n}^{2}\right) 
\nonumber \\
& +f_{n}[z;s)q_{n}(s)\text{{\LARGE )}},
\end{align}
with stipulated time-dependent masses $m_{n}\left( s\right) $, frequencies $%
\omega_{n}\left( s\right) ,$ and cross-terms $b_{n}\left( s\right) .$ The
function $f_{n}[z;s)$ is a functional of the system variables, and may be
left arbitrary. When the system and environment are initially factorizable ($%
\hat{\rho}_{zq}=\hat{\rho}_{z}\otimes\hat{\rho}_{q})$ the influence
functional is given by 
\begin{align}
F[z,z^{\prime}] & =\exp\text{{\LARGE \{}}\frac{-i}{\hbar}\sum_{n}\int
_{t_{i}}^{t_{f}}ds\int_{t_{i}}^{s}ds^{\prime} 
\label{QBM influence functional} \\
& \times\left( f_{n}[z;s)-f_{n}[z^{\prime};s)\right)  \nonumber \\
& \times\left( \zeta_{n}(s,s^{\prime})f_{n}[z,s^{\prime})+\zeta_{n}^{\ast
}(s,s^{\prime})f_{n}[z^{\prime},s^{\prime})\right) \text{{\LARGE \}}} 
\nonumber
\end{align}
where the nonlocal influence kernel $\zeta_{n}(s,s^{\prime})$ encodes the
complete influence of the environment on the system. In (\ref{QBM influence
functional}), the real ($\nu)$ and imaginary $\left( \mu\right) $ parts of
the kernel $\zeta$ are called the noise and dissipation kernel, respectively.

It is significant that only the imaginary part (the noise kernel) is
modified by the environment being at a non-zero temperature. The dissipation
kernel modifies the extremal solution of the path integral for the evolution
operator, and imparts a dissipative component in the classical equations of
motion. The noise determines the diffusion term in the master equation,
which is responsible for decoherence. It also determines the stochastic
properties in the Langevin description of the stochastic limit.

When the initial environment is a general Gaussian state, which may be
generated by any combination of squeeze, displacement, and rotation
operators acting on the vacuum state, the influence functional is, even for
the case of time-dependent coefficients, still given by the form (\ref{QBM
influence functional}), with the noise and dissipation kernels expressed in
terms of the Bogolubov coefficients which determine the time-dependence of
the field operators when coupled to system variables. This fact allows the
treatment of time-dependent background fields (e.g. electric fields) or
curved spacetime problems that lead to an effective time dependence of the
parameters in the Lagrangian. This explains the striking analogies between
quantum optics and cosmological particle productions \cite
{HuKangMataczSqueezedVaua95}. Hu and Matacz \cite{HuMataczQBMParametricOsc94}
derived a generalized influence functional and applied it to problems with
arbitrary, but generally fixed, time dependence. Here we review the case
where there are no cross-terms of the type $\varphi_{\alpha}\dot{\varphi}%
_{\alpha},$ and the initial state of the field is the squeezed state given
by 
\begin{align}
\hat{\rho}_{squeezed}\left( t_{i};\alpha\right) & =\prod_{n}\hat{S}%
_{n}\left( r\left( \alpha\right) ,\phi\left( \alpha\right) \right) \\
& \times\hat{\rho}_{thermal}\left( \alpha\right) \hat{S}_{n}^{\dagger
}\left( r\left( \alpha\right) ,\phi\left( \alpha\right) \right) ,  \nonumber
\end{align}
where $\hat{\rho}_{th}$ is the initial thermal state for each environment
mode: 
\begin{equation}
\hat{\rho}_{th}(\alpha)=\left( 1-\exp\left( \frac{-\omega_{\alpha}\hbar }{%
k_{B}T}\right) \right) \sum_{n}\exp\left( \frac{-n\hbar\omega_{\alpha}}{%
k_{B}T}\right) \left| n\right\rangle \left\langle n\right| .
\end{equation}
The squeeze operator is 
\begin{align}
& \hat{S}\left( r\left( \alpha\right) ,\phi\left( \alpha\right) \right) \\
& =\frac{1}{2}\exp\left\{ r\left( \alpha\right) \left( \hat{a}_{\alpha
}^{^{2}}e^{-2i\phi\left( \alpha\right) }-\hat{a}_{\alpha}^{\dagger^{2}}e^{2i%
\phi\left( \alpha\right) }\right) \right\} ,  \nonumber
\end{align}
where $\hat{a}_{\alpha},\hat{a}_{\alpha}^{\dagger}$ are the annihilation and
creation operators for each field mode (and polarization).

The influence functional then has the form (\ref{QBM influence functional}),
where the dissipation and noise kernels are given by 
\begin{align}
\mu\left( t,t^{\prime}\right) & =\frac{i}{2}\int d\omega I\left(
\omega\right) \\
& \times\text{{\LARGE \{}}\left[ \alpha_{\omega}\left( t\right)
+\beta_{\omega}\left( t\right) \right] ^{\ast}\left[ \alpha_{\omega
}(t^{\prime})+\beta_{\omega}\left( t^{\prime}\right) \right]  \nonumber \\
& -\left[ \alpha_{\omega}\left( t\right) +\beta_{\omega}\left( t\right) %
\right] \left[ \alpha_{\omega}\left( t^{\prime}\right) +\beta_{\omega
}\left( t^{\prime}\right) \right] ^{\ast}\text{{\LARGE \}}}  \nonumber
\end{align}
and 
\begin{align}
& \nu\left( t,t^{\prime}\right) \\
& =\frac{1}{2}\int d\omega I\left( \omega\right) \coth\left( \frac {%
\hbar\omega\left( t_{i}\right) }{2k_{b}T}\right) \times  \nonumber \\
& \text{{\LARGE \{}}\cosh2r\left( \omega\right) \left[ \alpha_{\omega
}\left( t\right) +\beta_{\omega}\left( t\right) \right] ^{\ast}\left[
\alpha_{\omega}(t^{\prime})+\beta_{\omega}\left( t^{\prime}\right) \right] 
\nonumber \\
& +\cosh2r\left( \omega\right) \left[ \alpha_{\omega}\left( t\right)
+\beta_{\omega}\left( t\right) \right] \left[ \alpha_{\omega}(t^{\prime
})+\beta_{\omega}\left( t^{\prime}\right) \right] ^{\ast}  \nonumber \\
& -\sinh2r\left( \omega\right) e^{-2i\phi\left( \omega\right) }\left[
\alpha_{\omega}\left( t\right) +\beta_{\omega}\left( t\right) \right] ^{\ast}
\nonumber \\
& \times\left[ \alpha_{\omega}(t^{\prime})+\beta_{\omega}\left( t^{\prime
}\right) \right] ^{\ast}-\sinh2r\left( \omega\right) e^{2i\phi\left(
\omega\right) }  \nonumber \\
& \times\left[ \alpha_{\omega}\left( t\right) +\beta_{\omega}\left( t\right) %
\right] \left[ \alpha_{\omega}(t^{\prime})+\beta_{\omega}\left(
t^{\prime}\right) \right] \text{{\LARGE \}.}}  \nonumber
\end{align}
The $\alpha_{\omega}\left( t\right) ,\beta_{\omega}\left( t\right) $ are the
Bogolubov coefficients determining the time-dependence of the field
annihilation and creation operators. There are found by the solutions to the
first order coupled equations 
\begin{align}
\dot{\alpha}\left( t\right) & =-iA\left( t\right) ^{\ast}\beta\left(
t\right) -iB\left( t\right) \alpha\left( t\right) ,  \nonumber \\
\dot{\beta}\left( t\right) & =iB\left( t\right) \beta\left( t\right)
+iA\left( t\right) \alpha\left( t\right) ,
\end{align}
where 
\begin{align}
A\left( t\right) & =\frac{1}{2}\left( \frac{m_{\alpha}\left( t\right)
\omega_{\alpha}^{2}\left( t\right) }{m_{\alpha}\left( t_{i}\right)
\omega_{\alpha}\left( t_{i}\right) }-\frac{m_{\alpha}\left( t_{i}\right)
\omega_{\alpha}\left( t_{i}\right) }{m_{\alpha}\left( t\right) }\right) , \\
B(t) & =\frac{1}{2}\left( \frac{m_{\alpha}\left( t_{i}\right)
\omega_{\alpha}\left( t_{i}\right) }{m_{\alpha}\left( t\right) }+\frac{%
m_{\alpha}\left( t\right) \omega_{\alpha}^{2}\left( t\right) }{%
m_{\alpha}\left( t_{i}\right) \omega_{\alpha}\left( t_{i}\right) }\right) . 
\nonumber
\end{align}

\section{Closed-Time-Path effective action}

\subsection{``In-in'' generating functional}

In this appendix, we review the CTP method and use it to find the ``in-in''
one-particle-irreducible (1PI) effective action, from which we derive the
real and causal equations of motion for the mean field. One may also use the
CTP method to derive the ``in-in'' N particle irreducible (NPI) effective
action, ($N\rightarrow\infty$ is called the master effective action), from
which the Dyson-Schwinger equations of motion for the correlation hierarchy
may be derived \cite{Calzetta-Hu-cddn,Calzetta-Hu(00):stobol}. This is a
powerful tool for addressing problems where nonperturbative effects from
higher order correlations need be treated in a self-consistent manner, and
it may be applied to the problems we address here if dissipation and
fluctuations are to significantly affect the mean-trajectory solutions.

To formulate the CTP method, we define a ``doubled'' spacetime manifold $%
M_{CTP}=M\times\left\{ 1,2\right\} $ \cite{Ramsey97(thesis)}. $M$ is the
Minkowski (or curved) spacetime manifold in between the initial time
hypersurface $\Sigma(t_{i})$ and the final time hypersurface $\Sigma(t_{f}) $%
, where the two copies of $M$ are joined. Spacetime points ($x^{a})$ carry a
CTP index $a,b,...$ indicating whether they live on $M^{1}$ or $M^{2} $.At $%
\Sigma_{f}$, where $M^{1}$ is joined to $M^{2}$, the points $%
(x^{1},t_{f}^{1})$ and $(x^{2},t_{f}^{2})$ are identified (see Figure 3). We
will usually write $\varphi^{a}(x)$ instead of $\varphi(x^{a}).$ $%
M^{1}(M^{2})$ is often called the positive (negative) time-branch because
positive frequency modes living in $M^{1}(M^{2})$ evolve forward (backward)
in time. The direction of time of positive frequency mode propagation is
determined by the sign in the exponential in (\ref{Z in in}). One can keep
track of this sign by introducing a CTP metric $c^{ab}=\delta^{a1}%
\delta^{b1}-\delta^{a2}\delta^{b2};$ $c^{ab}$ will always be used to
contract CTP indices (e.g. $\hat{\varphi}^{a}h_{a}=\hat{\varphi}^{1}h_{1}-%
\hat{\varphi}^{2}h_{2})$. We define the CTP time-ordering operator $\bar{T}$
so that operators on $M^{1}$ are time-ordered, operators on $M^{2}$ are
anti-time-ordered, and all operators on $M^{2}$ are ordered left of
operators on $M^{1}.$ Histories are to be thought of as beginning on $M^{1}$
at $t_{i},$ moving forward in time to $t_{f}$ where $M^{1}$ joins $M^{2}$ at 
$\Sigma(t_{f}),$ and then going backward in time to $t_{i}^{\prime}$ on $%
M^{2}$, thus the name 'closed-time-path'.

We first consider a single free field. The ``in-in'' generating functional
is given by

\begin{align}
Z_{\text{in-in}}[h^{a}] & =\int d\varphi_{f}\,\left\langle \varphi
_{i}|\varphi_{f}\right\rangle _{h_{2}}\left\langle \varphi_{f}|\varphi
_{i}\right\rangle _{h_{1}}  \label{Z in in} \\
& =\text{Tr}\left[ \bar{T}\,\text{$\exp$}\left\{ \frac{i}{\hbar}\int dx\,%
\hat{\varphi}_{H}^{a}(x)\,h_{a}(x)\right\} \hat{\rho}_{i}\right]  \nonumber
\end{align}
The subscript $H$ indicates that the operators are in the Heisenberg picture
in this expression. It follows that the CTP time-ordered expectation values
are given by 
\begin{align}
& \left\langle \bar{T}\hat{\bar{\varphi}}_{H}^{a}(x_{1})\cdots\hat{\varphi }%
_{H}^{b}(x_{n})\right\rangle _{J_{a}} \\
& \equiv\text{Tr}\left\{ \bar{T}\hat{\bar{\varphi}}_{H}^{a}(x_{1})\cdots 
\hat{\varphi}_{H}^{b}(x_{n})\hat{\rho}_{i}\right\} _{J_{a}}  \nonumber \\
& =\left( \frac{-i}{\hbar}\right) ^{n}Z[h_{a}]^{-1}\left( \frac{\delta
^{n}Z[h_{a}]}{\delta h_{a}(x_{1})\cdots\delta h_{b}(x_{n})}\right) \,. 
\nonumber
\end{align}
The generating functional has the path integral representation 
\begin{equation}
Z[h_{a}]=\int_{CTP}D\varphi^{a}\,\text{$\exp$}\left\{ \frac{i}{\hbar}\left(
S_{\varphi}[\varphi^{a}]+\varphi^{a}h_{a}\right) \right\}
\,\rho_{i}(\varphi_{i}^{a}). 
\label{Z[Ja]=d[p]exp(S[pa]+pJ) CTP functional integral}
\end{equation}
In the last line we used an abbreviated notation where $S_{\varphi}[%
\varphi^{a}]=S_{\varphi}[\varphi^{1}]-S_{\varphi}\left[ \varphi^{2}\right] ,$
$\rho_{i}(\varphi_{i}^{a})=\left\langle \varphi_{i}^{1}\right| \hat{\rho }%
_{i}\left| \varphi_{i}^{2}\right\rangle ,$ and $\varphi^{a}h_{a}=\int
dx\varphi^{a}(x)h_{a}(x)$. The CTP subscript on the integral implies the CTP
boundary conditions on integrations over the initial density matrix. We will
assume that the initial state is approximated by the Gaussian

\begin{equation}
\rho_{i}(\varphi_{i}^{a})=A\,\text{$\exp$}\left\{
-\int\varphi_{i}^{a}(x)K_{ab}(x;x^{\prime})\varphi_{i}^{b}(x^{\prime})\right%
\} ,   \label{CTP Gaussian state}
\end{equation}
where the quadratic kernel 
\begin{equation}
K_{ab}(x^{1};x^{2})=K_{ab}(t_{i},{\bf x}^{1};t_{i},{\bf x}^{2})
\end{equation}
vanishes everywhere except on $\Sigma(t_{i}).$ Any specific initial
condition makes the evolution special, as it carries the particular
information at $t_{i}$. It is in this restricted sense that Lorentz
invariance of the theory is broken (unless $\hat{\rho}$ is the Lorentz
invariant vacuum state for the full theory). Any non-vacuum state picks out
a reference point, as is the case for the influence functional constructed
with a specified initial condition of the system and environment. There are
many physical situations where the information of a chosen initial state or
a fiducial time is encoded in the dynamics, and affects the attributes of a
system. Most nonequilibrium processes are of such a nature, while chaotic
systems have particular sensitivity. For instance, if (\ref{CTP Gaussian
state}) is a thermal state defined on $\Sigma(t_{i})$, the thermal
fluctuations will only be isotropic in the rest frame determined by the time
coordinate $t.$ Another example which picks out a particular reference frame
is when the system and field environment are initially uncorrelated. For
interacting theories, exact factorizability holds only on a single spacelike
surface since interactions inevitably lead to correlations both before and
after such a special surface. For this reason, field-environment induced
noise in a (particle) subsystem is not Lorentz invariant when one assumes
that systems A and B factorize at the initial time\footnote{%
This observation is more generally true, of course. Only the dressed vacuum
state of the interacting theory is Lorentz invariant, and in that case the
influence functional cannot easily be found. So generically, the noise in
the stochastic action will not be Lorentz invariant. Our belaboring on this
general point arose from Prof. Ted Jacobson's insistence on clarity to his
queries.}.

We now consider the case where the field action $S_{\varphi }\left[ \varphi %
\right] $ has the quadratic form 
\begin{align}
S_{\varphi }^{a}\left[ \varphi \right] & =\frac{1}{2}\int dx\left( \partial
_{\mu }\varphi ^{a}\partial ^{\mu }\varphi _{a}+m^{2}\varphi _{a}\varphi
^{a}\right)   \label{Quadratic field action} \\
& =\frac{1}{2}\int dxdx^{\prime }\varphi ^{a}\left( x\right) A_{ab}\left(
x,x^{\prime }\right) \varphi ^{b}\left( x^{\prime }\right) +B.T.,  \nonumber
\end{align}
where 
\[
A_{ab}\left( x,x^{\prime }\right) =c_{ab}\left( -\partial ^{2}+m^{2}\right)
\delta \left( x,x^{\prime }\right) 
\]
and $B.T.$ is the boundary term. Further assuming a Gaussian initial state (%
\ref{CTP Gaussian state}), the functional integral in (\ref
{Z[Ja]=d[p]exp(S[pa]+pJ) CTP functional integral}) may be solved exactly
giving 
\begin{align}
Z[h_{a}]& =\left( \text{$\det $}\left( G_{ab}\right) \right) ^{-1/2} \\
& \times \text{$\exp $}\left\{ \frac{-i}{2\hbar }\int dxdx^{\prime
}h_{a}(x)G^{ab}(x,x^{\prime })h_{b}(x^{\prime })\right\} ,  \nonumber
\end{align}
where the Green's functions 
\begin{align}
G_{ab}(x,x^{\prime })& =A_{ab}^{-1}\left( x,x^{\prime }\right)  \\
& =\frac{\delta ^{2}S[\bar{\varphi}_{a}]}{\delta \varphi ^{a}(x)\delta
\varphi ^{b}(x^{\prime })}  \nonumber
\end{align}
are also given by 
\begin{align}
\left( 
\begin{array}{cc}
G^{11} & G^{12} \\ 
G^{21} & G^{22}
\end{array}
\right) & =\left( 
\begin{array}{cc}
-i\left\langle T^{+}\hat{\varphi}_{x}\hat{\varphi}_{x^{\prime
}}\right\rangle  & -i\left\langle \hat{\varphi}_{x}\hat{\varphi}_{x^{\prime
}}\right\rangle  \\ 
-i\left\langle \hat{\varphi}_{x^{\prime }}\hat{\varphi}_{x}\right\rangle  & 
-i\left\langle T^{-}\hat{\varphi}_{x}\hat{\varphi}_{x^{\prime
}}\right\rangle 
\end{array}
\right)   \nonumber \\
& =\left( 
\begin{array}{cc}
G_{F}(x,x^{\prime }) & G_{+}(x,x^{\prime }) \\ 
G_{-}(x,x^{\prime }) & G_{\bar{F}}(x,x^{\prime })
\end{array}
\right) .
\end{align}
$G_{F}$ is the Feynman and $G_{\bar{F}}$ is the anti-Feynman (also known as
the Dyson) Green's function and $G^{\left( \pm \right) }$ are the positive
(or negative) frequency Wightman functions. $T^{+}$ denotes time-ordering; $%
T^{-}$ denotes anti-time ordering. In the equations of motion the retarded ($%
G^{R}),$ advanced $\left( G^{A}\right) ,$ and Hadamard $\left( G^{H}\right) $
Green's functions play a prominent role. They are given by 
\begin{align}
G_{R}(x,x^{\prime })& =-i\theta (x,x^{\prime })\langle \lbrack \hat{\phi}%
_{x},\hat{\phi}_{x^{\prime }}]\,\rangle =G_{F}-G_{+} \\
G_{A}(x,x^{\prime })& =-i\theta (x^{\prime },x)\left\langle [\hat{\phi}_{x},%
\hat{\phi}_{x^{\prime }}]\right\rangle \,=G_{F}-G_{-}  \nonumber \\
G_{H}(x,x^{\prime })& =-i\left\langle \{\hat{\phi}_{x},\hat{\phi}_{x^{\prime
}}\}\right\rangle =G_{F}+G_{\bar{F}}=G_{+}+G_{-},  \nonumber
\end{align}
These various Green's functions are not independent. They satisfy the
relation 
\begin{equation}
G_{F}+G_{\bar{F}}=G_{+}+G_{-}.
\end{equation}
Defining sum and difference variables $h^{+}\equiv
(h^{1}+h^{2})/2\,,\;\;h^{-}\equiv (h^{1}-h^{2})$ and $\varphi ^{+}\equiv
(\varphi ^{1}+\varphi ^{2})/2,\;\;\varphi ^{-}\equiv (\varphi ^{1}-\varphi
^{2}),$ we can express the generating functional in terms of retarded and
Hadamard Green's functions as 
\begin{align}
Z[h^{\pm }]& =\left( \det G^{ab}\right) ^{-1/2}\exp \text{{\LARGE \{}}-\frac{%
i}{2\hbar }\int dxdx^{\prime }  \nonumber \\
& \times \lbrack 2h^{-}(x)G_{R}(x,x^{\prime })h^{+}(x^{\prime })  \nonumber
\\
& +ih^{-}(x)G_{H}(x,x^{\prime })h^{-}(x^{\prime })]\text{{\LARGE \}}}.
\label{Z[J+,J-]=exp(J- Gr J+  +J- Gh J-) CTP}
\end{align}
The close relationship between the generating functional $Z\left[ h^{\pm }%
\right] $ and the influence functional $F\left[ j^{\pm }\right] $ is now
apparent; in fact, 
\begin{equation}
F\left[ j^{\pm }\right] =\left. \left( \det G^{ab}\right) ^{1/2}Z\left[
h^{\pm }\right] \right| _{h^{\pm }=j^{\pm }\left[ z^{\pm }\right] },
\end{equation}
when the source $h$ is identified with the physical current $j$ produced by
the particles $z^{\mu }\left( \tau \right) .$

The generating functional for normalized expectation values, or connected
correlation functions in the language of Feynman diagrams, is 
\begin{align}
W[h_{a}] & =-i\hbar\ln Z[h_{a}] \\
& =-\frac{1}{2}\int dxdx^{\prime}h_{a}(x)G^{ab}(x,x^{\prime})h_{b}(x^{\prime
})  \nonumber \\
& -\frac{i\hslash}{2}\ln\det G^{ab},  \nonumber
\end{align}
or in terms of the sum and difference variables $h^{\pm}$%
\begin{align}
W[h_{a}] & =-\int dxdx^{\prime}\text{{\LARGE \{}}h^{-}(x)G_{R}(x,x^{\prime
})h^{+}(x^{\prime})  \nonumber \\
& +\frac{i}{2}h^{-}(x)G_{H}(x,x^{\prime})h^{-}(x^{\prime})\text{{\LARGE \}}}-%
\frac{i\hslash}{2}\ln\det G^{ab}  \nonumber \\
& ==S_{IF}\left[ h^{\pm}\right] -\frac{i\hslash}{2}\ln\det G^{ab}.
\end{align}

Thus, $W\left[ h^{a}\right] $ is just the Feynman-Vernon influence action,
up to the log-det term. But for the quadratic action and initial state we
are considering, $G^{ab}$ is independent of $h,$ and so the log-det term is
just a constant. Noting that $h^{a}\varphi_{a}=h^{+}\varphi_{-}+h^{-}%
\varphi_{+},$ we may use (\ref{Z[J+,J-]=exp(J- Gr J+ +J- Gh J-) CTP}) to
find the mean-field and symmetrized Green's function as 
\begin{align}
\left\langle \varphi(x)\right\rangle _{h} & =\left\langle \varphi
^{+}(x)\right\rangle _{h^{-}=0}=\left. \frac{\delta W[h^{+},h^{-}]}{\delta
h^{-}(x)}\right| _{h^{-}=0}  \label{<phi>=G(x,x')J(x'), mean field phi} \\
& =\int dxG_{R}(x,x^{\prime})h(x^{\prime})  \nonumber
\end{align}
and 
\begin{equation}
\left\langle \left\{ \varphi(x),\varphi(x^{\prime})\right\} \right\rangle
_{h}=\frac{\delta^{2}W[h^{+},h^{-}]}{\delta h^{-}(x)\delta h^{-}(x^{\prime})}%
=G_{H}(x,x^{\prime}).
\end{equation}
We will follow the standard terminology and call $\bar{\varphi}_{a}[h^{a}]$
the classical fields, defined by 
\begin{equation}
\bar{\varphi}_{a}[h^{a}]=\frac{\delta W[h^{a}]}{\delta h^{a}},
\end{equation}
This terminology is formal, since the physical (real-situation) meaning of
classicality involves decoherence which requires coarse-graining.

The one-particle-irreducible effective action is the Legendre transformation
of the source dependent functional $W[h^{a}]$ with respect to the classical
field: 
\begin{equation}
\Gamma \lbrack \bar{\varphi}_{a}(x)]=W[h^{a}]-\int dxh^{a}(x)\bar{\varphi}%
_{a}(x).  \label{Free phi effective action}
\end{equation}
The real and causal equations of motion for $\bar{\varphi}_{a}$ are found
from 
\begin{equation}
\frac{\delta \Gamma \lbrack \bar{\varphi}_{a}]}{\delta \bar{\varphi}_{a}(x)}%
=-h^{a}(x).  \label{dG/dp=J (eq motion for phi)}
\end{equation}
For the quadratic case we are treating it is simple to invert the equations
of motion giving 
\begin{align}
h_{a}(x)& =\int dx^{\prime }G_{ab}^{-1}(x,x^{\prime })\bar{\varphi}%
^{b}(x^{\prime }) \\
& =\int dx^{\prime }A_{ab}(x,x^{\prime })\bar{\varphi}^{b}(x^{\prime }). 
\nonumber
\end{align}
Substituting this result for $h_{a}\left( x\right) $ into (\ref{Free phi
effective action}) gives 
\begin{align}
\Gamma \lbrack \bar{\varphi}_{a}(x)]& =\frac{1}{2}\int dxdx^{\prime }\bar{%
\varphi}^{a}(x^{\prime })A_{ab}(x,x^{\prime })\bar{\varphi}^{b}(x^{\prime })
\nonumber \\
& -\frac{i\hslash }{2}\ln \det G^{ab} \\
& =S_{\varphi }^{a}\left[ \bar{\varphi}^{a}\right] -\frac{i\hslash }{2}\ln
\det G^{ab}.  \nonumber
\end{align}
Hence, we see that the effective action is just the classical action
evaluated in terms of the solutions $\bar{\varphi}^{a},$ plus a constant.
This result is only the case for quadratic theories for which quantum
corrections don't change the mean-field equations of motion.

We can also proceed more formally, and generalize these results to
non-quadratic (nonlinear) field actions. First, we substitute (\ref{dG/dp=J
(eq motion for phi)}) into (\ref{Free phi effective action}). Using the path
integral representation (\ref{Z[Ja]=d[p]exp(S[pa]+pJ) CTP functional
integral}) for $Z[h^{a}]$, we find the functional integrodifferential
equation for the effective action 
\begin{align}
\Gamma \left[ \bar{\varphi}_{a}\right] & =-i\hbar \text{$\ln ${\LARGE $\{$}}%
\int_{CTP}D\varphi _{a}\text{$\exp $}\frac{i}{\hbar }(S[\varphi _{a}] 
\nonumber \\
& -\int dx\frac{\delta \Gamma }{\delta \bar{\varphi}_{a}}(\varphi _{a}-\bar{%
\varphi}_{a}))\rho _{i}(\varphi _{a})\text{{\LARGE \}}}.
\end{align}
Defining the fluctuation field $\tilde{\varphi}_{a}(x)=\varphi _{a}(x)-\bar{%
\varphi}_{a}(x),$ and using $D\varphi _{a}=D\tilde{\varphi}_{a},$ we may
rewrite this as the functional integral of the fluctuations around the
classical fields $\bar{\varphi}_{a}$: 
\begin{align}
\Gamma \left[ \bar{\varphi}_{a}\right] & =-i\hbar \text{$\ln ${\LARGE \{}}%
\int_{CTP}D\tilde{\varphi}_{a}\text{$\exp $}\frac{i}{\hbar }(S[\tilde{\varphi%
}_{a}+\bar{\varphi}_{a}]  \label{1PI equation} \\
& -\int dx\frac{\delta \Gamma }{\delta \bar{\varphi}_{a}}\tilde{\varphi}%
_{a})\rho _{i}(\tilde{\varphi}_{a}+\bar{\varphi}_{a})\text{{\LARGE \}}}. 
\nonumber
\end{align}
Equation (\ref{1PI equation}) has the formal solution 
\begin{equation}
\Gamma \lbrack \bar{\varphi}_{a}(x)]=S[\bar{\varphi}_{a}]-\frac{i\hbar }{2}%
\text{$\ln $\thinspace $\det $}\left( G^{ab}(x,x^{\prime })\right) +i\hbar
\Gamma _{1}[\bar{\varphi}_{a}],  \label{formal solution}
\end{equation}
where 
\begin{equation}
G^{ab}(x,x^{\prime })_{\bar{\varphi}}=\left. \frac{\delta ^{2}S_{\varphi
}^{a}\left[ \varphi \right] }{\delta \varphi _{a}\left( x\right) \delta
\varphi _{b}\left( x^{\prime }\right) }\right| _{\varphi _{a}=\bar{\varphi}%
_{a}}
\end{equation}

The $G^{ab}(x,x^{\prime})_{\bar{\varphi}}$ are the one-loop propagators for
the fluctuation fields $\tilde{\varphi}_{a},$ but now they are functionals
of the classical fields $\bar{\varphi}_{a}$. The functional $\Gamma_{1}$ is
given by the sum of one-particle-irreducible (1PI) graphs with external legs
determined by $\hat{\rho}_{i}(\varphi),$ and with propagators given by the
one-loop fluctuation field propagators. These quantum corrections begin at
two-loop. Vertices are given by the shifted interaction term for the
fluctuation fields $\tilde{\varphi}:$%
\begin{align}
S_{int}[\tilde{\varphi}_{a}] & =S[\tilde{\varphi}_{a}+\bar{\varphi}_{a}]-S[%
\bar{\varphi}_{a}]-\int dx\frac{\delta S[\bar{\varphi}_{a}]}{%
\delta\varphi_{a}}\tilde{\varphi}_{a} \\
& -\frac{1}{2}\int dxdx^{\prime}\frac{\delta^{2}S[\bar{\varphi}_{a}]}{%
\delta\varphi_{a}(x)\delta\varphi_{b}(x^{\prime})}\tilde{\varphi}_{a}(x)%
\tilde{\varphi}_{b}(x^{\prime}).  \nonumber
\end{align}
The one-loop contribution to the effective action is given by the
log-determinant term, but now, unlike for the quadratic action case, this
term is not independent of $\bar{\varphi}^{a}.$ For one-loop calculations
one neglects $\Gamma_{1}$, however, under some circumstance important
nonperturbative (with respect to the coupling constant) effects only begin
at two-loop \cite{Jackiw74}. For instance, the time-dependent Hartree-Fock
approximation involves two-loop corrections.

Both the influence functional and CTP approaches may be used to find the
full quantum theory of the particles, and not just the semiclassical limit.
Within the CTP framework, this may be done in the context of the
N-particle-irreducible effective action, or its full extension ($%
N\rightarrow \infty$), the master effective action \cite{Calzetta-Hu-cddn}.
The full quantum theory is equivalent in content to the Dyson-Schwinger
hierarchy of correlation functions \cite{Calzetta-Hu-cddn}. By using the
master effective action, this hierarchy may be truncated consistently at
finite order by slaving higher order correlation functions to some finite
number of lower order correlation functions. Within the influence functional
approach, one may derive the finite time propagator for the system's (in
these studies, particle's) reduced density matrix. This propagator then
provides a description of the effective dynamics of the particles with
backreaction from quantum fields. This may then be used directly, or as an
intermediate step, in deriving the master equation. Solving for the
propagator requires evaluating the functional integrals in (\ref{evolution
kernel}). For linear theories, this may be done exactly. For nonlinear
theories, one may either expand the functional integrals around the
stationary phase solution giving the loop expansion (in order of $\hbar$)
for the propagator, or one may expand the influence functional in the
coupling strength (\ref{ordinary perturbation theory}), giving coupling
constant perturbation theory for the reduced density matrix propagator.

\subsection{CTP coarse-grained effective action}

The coarse-grained effective action (CGEA) was first introduced by Hu and
Zhang \cite{Hu91-silarg} to treat the backreaction of one subsystem on
another, where their separation is justified by the existence of a physical
parameter (e.g., heavy-light mass, fast-slow time, high-low modes) which one
uses to carry out a saddle-point expansion for the action of the open
system. It includes the ordinary (n-loop) effective action as a special case
where the expansion parameter is $\hbar$. Here we apply CGEA to interacting
particles and fields\footnote{%
This Appendix is formal, in that we assume a worldline restricted-path
integral structure without developing the full details, which is the topic
of our second series.}. In this sub-appendix, we extend the Closed-Time-Path
method to the particle plus field model.

To produce a generating functional for both particle and field correlation
functions we couple both particle and field variables to independent
sources, adding the term 
\begin{equation}
\sum_{n}\int d\tau J_{n\mu}\left( \tau\right) z_{n}^{\mu}\left( \tau\right)
+\int dxh\left( x\right) \varphi\left( x\right)
\end{equation}
to the action. For $N$ particles (indexed by $n)$ in $D$ spacetime
dimensions (indexed by $\mu)$ we must have $N\times D$ sources $J_{n\mu}$
independently coupled to the spacetime coordinates of each particle.
However, rather than considering the full generating functional for all
particle and field correlation functions, we shall study the open system
dynamics of particles with the influence functional obtained by
coarse-graining (integrating out) the field degrees of freedom. We do this
by not including a field-source $h\left( x\right) .$

The exact ``in-in" connected generating functional (for particle correlation
functions) is then given by

\begin{align}
W[J^{a}] & =-i\hbar\text{$\ln${\LARGE [}}\int_{CTP}D\varphi^{a}Dz^{a}\text{$%
\exp${\Large \{}}\frac{i}{\hbar}S^{a}[\varphi^{a},z^{a}]  \nonumber \\
& +\sum_{n}\int d\tau J_{\mu}^{a}z_{z,a}^{\mu}\text{{\Large \}}}\rho
(\varphi_{i}^{a};z_{i}^{a})\text{{\LARGE ]}}.
\end{align}
where 
\begin{align}
S[\varphi^{a},z^{a}] & =S_{\varphi}^{a}\left[ \varphi\right] +S^{a}\left[ z%
\right] +S_{int}^{a}\left[ \varphi,z\right]  \nonumber \\
S_{int}^{a}\left[ \varphi,z\right] & =\int dxj^{a}\left( x\right)
\varphi_{a}\left( x\right)  \nonumber \\
S^{a}\left[ z\right] & =S\left[ z\right] -S\left[ z^{\prime}\right] ,
\end{align}
and $S_{\varphi}^{a}\left[ \varphi\right] $ is the quadratic field action
given in (\ref{Quadratic field action}). We are considering cases like QED,
where the nonlinearity arise only through interactions between the particles 
$z$ and field $\varphi.$ As per our earlier discussion (see Appendix A), the
path integrals over the worldlines are restricted to $t_{i}\leq z^{0}.$ The
CTP temporally-ordered correlation functions $\left\langle \bar{T}\hat{z}_{1}%
\hat{z}_{2}...\hat{z}_{n}\right\rangle $ are found from the functional
derivatives of $W[J^{a}]$ with respect to $J^{a}(\tau).$ We define the
``classical'' particle solutions to be 
\begin{equation}
\bar{z}_{a}[J^{a};\tau)=\frac{\delta W[J^{a}]}{\delta J^{a}(\tau)}. 
\label{mean field ctp}
\end{equation}
The solutions $\bar{z}_{a}[J^{a};y)$ are functionals of the sources $J^{a}.$
As we mentioned earlier, calling $\bar{z}^{a}$ the classical solutions is a
formal terminology since classicality involves other issues like
decoherence. The CTP coarse-grained effective action is defined by the
Legendre transform 
\begin{equation}
\Gamma\lbrack\bar{z}_{a}(\tau)]=W[J^{a}]-\int d\tau\,J^{a}(\tau)\bar{z}%
_{a}(\tau).   \label{G=W-gz legendre transfomation}
\end{equation}
$\Gamma\lbrack\bar{z}_{a}(\tau)]$ is the coarse-grained
one-particle-irreducible generating functional for particle correlation
functions. We find the real and causal equations of motion for $\bar{z}_{a}$
from 
\begin{equation}
\frac{\delta\Gamma\lbrack\bar{z}_{a}]}{\delta\bar{z}^{a}(\tau)}=J_{a}(\tau). 
\label{dG/dz=g (CTP equations of motion)}
\end{equation}
When $J^{1}=J^{2},$ the solutions on the two time-branches $M^{a}$ are
equal. The physical solutions are found by setting $J^{a}=0.$

We may now use the results from the previous Appendix to integrate out the
field $\varphi ^{a}$ keeping $z^{a}$ fixed, and assuming that the initial
field and particle states are factorizable. We simply replace the source $%
h^{a}\left( x\right) $ with the physical current $j^{a}\left( x\right) .$
The connected generating functional is then given by 
\begin{align}
W[J^{a}]& =-i\hbar \ln \text{{\LARGE [}}\int_{CTP}Dz^{a}\text{$\exp \frac{i}{%
\hbar }${\LARGE \{}}S^{a}[z]+S_{IF}\left[ j^{a}\right]   \nonumber \\
& -\frac{1}{2}\text{Tr}\ln G^{ab}+\sum_{n}\int d\tau J_{\mu
}^{a}z_{z,a}^{\mu }\text{{\LARGE \}}}\rho (z_{i}^{a})\text{{\LARGE ]}}.
\end{align}
The trace-log term just comes from the normalization factor $\left( \det
G^{ab}\right) ^{-1/2}$ which is independent of $z$.

Substituting (\ref{dG/dz=g (CTP equations of motion)}) into (\ref{G=W-gz
legendre transfomation}) gives an integrodifferential equation for the
effective action 
\begin{align}
\Gamma\lbrack\bar{z}_{a}] & =-i\hbar\ln\text{{\LARGE [}}\int_{CTP}Dz^{a}%
\text{$\exp$}\frac{i}{\hbar}\text{{\Large \{}}S^{a}[z]+S_{IF}\left[ j^{a}%
\right]  \nonumber \\
& -\frac{1}{2}\text{Tr}\ln G^{ab}  \label{1PI effective action1} \\
& -\int d\tau\,\frac{\delta\Gamma}{\delta\bar{z}^{a}}(z^{a}-\bar{z}^{a})%
\text{{\Large \}}}\rho(z_{i}^{a})\text{{\LARGE ]}}.  \nonumber
\end{align}
The difference $z^{a}-\bar{z}^{a}$ is the deviation of the particles path
integration histories from the classical trajectory $\bar{z}^{a}.$ Shifting
variables by defining the fluctuation coordinate 
\begin{equation}
\breve{z}^{a}(\tau)=z^{a}(\tau)-\bar{z}^{a}(\tau),
\end{equation}
and using the invariance of the functional measure under translation: $%
Dz^{a}=D\breve{z}^{a},$ we may express (\ref{1PI effective action1}) as the
functional integral of fluctuations $\breve{z}^{a}$ around $\bar{z}^{a}.$ We
obtain 
\begin{align}
\Gamma\lbrack\bar{z}^{a}] & =-i\hbar\ln\text{{\LARGE [}}\int_{CTP}D\breve {z}%
^{a}\text{$\exp$}\frac{i}{\hbar}\text{{\Large \{}}S_{CGEA}^{a}[\bar{z}^{a}+%
\breve{z}^{a}]  \nonumber \\
& -\frac{1}{2}\text{Tr}\ln G^{ab}-\int d\tau\frac{\delta\Gamma}{\delta\bar {z%
}^{a}}\breve{z}^{a}\text{{\Large \}}}\rho(z_{i}^{a})\text{{\LARGE ]}}, 
\label{Effective action}
\end{align}
where $S_{CGEA}^{a}[\bar{z}^{a}+\breve{z}^{a}]=S_{z}^{a}\left[ z^{a}\right]
+S_{IF}[\bar{z}^{a}+\breve{z}^{a}]$ is the coarse-grained effective action
written in terms of the new variables $\breve{z}$ and $\bar{z}.$

Now the restriction on the particle-worldline path integrals means that even
the one-loop approximation goes beyond the exact Gaussian integrations. This
is a finite-size effect, which we discussed earlier. So that we can avoid
considering finite-size corrections here, we will assume that the initial
particle state is defined in the infinite past, and hence the path
integrations over the time coordinate $z^{0}\left( \tau \right) $ have the
full range from minus to plus infinity. Then the effective action has the
formal solution 
\begin{align}
\Gamma \left[ \bar{z}^{a}\right] & =S_{CGEA}\left[ \bar{z}^{a}\right] -\frac{%
i\hslash }{2}\ln \det G^{ab}-\frac{i\hslash }{2}\ln \det \left(
B_{ab}^{-1}\right)   \nonumber \\
& +\Gamma _{1}\left[ \bar{z}^{a}\right] 
\end{align}
where

\begin{equation}
iB_{ab}(\tau ,\tau ^{\prime })_{\bar{z}}=\frac{\delta ^{2}S_{CGEA}}{\delta
z^{a}(\tau )\delta z^{b}(\tau ^{\prime })}[\bar{z}^{a}]
\label{one loop propagator}
\end{equation}
is the one-loop propagator for the particle fluctuations $\tilde{z}^{a}$
around the classical solutions. $S_{CGEA}\left[ \bar{z}^{a}\right] $ is the
same effective action found from the influence functional method. $\Gamma
_{1}[\bar{z}^{a}]$ contains two-loop and higher quantum particle corrections
that arise in nonlinear theories. It is given by the sum of all
one-particle-irreducible graphs with external legs determined by $\hat{\rho}%
_{i}(z^{a})$, internal propagators given by $B_{ab}(\tau ,\tau ^{\prime })_{%
\bar{z}},$ and vertices given by the shifted interaction term for the
fluctuation coordinate $\tilde{z},$

\begin{align}
S_{int}[\breve{z}^{a}] & =S_{CGEA}[\breve{z}^{a}+\bar{z}^{a}]-S_{CGEA}[\bar{z%
}^{a}]  \nonumber \\
& -\int d\tau\left( \frac{\delta S_{CGEA}}{\delta z^{a}}\left[ \bar{z}^{a}%
\right] \right) \breve{z}^{a}\left( \tau\right)  \nonumber \\
& -\frac{1}{2}\int d\tau\int d\tau^{\prime}\left( \frac{\delta^{2}S_{CGEA}}{%
\delta z^{a}(\tau)\delta z^{b}(\tau^{\prime})}\left[ \bar{z}^{a}\right]
\right)  \nonumber \\
& \times\breve{z}^{a}(\tau)\breve{z}^{b}(\tau^{\prime}). 
\label{Particle vertices}
\end{align}

From (\ref{one loop propagator}) and (\ref{Particle vertices}), it follows
that the sum of graphs for $\Gamma_{1}$ have vertices and propagators that
depend on $\bar{z}^{a}.$ In terms of Feynman diagrams, the graphs that are
included in $\Gamma_{1}$ have only particle propagator (\ref{one loop
propagator}) lines. The implicit dependence in the graphs on the integrated
field variables comes about through the dependence of the vertices on the
solutions $\bar{z}^{a},$ which in turn depend implicitly on the average
field properties through $\bar{z}$'s defining equation (\ref{dG/dz=g (CTP
equations of motion)}).

\section{Correspondence between Langevin and Heisenberg equations of motion}

The construction of the stochastic effective action made use of a Gaussian
identity to rewrite the real part of the influence functional as the
stochastic average of the characteristic function of a noise. But we might
still ask if all aspects of the field's quantum fluctuations are encoded in
the statistics of the noise, or if different states of the quantum field can
give the {\it same }noise in the stochastic limit. This possibility has been
pointed out by Raval in \cite{Raval96(Thesis)}. For a Brownian particle
linearly coupled to a scalar field, the Heisenberg equations of motion are 
\begin{equation}
\frac{d^{2}}{dt^{2}}\hat{q}(t)+\omega_{0}^{2}\hat{q}(t)^{2}=e\hat{\varphi }%
(x(t))
\end{equation}
and 
\begin{equation}
\partial^{\mu}\partial_{\mu}\hat{\varphi}({\bf x(}t),t)=-\hat{q}(t)\delta(%
{\bf x-x(t)),}   \label{Field Heseinberg equations of motion}
\end{equation}
where the particle, with internal degree of freedom $\hat{q},$ moves on a
fixed trajectory ${\bf x(}t).$ The causal solution to (\ref{Field Heseinberg
equations of motion}) in terms of the retarded Green's function plus a
homogeneous free field solution gives 
\begin{equation}
\frac{d^{2}}{dt^{2}}\hat{q}(t)-e\int dt^{\prime}G^{R}(t,t^{\prime})\hat {q}%
(t^{\prime})+\omega_{0}^{2}\hat{q}(t)=\hat{\varphi}_{0}({\bf x}(t),t). 
\label{Heseinberg eq of motion for  q}
\end{equation}
When $G^{R}(t,t^{\prime})$ is local (as is the case for massless fields), we
may integrate by parts and write 
\begin{equation}
\frac{d^{2}}{dt^{2}}\hat{q}(t)+ie\frac{d}{dt}\hat{q}(t^{\prime})+\omega
_{0}^{2}\hat{q}(t)=\hat{\varphi}_{0}({\bf x}(t),t)
\end{equation}
with the solution 
\begin{align}
\hat{q}(\omega) & =\left( \omega^{2}+ie\omega-\omega_{0}^{2}\right) ^{-1}%
\hat{\varphi}_{0}(\omega,{\bf x(}t))+\hat{q}_{0}(\omega )
\label{Heisenberg equations of motion in w} \\
& \equiv g(\omega)\hat{\varphi}_{0}(\omega,{\bf x(}t))+\hat{q}_{0}(\omega), 
\nonumber
\end{align}
where 
\begin{equation}
\hat{\varphi}_{0}(\omega,{\bf x}(t))=\int dt\,e^{i\omega t}\hat{\varphi }%
_{0}({\bf x}(t),t).
\end{equation}
We may now use (\ref{Heisenberg equations of motion in w}) to find both the
particle commutators (neglecting the homogeneous solutions $\hat{q}%
_{0}(\omega)),$%
\begin{align}
\left\langle \left[ \hat{q}(\omega^{\prime}),\hat{q}(\omega)\right]
\right\rangle & =g(\omega^{\prime})\,g(\omega) \\
& \times\left\langle \left[ \hat{\varphi}(\omega^{\prime},{\bf x(}t)),\hat{%
\varphi}(\omega,{\bf x(}t))\right] \right\rangle  \nonumber
\end{align}
and the particle anticommutators, 
\begin{align}
\left\langle \left\{ \hat{q}(\omega^{\prime}),\hat{q}(\omega)\right\}
\right\rangle & =g(\omega^{\prime})\,g(\omega) \\
& \times\left\langle \left\{ \hat{\varphi}(\omega^{\prime},{\bf x(}t)),\hat{%
\varphi}(\omega,{\bf x(}t))\right\} \right\rangle .  \nonumber
\end{align}
The corresponding Langevin equation replaces the homogeneous field operator
solution $\hat{\varphi}_{0}$ in (\ref{Heseinberg eq of motion for q}) with
the noise $\xi(s),$ and the quantum operators $\hat{q}(t)$ are replaced with
the stochastic variables $q_{\xi}(t)$. The solution otherwise follows as
above, with 
\begin{equation}
q_{\xi}(\omega)=g(\omega)\xi(\omega,{\bf x(}t)).
\end{equation}
Again, we drop the homogeneous solution to $q_{\xi}(\omega).$ The
symmetrized commutator of $q_{\xi}(\omega)$ is then given by 
\begin{align}
\left\langle \left\{ q_{\xi}(\omega^{\prime}),q_{\xi}(\omega)\right\}
\right\rangle & =g(\omega^{\prime})\,g(\omega)\,  \nonumber \\
& \times\left\langle \left\{ \xi(\omega^{\prime},{\bf x(}t)),\xi (\omega,%
{\bf x(}t))\right\} \right\rangle  \nonumber \\
& =g(\omega^{\prime})\,g(\omega)\,  \label{symmetrized fields} \\
& \times\left\langle \left\{ \hat{\varphi}(\omega^{\prime},{\bf x(}t)),\hat{%
\varphi}(\omega,{\bf x(}t))\right\} \right\rangle ,  \nonumber
\end{align}
where in going from the second to the last step we have used the equality of
the symmetrized noise and field operator commutators. But 
\begin{align}
\left\langle \left[ q_{\xi}(\omega^{\prime}),q_{\xi}(\omega)\right]
\right\rangle & =g(\omega^{\prime})\,g(\omega)\,\left\langle \left[
\xi(\omega^{\prime},{\bf x(}t)),\xi(\omega,{\bf x(}t))\right] \right\rangle 
\nonumber \\
& =0  \label{eq 2} \\
& \neq\left\langle \left[ \hat{q}(\omega^{\prime}),\hat{q}(\omega)\right]
\right\rangle .  \nonumber
\end{align}
Equation (\ref{eq 2}) follows as a consequence of 
\begin{equation}
\left\langle \lbrack\xi(t),\xi(t^{\prime})]\right\rangle =0.
\end{equation}
Therefore, it appears to be formally possible for the Heisenberg equations
of motion to differ from the Langevin equations by terms proportional to the
field commutator. The origin of this possible discrepancy between the
stochastic and Heisenberg correlation functions remains unclear. Raval has
speculated that it may be related in some way to the extremization principle
used to find the stochastic effective action \cite{Raval96(Thesis)}. Because 
$\left\langle \hat{\varphi}_{0}\right\rangle =\left\langle \xi\right\rangle
=0,$ the mean equations of motion will exactly agree, so any difference
between the stochastic and quantum theory will appear in the higher order
fluctuations.

%
%
%


\begin{figure}[tbh]
\begin{center}
\epsfig{file=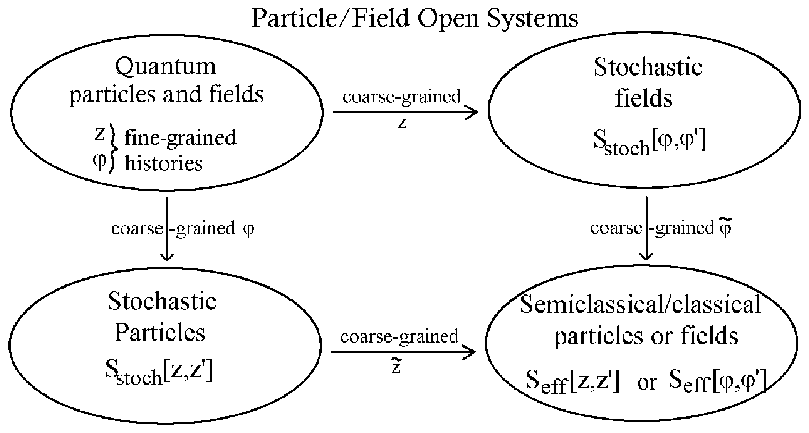, width=3 in}
\end{center}
\caption{A schematic depiction of the relationship between the classical,
semiclassical, stochastic, and quantum regimes.}%
\label{label1}%
\end{figure}

\begin{figure}[tbh]
\begin{center}
\epsfig{file=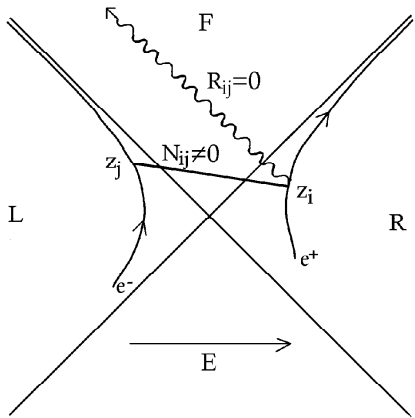, width=3 in}
\end{center}
\caption{Example of two oppositely charged particles in a background electric
field, whose semiclassical motions $\bar{z}_{e^{\pm}}$ are in causally
disjoint regions of spacetime. The radiation force $R_{12}$ vanishes, but the
correlation-noise $N_{12}$ does not (for explanation see Sec. 5).}%
\label{label2}%
\end{figure}

\begin{figure}[tbh]
\begin{center}
\epsfig{file=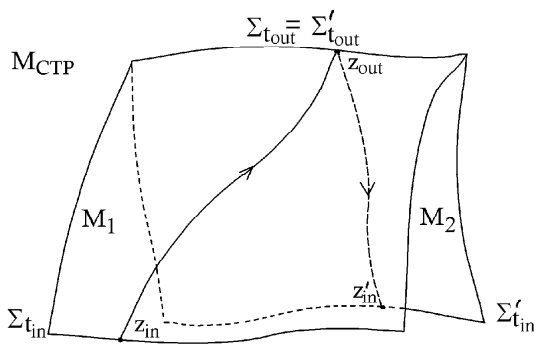, width=3 in}
\end{center}
\caption{The closed-time-path manifold $M_{CTP}$ is composed of two copies of
the ordinary spacetime manifold, $M_{1}$ and $M_{2},$ joined at a final-time
hypersurface $\Sigma\left(  t_{f}\right)  .$}%
\label{label3}%
\end{figure}

\end{document}